\documentclass{aa}
\usepackage{natbib}
\bibpunct{(}{)}{;}{a}{}{,} 
\usepackage{graphicx}
\usepackage{hyperref}
\usepackage{xcolor}
\usepackage{txfonts}

\usepackage{amsmath}	
\usepackage{orcidlink}
\usepackage{longtable}
\usepackage{comment}

\makeatletter
\renewcommand*\aa@pageof{, page \thepage{} of \pageref*{LastPage}}
\makeatother

\newcommand{\msolar}{M$_{\odot}$}

\begin{document} 

\title{SN 2026dix: a nearby, transitional Type IIb/Ib supernova consistent with a warm supergiant progenitor}

\author{
\orcidlink{0000-0003-4610-1117}T. Szalai\inst{1,2} 
\and
\orcidlink{0000-0001-9038-9950}S.~D.~Van Dyk\inst{3} \and
\orcidlink{0009-0000-9929-7518}D. B{\'a}nhidi\inst{1,4,5} \and 
\orcidlink{0000-0002-9324-3903}A.~P. Nagy\inst{1} \and
\orcidlink{0000-0002-8770-6764}R. K\"onyves-T\'oth\inst{6,7} \and
\orcidlink{0000-0003-3460-0103}A.~V. Filippenko\inst{8} \and
\orcidlink{0000-0001-5955-2502}T.~G. Brink\inst{8} \and
\orcidlink{0000-0002-2636-6508}W. Zheng\inst{8} \and
\orcidlink{0000-0001-9061-2147}I.~B. B{\'i}r{\'o} \inst{4,5} \and
\orcidlink{0009-0003-3711-6226}I. Cs{\'a}nyi\inst{4} \and
\orcidlink{0000-0001-6232-9352}Zs. Bora\inst{6,7,9} \and
\orcidlink{0009-0002-0157-4228}Zs. Horv{\'a}th\inst{6,7,9} \and
\orcidlink{0009-0008-2052-8474}A. Horti-D{\'a}vid\inst{6,7,9} \and
\orcidlink{0000-0001-5203-434X}A.~P. Jo{\'o} \inst{6,7,9} \and
\orcidlink{ 0009-0008-3094-5060}D. Koller\inst{6,7,9} \and
\orcidlink{0000-0002-1792-546X}L. Kriskovics\inst{6,7} \and
\orcidlink{0009-0007-3760-515X}K. Lelkes\inst{6,7,9} \and
K. Nagy\inst{6,7,9} \and
\orcidlink{0000-0003-0926-3950}K. S{\'a}rneczky \inst{6,7} \and
\orcidlink{0000-0001-7806-2883}{\'A.} S{\'o}dor \inst{6,7} \and
\orcidlink{0000-0002-1698-605X}R. Szak{\'a}ts \inst{6,7} \and
\orcidlink{0000-0001-5449-2467}A. P{\'a}l \inst{6,7} \and
\orcidlink{0000-0001-8764-7832}J. Vink{\'o} \inst{1,6,7,9}
}
     
\institute{
Department of Experimental Physics, Institute of Physics, University of Szeged, D{\'o}m t{\'e}r 9, 6720 Szeged, Hungary \\ 
\email{szaszi@titan.physx.u-szeged.hu}
\and
MTA-ELTE Lend\"ulet ``Momentum'' Milky Way Research Group, Szent Imre H. st. 112, 9700 Szombathely, Hungary 
\and
Caltech/IPAC, Mailcode 100-22, Pasadena, CA 91125, USA 
\and
Baja Astronomical Observatory of University of Szeged, Szegedi {\'u}t, Kt. 766, 6500 Baja, Hungary 
\and
HUN-REN--SZTE Stellar Astrophysics Research Group, Szegedi {\'u}t, Kt. 766, 6500 Baja, Hungary 
\and
HUN-REN Research Centre for Astronomy and Earth Sciences, Konkoly Observatory, Konkoly Th. M. {\'u}t 15-17., 1121 Budapest, Hungary 
\and 
CSFK, MTA Centre of Excellence, Konkoly Thege Mikl{\'o}s {\'u}t 15-17, 1121 Budapest, Hungary 
\and
Department of Astronomy, University of California, Berkeley, CA 94720-3411, USA 
\and
ELTE E{\"o}tv{\"o}s Lor{\'a}nd University, Institute of Physics and Astronomy, P{\'a}zm{\'a}ny P{\'e}ter s{\'e}t{\'a}ny 1/A, Budapest, 1117, Hungary 
}

\date{Accepted XXX. Received YYY; in original form ZZZ}

  \abstract
   {Understanding the final evolution stages of (very) massive stars and their explosive outcomes, stripped-envelope supernovae (SESNe), represent a long-term challenge for astrophysics. The latest results support a continuous distribution within the traditional SESN subclasses (IIb, Ib, Ic).}
   {The nearby ($D \approx 17.5$ Mpc) SN~2026dix seems to be another member of the recently identified group of transitional Type IIb/Ib explosions. A point source is located at the SN position  within the uncertainties on multiple pre-explosion {\it HST} images. Together with post-explosion photometry and spectroscopy, it provides a good opportunity to study a rare type of SN in detail.}
   {We carried out a thorough comparative light-curve (LC) and spectral analysis of SN~2026dix. We also constructed its bolometric LC and modeled it semi-analytically. In addition, we  constructed the spectral energy distribution of the presumed progenitor and compared this to model stellar atmospheres and binary stellar evolution tracks.}
   {We infer that SN~2026dix arose from an explosion in an interacting binary system, consistent with the properties of a warm ($T_\textrm{eff} \approx 6750$--7750 K;  spectral type F2 to A7), luminous ($\log(L_{\rm bol}/L_{\odot})\approx 4.9$--5.3) supergiant primary and a less luminous, less massive hot dwarf companion. The nature of the identified progenitor closely resembles that of some other known cases of SNe IIb 
   and also well aligns with both the results of our semi-analytical LC modeling and the comparison of the LCs and spectra of SN~2026dix with the output of radiative-transfer models of an exploding star with $M_\textrm{ini}$ = 18 \msolar.}
 {}

   \maketitle
   \nolinenumbers
\section{Introduction}\label{sec:intro}

Core-collapse supernovae (CCSNe), cataclysmic events that are the consequences of the gravitational collapse of iron cores of massive ($> 8-10$ \msolar) stars \citep{Boccioli_2024_rev}, show a large variability in their observed properties. As revealed in the last few decades, the type and properties of a CCSN depend mainly on the degree of preservation of the outer layers during the pre-explosion mass-loss processes. The late-time evolution of stars depends primarily on their initial masses, but it may also be strongly affected by various means, such as metallicity or interactions with a companion star in a binary \citep[e.g.,][]{Smartt_2009,Smith_2014_rev,Limongi_2017_HSN,Laplace_2021}.

Understanding the final evolution stages and the explosion mechanism of (very) massive stars is especially challenging. These stars lose most of their outer layers before the explosion; thus, their final outcomes are called stripped-envelope (SE) SNe. In recent years, serious efforts have been made both in mapping the observable properties of SESNe \citep[e.g.,][]{Modjaz_2014,Bianco_2014,Liu_2016,Taddia_2018,Prentice_2019,Shivvers_2019,Fang_2022,Stritzinger_2023,Holmbo_2023} and in revealing the potential progenitor channels of these events \citep[e.g.,][]{Dessart_2015,Dessart_2016,Sukhbold_2016,Yoon2017,Woosley_2021,Dessart2024}.
Although traditionally SESNe are divided into three main observational subclasses (IIb: H-poor and He-rich, Ib: H-free but He-rich, and, Ic: H- and He-free, metal-rich events, see e.g. \citealp{Filippenko_97_rev}), the latest results support a continuous distribution of SESNe that may also extend further into interaction-dominated and superluminous events \citep[see reviews by][]{Gal-Yam_2017_HSN,Gangopadhyay_2026}.
SNe~IIb, first identified by \citet{Filippenko_1988}, are known to be heterogeneous.
As introduced by \cite{Chevalier_Soderberg_2010}, these events can be divided into two subgroups: SN~1993J and some similar explosions discovered later \citep[see, e.g.,][]{Szalai_2016} are thought to arise from massive stars with extended H envelopes (some hundreds of solar radii), while progenitors of other SNe~IIb (e.g., 2001ig, 2003bg, 2008ax) seem to have been much more compact (a few tenths of the solar radius). Members of these two subgroups are sometimes called SN~eIIb (extended IIb) and SN~cIIb (compact IIb), respectively (note, however, that intermediate cases, such as the famous SN~2011dh, are also known; \citealp{Arcavi_2011}). More recently, ``transitional'' events have been identified that seem to connect the cIIb and Ib subclasses \citep[see, e.g.,][]{Medler_2022,Gangopadhyay_2023,Dong_2024}.

The nearby ($D \approx 17.5$ Mpc) SN~2026dix seems to be another transitional Type IIb/Ib SN. It was discovered on 2026 February 16.39  (UTC dates are used throughout this paper; 61087.39 MJD) by the MASTER program \citep{MASTER_disc} in the face-on spiral galaxy NGC~3913 (at coordinates of [($\alpha$(J2000) = $11^{\rm hr}50^{\rm m}37.43^{\rm s}$, $\delta$(J2000) = $+55^\circ21'12.92''$], at redshift $z \approx 0.003$, adopted from NASA/IPAC Extragalactic Database (NED)\footnote{\href{https://ned.ipac.caltech.edu}{https://ned.ipac.caltech.edu}}). This is the third SN discovered in this galaxy after SNe 1963J and 1979B \citep{Barbon_1982}.
Very soon thereafter, SN~2026dix was classified as a young Type IIb SN, showing H and \ion{He}{i} absorption lines at $\sim$ 20,000 km$^{-1}$ in its early spectrum \citep{TNS_classific}.

Here we present the results of our detailed photometric and spectroscopic analysis of SN~2026dix and on its assumed progenitor star identified in pre-explosion {\it Hubble Space Telescope} ({\it HST}) images. The steps of data collection and reduction are described in Sec. \ref{sec:obs}. Sec. \ref{sec:anal}  presents the details of light-curve (LC) and spectral analysis, as well as analysis of the spectral energy distribution (SED) of the presumed progenitor star. We summarize our conclusions in Sec \ref{sec:concl}.

\section{Observations and data reduction}\label{sec:obs}

\subsection{Photometry}

\begin{figure}[h!]
\centering
\includegraphics[width=\columnwidth]{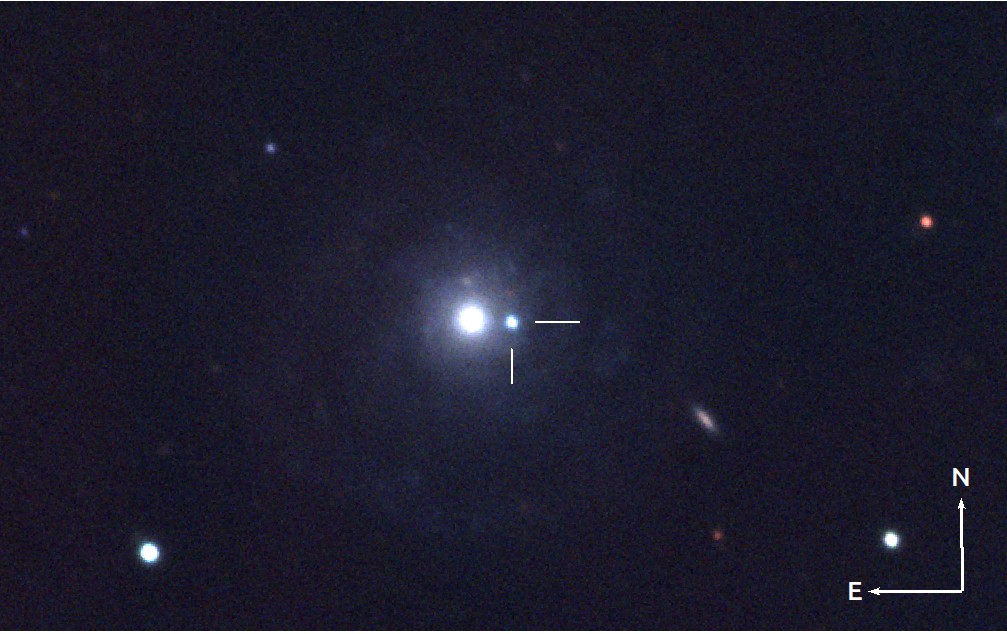}
\caption{Color composite image of SN~2026dix (in NGC~3913) from our BRC80 $gri$ images obtained on 2026-02-18 (2 days after discovery).}
\label{fig:sn_img}
\end{figure}

SN~2026dix was observed using two identical 0.8\,m Ritchy--Chrétien telescopes located at the Baja Observatory of the University of Szeged (hereafter BRC80) and at the Konkoly Observatory, Hungary (hereafter Konkoly). The telescopes were manufactured by the AstroSysteme Austria, and are each equipped with a back-illuminated, 2048 $\times$ 2048 pixel FLI PL230 CCD chip having a scale of 0.55\arcsec\ pixel$^{-1}$, and Johnson-Cousins {\it BV} and Sloan {\it griz} filters. With this telescope, photometric observations of nearby ($z < 0.1$) SNe can be obtained with a signal-to-noise ratio (S/N) $\gtrsim 10$.
 
Photometric data were processed with standard Image Reduction and Analysis Facility  \citep[IRAF;][]{Tody1986,1993ASPC...52..173T} routines, including basic corrections. Then we coadded three images per filter per night aligned with the {\tt wcsxymatch}, {\tt geomap}, and {\tt geotran} tasks. 
The instrumental magnitudes of the SN are calculated with the image-subtraction method using a pre-explosion Pan-STARRS1 Data Release 1 (PS1 DR1)\footnote{\href{https://catalogs.mast.stsci.edu/panstarrs/}{https://catalogs.mast.stsci.edu/panstarrs/}} image as the template.
This template image is properly scaled in size, flux, and full width at half-maximum intensity (FWHM) of the star profiles’ point-spread function (PSF)
with the {\tt geomap}, {\tt gregister}, {\tt psfmatch}, and {\tt linmatch} IRAF tasks.
Photometric calibration was carried out using stars from PS1 DR1.
To obtain reference magnitudes for our $B$- and $V$-band frames, the PS1 magnitudes were transformed into the Johnson $BVRI$ system based on equations and coefficients from \cite{Tonry12}. Finally, the instrumental magnitudes were transformed into standard $BVgriz$ magnitudes by applying a linear color term (using $g-i$) and wavelength-dependent zero points. Since the reference stars fell within a few arcminutes around the target, no atmospheric extinction correction was necessary. K- and S-corrections (\citealp{Blanton_2007}, \citealp{Stritzinger_2002}, respectively) were not applied. Further technical details are described by \cite{Banhidi_2025}.

\begin{figure*}[h!]
\centering
\includegraphics[width=0.85\textwidth]{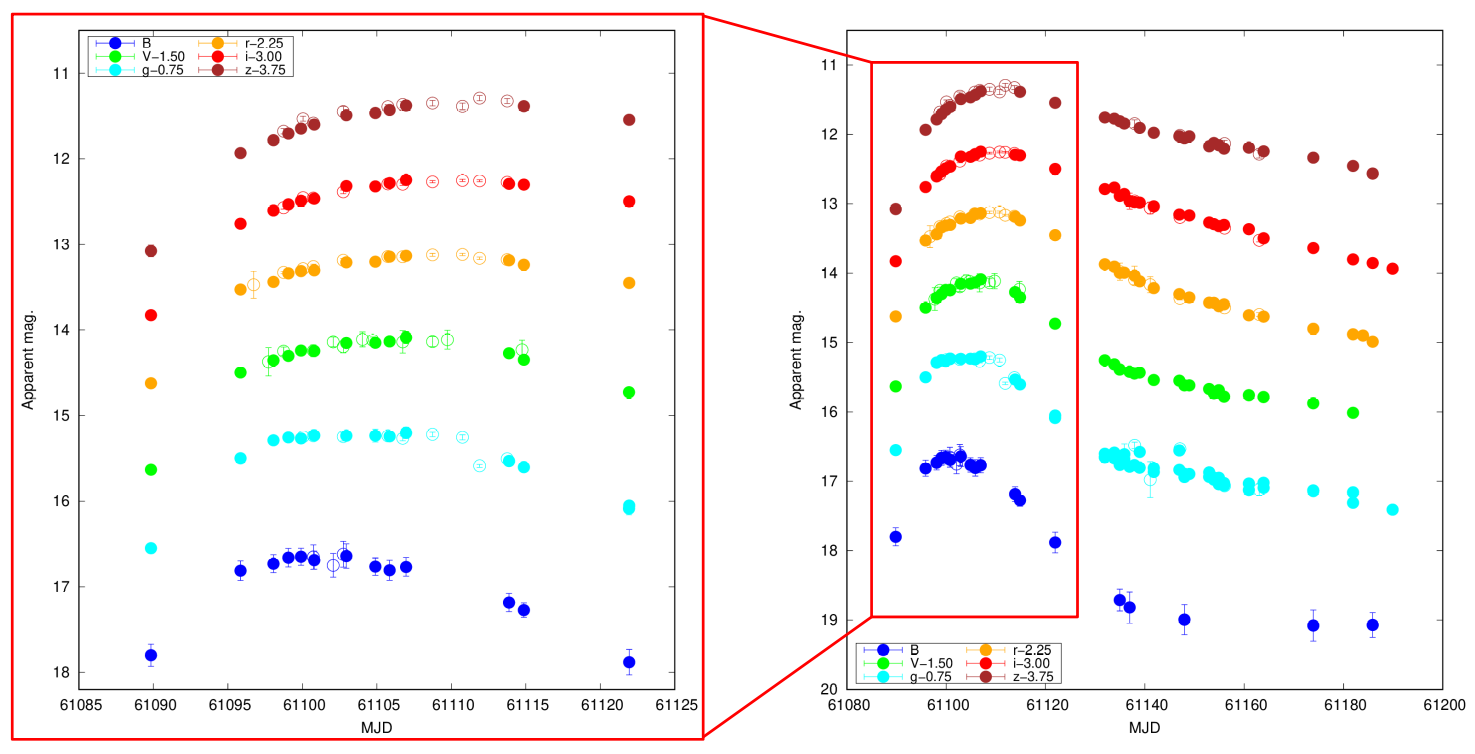}
\caption{$BVgriz$ photometry of SN~2026dix. Filled and open circles indicate data obtained at Baja and Konkoly Observatories, respectively.}
\label{fig:lcs}
\end{figure*}

A composite color image of SN~2026dix (obtained $\sim$ 2 days after discovery) is shown in Fig. \ref{fig:sn_img}.
$BVgriz$ data and LCs of SN~2026dix are presented in Tables \ref{tab:phot_data_Baja} and \ref{tab:phot_data_Konkoly}, and are shown in Fig. \ref{fig:lcs}.

\subsection{Spectroscopy}\label{sec:spec}

\begin{figure}[h!]
\centering
\includegraphics[width=\columnwidth]{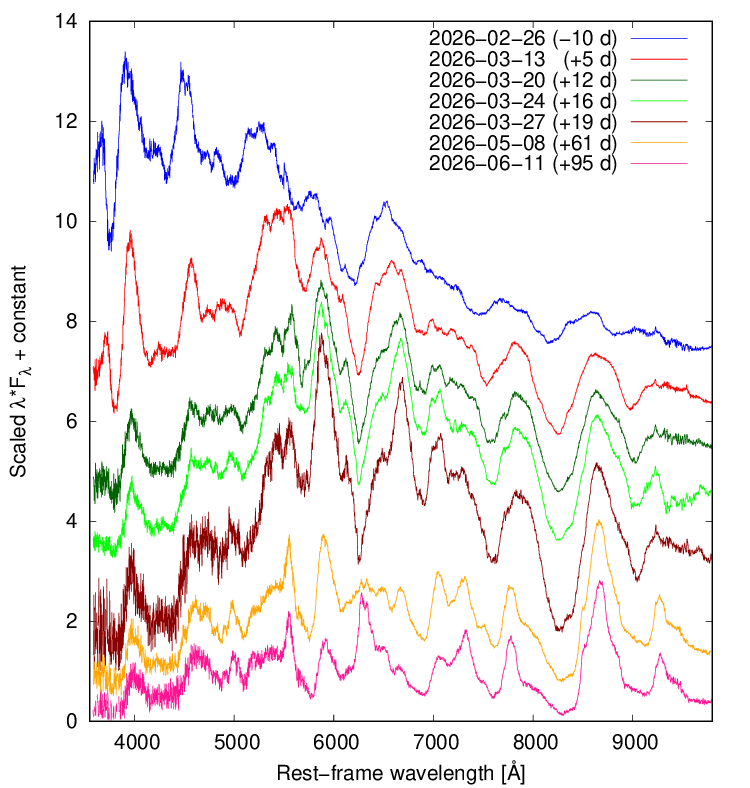}
\caption{De-reddened Lick/Shane+Kast optical spectra of SN~2026dix. Phases are given relative to the epoch of {\it V}-band maximum light (61106.9 MJD, 2026-03-08.9).}
\label{fig:sp_all}
\end{figure}

Spectra of SN~2026dix were obtained with the Kast double spectrograph on the Shane 3\,m telescope at Lick Observatory, California (hereafter Lick/Shane+Kast), at seven different epochs (see Table \ref{tab:spec} and Fig. \ref{fig:sp_all}); we used a 2\arcsec-wide slit, the D57 dichroic, the 600/4310 grism, and the 300/7500 grating. 
This configuration has a combined wavelength range of $\sim 3600$--10,700 \AA\ and a spectral resolving power $ R \approx 800$. The slit was oriented at or near the parallactic angle in order to minimize slit losses caused by atmospheric dispersion \citep{Filippenko1982}.

Data were reduced following standard techniques for CCD processing and spectrum extraction based on standard {\tt IRAF} routines and custom Python and {\tt IDL} software\footnote{ \href{https://github.com/ishivvers/TheKastShiv}{https://github.com/ishivvers/TheKastShiv}} \citep{Silverman2012}. Low-order polynomial fits to comparison-lamp spectra were established to calibrate the wavelength scale, and small adjustments attained from night-sky lines in the target frames were applied. Spectra were flux-calibrated via observations of spectrophotometric standard stars observed on the same nights at similar airmasses, and in an identical instrumental configuration. The standard-star spectra were also used to remove telluric absorption.

\begin{table}
\begin{center}
\caption{Log of spectroscopic observations obtained with Lick/Shane+Kast.}
\label{tab:spec}
\begin{tabular}{ll}
\hline
\hline
Date & Phase$^a$ \\
~ & (days) \\
\hline
2026-02-26 & $-10$ \\
2026-03-13 & $+5$ \\
2026-03-20 & $+12$ \\
2026-03-24 & $+16$ \\
2026-03-27 & $+19$ \\
2026-05-08 & $+61$ \\
2026-06-10 & $+95$ \\
\hline
\end{tabular}
\end{center}
\smallskip
{\bf Notes.}$^a$ With respect to the {\it V}-band maximum. All spectra have a range of 3600$-$10\,700 \AA\ and a resolution of $R=800$.
\end{table}

\section{Analysis} \label{sec:anal}

\subsection{Distance \& reddening estimations}\label{sec:anal_d_ebv}

There are several recent distance estimates of the host galaxy (NGC~3913) in the literature.
\cite{Leroy_2019} presented a multiwavelength study of local galaxies and also made efforts to synthesize their distance determinations by adopting and combining redshift-independent and Hubble-flow distances from large-scale galaxy compilations -- the Lyon/Meudon Extragalactic Database \citep[LEDA;][]{Paturel_2003a,Paturel_2003ab,Makarov_2014}, the Extragalactic Distance Database \citep[EDD;][]{Tully_2009}, and the CosmicFlows project \citep{Courtois_2012,Tully_2016}. They  corrected for local motions toward the Virgo cluster, giving a distance of $17.10\pm0.12$ Mpc for NGC~3913. In later studies, we found luminosity-distance estimates of $17.8\pm0.4$ Mpc \citep{Abreu_2022} and $17.9\pm7.5$ Mpc \citep{Davis_2024}.
By combining all these estimates, here we adopt a value of $D=17.5\pm0.5$ Mpc ($\mu=31.22\pm0.06$ mag).

The Galactic extinction toward SN~2026dix is quite low ($E(B-V)_\textrm{gal}=0.01$ mag, based on \citealp{SF11}). Owing to the lack of early-time high-resolution spectra, we cannot estimate the amount of host extinction from the equivalent width of the \ion{Na}{i} D line. Instead, we used the SESN color-curve templates from \cite{Stritzinger_2018} to estimate the total extinction. We compared the $B-V$, $V-r$, and $V-i$ color curves of SN~2026dix to both SN~IIb and SN~Ib templates, finding that $E(B-V)_\textrm{tot}=0.6\pm 0.1$ mag is consistent for all three curves (see Fig. \ref{fig:color_ebv}). Thus, we applied this value for the total extinction during our work.

\subsection{Spectral analysis}\label{sec:anal_sp}

As a first step, we corrected our observed spectra for redshift and extinction, and compared them to a large sample of SN templates in SNID-SAGE (SuperNova IDentification-Spectral Analysis and Guided Exploration; \citealp{SNID-SAGE_2026}). According to the original spectral classification, the early-phase spectra of SN~2026dix show the best match with those of Type IIb events (see also Sec. \ref{sec:anal_sp_syn++)}). Nevertheless, around the epoch of {\it V}-band maximum, the spectral evolution of SN~2026dix changes; after a while, its spectra seem to better resemble those of SN~Ib. 

Such transitions between the spectral evolution of Type IIb and Type Ib SNe have already been revealed, as in the cases of SN 2020acat \citep{Medler_2022,Ergon_2024} and SN 2022crv \citep{Gangopadhyay_2023,Dong_2024}. Thus, we downloaded the available spectra of these events from the WISeREP\footnote{\href{https://www.wiserep.org/}{https://www.wiserep.org/}} \citep{Yaron_2012} catalog to perform a comparative spectral and LC analysis (see the latter in Section \ref{sec:anal_lc}). We also used observations of the ``normal'' SN IIb 2011dh and SN Ib 2009jf. From our SNID-SAGE runs, we found that spectra of SN 2009mg \citep[][originally classified as an SN~IIb]{Oates_2012} closely resemble those of SN~2026dix; thus, we suggest reclassifying this object as an SN~IIb/Ib transitional event.
Basic data for all of these SNe are found in Table \ref{tab:sne}. For the spectral comparison plots, we give the phases according to the epoch of the {\it V}-band maximum (61106.9 MJD, 2026-03-08.9; see Sec. \ref{sec:anal_lc}).

\begin{table*}
\begin{center}
\caption{SN data used for our comparative analysis with SN~2026dix.}
\label{tab:sne}
\begin{tabular}{lcccccc}
\hline
\hline
Name & Type & Explosion Date & $z^b$ & $D$ & $E(B-V)_\textrm{tot}$ & Sources$^c$ \\
~ & ~ & (MJD) & ~ & (Mpc) & (mag) & ~ \\
\hline
{\bf SN~2026dix} & {\bf IIb/Ib} & {\bf 61086.45 $\pm$ 0.45} & {\bf 0.003} & {\bf 17.5 $\pm$ 0.5} & {\bf 0.6 $\pm$ 0.1} & {\bf This work} \\
SN~2009jf & Ib & 55102.3$^a$ & 0.008 & 34.3 $\pm$ 4.2 & 0.11 & 1 \\
SN~2009mg & IIb/Ib(?) & 55172.9$^a$ & 0.008 & 32.8 $\pm$ 2.3 & 0.14 & 2 \\
SN~2011dh & IIb & 55712.5 $\pm$ 0.3 & 0.002 & 8.05 $\pm$ 0.35 & 0.035 & 3-5 \\
SN~2016gkg & IIb & 57651.7 $\pm$ 0.1 & 0.005 & 26.4 $\pm$ 5.3 & 0.09 & 6, 7 \\
SN~2020acat & IIb/Ib & 59191.6 $\pm$ 1.0 & 0.008 & 35.3 $\pm$ 4.4 & 0.02 & 8, 9 \\
SN~2022crv & IIb/Ib & 59627.8 $\pm$ 0.5 & 0.008 & 34.4 $\pm$ 2.4 & 0.22 & 10, 11 \\
SN~2024abfo & IIb & 60628.28 $\pm$ 0.02 & 0.004 & 10.85 $\pm$ 1.30 & 0.01 & 12-14 \\
\hline
\end{tabular}
\end{center}
\smallskip
{\bf Notes.} $^a$Estimated from $B$- and $V$-band maxima. $^b$Redshifts are adopted from NED. $^c$Data sources: (1) \cite{Sahu_2011}; (2) \cite{Oates_2012}; (3) \cite{Sahu_2013}; (4) \cite{Ergon_2014}; (5) \cite{Marion_2014}; (6) \cite{Tartaglia2017}; (7) \cite{Bersten2018}; (8) \cite{Medler_2022}; (9) \cite{Ergon_2024}; (10) \cite{Gangopadhyay_2023}; (11) \cite{Dong_2024}; (12) \cite{Niu2025}; (13) \cite{Reguitti2025}; (14) \cite{deWet2025}. 
\end{table*}

\begin{figure*}[h!]
\centering
\includegraphics[width=0.45\textwidth]{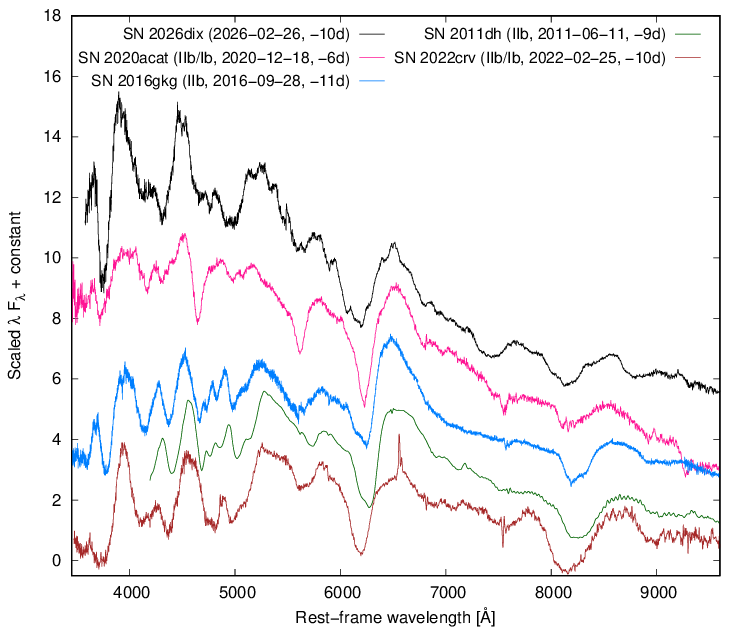}
\includegraphics[width=0.45\textwidth]{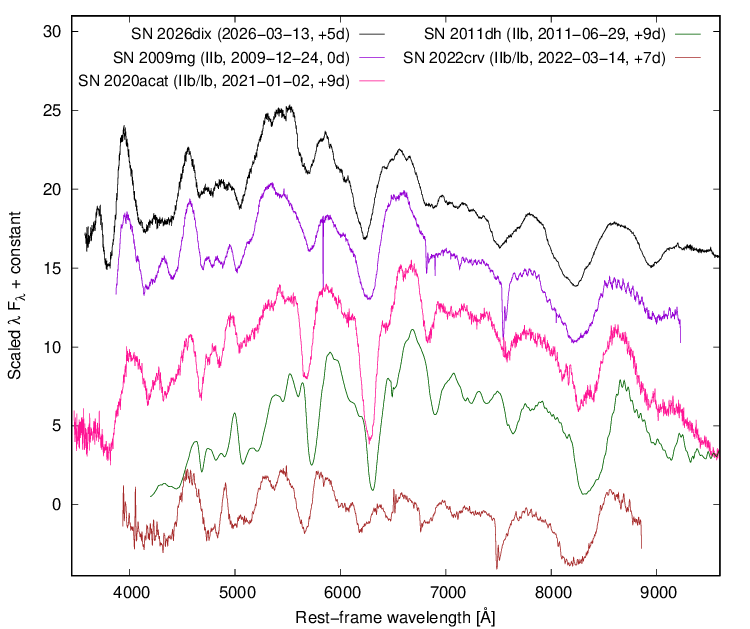}
\includegraphics[width=0.45\textwidth]{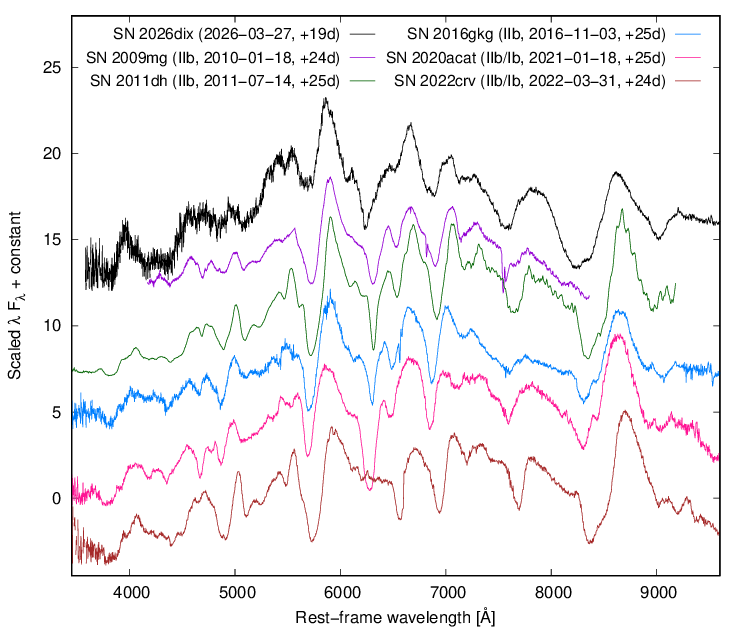}
\includegraphics[width=0.45\textwidth]{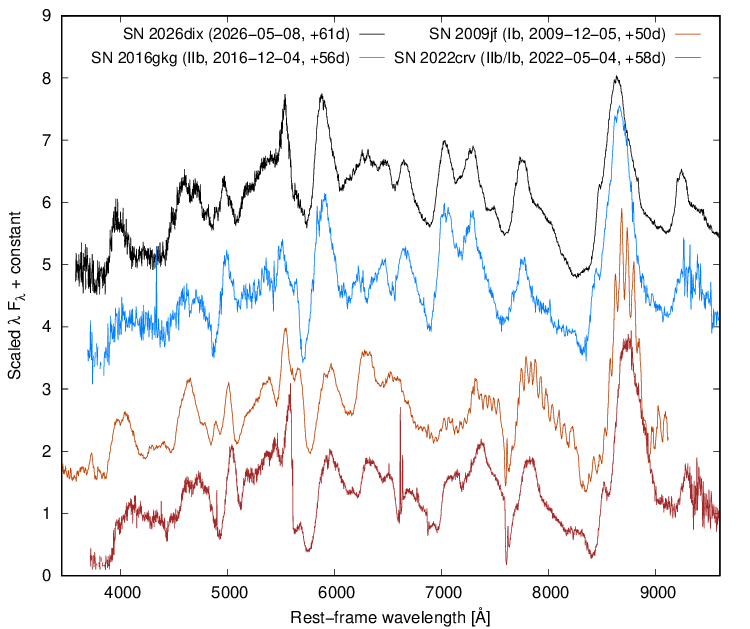}
\includegraphics[width=0.45\textwidth]{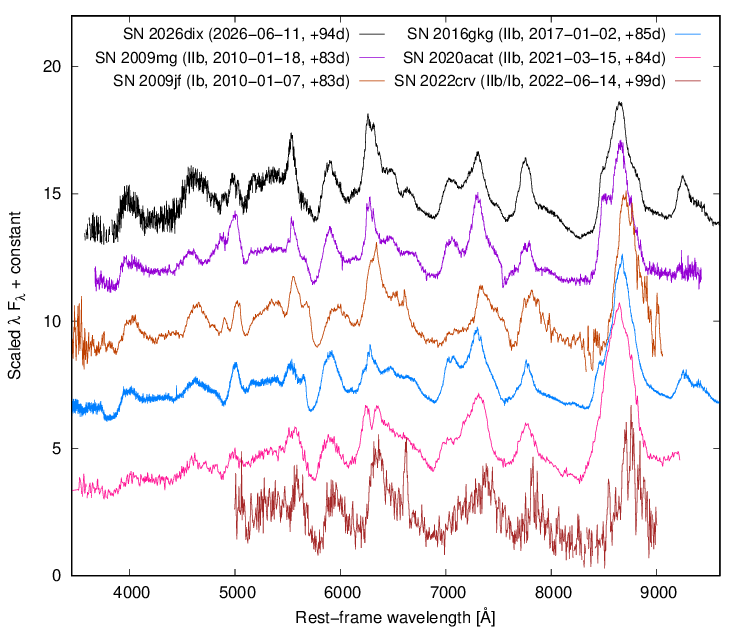}
\caption{Comparison of dereddened spectra of SN~2026dix to those of ``normal'' SNe IIb 2011dh and 2016gkg, of SN Ib 2009jf, as well as of IIb/Ib transitional SNe 2009mg, 2020acat, and 2022crv. Phases are given with respect to {\it V}-band maximum light. Spectra of the comparison SNe were downloaded from WISeREP; sources of data are given in Table \ref{tab:sne}. The +56d spectrum of SN~2016gkg, as well as the +58d and +99d spectra of SN~2022crv are binned (with a window of 5, 2, and 5 \AA, respectively).}
\label{fig:sp_comp}
\end{figure*}

\subsubsection{SYN++ analysis}\label{sec:anal_sp_syn++)}

We applied the parameterized resonance-scattering code {\tt SYN++}\footnote{\href{https://c3.lbl.gov/es/}{https://c3.lbl.gov/es/}} \citep{thomas11} for the analysis of the spectra of SN~2026dix (all the observed data were corrected for both extinction and redshift before starting the analysis). The code allows estimates of some global parameters, such as photospheric temperature ($T_{\rm ph}$) and velocity  ($v_{\rm ph}$). The contribution of the single ions can be taken into account by setting some local parameters for each identified ion: the optical depth ($\tau$), the minimum and maximum velocity of the line-forming region ($v_{\rm min}$ and $v_{\rm max}$, respectively), the scale height of the optical depth above the photosphere (aux), and the excitation temperature ($T_{\rm exc}$).

We ran a range of models around each spectrum individually and then established the best comparison model.
$v_{\rm ph}$ values were searched in the ranges of 10,000$-$20,000 km s$^{-1}$, of 6,000$-$12,000 km s$^{-1}$, and of 3,000$-$8,000 km s$^{-1}$ up to +5 days, between +12 and +19 days, and after +61 days, respectively. The checked $T_{\rm ph}$ range for the first spectrum was 10,000$-$20,000 K, while we examined the 3,000$-$10,000 K range for the later spectra. 
Fig. \ref{fig:syn++} shows the final results of {\tt SYN++} modeling.


The first spectrum, obtained on 2026-02-26 ($-10$ days relative to {\it V}-band maximum light), was modeled using \ion{Fe}{ii}, \ion{Ca}{ii}, \ion{Si}{ii}, \ion{Mg}{ii}, \ion{He}{i}, and \ion{H}{i} lines, exhibiting characteristics typical of Type IIb SNe. Both photospheric velocity and temperature were found to be exceptionally high, with $v_{\rm phot} \approx 16,000$ km s$^{-1}$ and $T_{\rm phot} \approx 14,000$ K.

By the next observation on 2026-03-13 ($+5$ days), the photospheric velocity had decreased to $\sim 12,000$ km s$^{-1}$ and the temperature to 9000 K. Although the best-match model utilized the same ions as in the first epoch, this spectrum required an \ion{H}{i} component with a high velocity of $16,000$ km s$^{-1}$.

The best-fit {\tt SYN++} model for the third spectrum, observed on 2026-03-20 ($+12$ days), includes \ion{Sc}{ii} in addition to the previously listed ions. This model yields $v_{\rm phot} \approx 11,000$ km s$^{-1}$ and $T_{\rm phot} = 6000$ K, with the \ion{H}{i} line velocity at $\sim 15,000$ km s$^{-1}$. The spectrum taken four days later, on 2026-03-24 (phase $+16$ days), is best fit by the same elemental composition, resulting in $v_{\rm phot} = 10,000$ km s$^{-1}$ and $T_{\rm phot} \approx 5600$ K, still retaining a high-velocity \ion{H}{i} component. By 2026-03-27 ($+19$ days), the photospheric velocity dropped to $\sim 7000$ km s$^{-1}$, while the temperature increased slightly to $\sim 6000$ K.

The subsequent spectrum, taken on 2026-05-08 ($+61$ days), reveals significant evolution: the \ion{He}{i} lines strengthened, while H I features disappeared, rendering the object spectroscopically similar to SNe~Ib. The best-match model for this epoch includes \ion{Fe}{ii}, \ion{Sc}{ii}, \ion{Ca}{ii},\ion{Mg}{ii}, \ion{O}{i}, and \ion{He}{i}, which yields $v_{\rm phot} \approx 6000$ km s$^{-1}$ and $T_{\rm phot} \approx 6000$ K.

The final observed spectrum, obtained on 2026-06-11 ($+95$ days), implies a further decline to $v_{\rm phot} \approx 5000$ km s$^{-1}$ and $T_{\rm phot} \approx 5500$ K. At this stage, \ion{He}{i} and \ion{Fe}{ii} lines became increasingly prominent.

In conclusion, our {\tt SYN++} modeling also shows that SN~2026dix initially resembled a Type IIb SN, but gradually evolved to a Type Ib SN.
However, its $v_\textrm{phot}$ values are considerably higher than those of typical SNe IIb during the pre-maximum phases and remain elevated until $\sim20$ days after maximum light, at which time they converge to velocities similar to those of the SN Ib 2009jf and of SN IIb/Ib transitional SN 2009mg (see bottom-right panel of Fig. \ref{fig:syn++}).
A similar comparison for $T_\textrm{phot}$ values is much more complicated, since there are only blackbody temperatures for most of the SNe in our comparison sample. Nevertheless, after maximum light, the evolution of $T_\textrm{phot}$ seems to agree with the published temperatures of SNe~IIb and SN~IIb/Ib-like objects.


\subsection{Light-curve and color evolution}\label{sec:anal_lc}

First, we estimated the date of first light ($t_0$). Looking at the pre-explosion photometry obtained by the BHTOM.space Global Telescope Network \citep{Majumdar_2026_BHTOM}, the epoch of the last nondetection was $\sim$10 days before the discovery. Nevertheless, photometric data published on the TNS website of the object\footnote{\href{https://www.wis-tns.org/object/2026dix}{https://www.wis-tns.org/object/2026dix}} show nondetections up to 61086.0 MJD and the first detection (in ZTF $r$ band) at 61086.9 MJD (half a day before discovery). Thus, we constrain $t_0 = 61086.45 \pm 0.45$ MJD as the date of explosion.

Next, we fit simple quadratic functions to the brightness values around the $BVgriz$ LC peaks of SN~2026dix to provide proper estimations of the peak magnitudes, peak times ($t_\textrm{peak}$), and rise times ($t_\textrm{rise}$; see Table \ref{tab:LC_peaks}). 

\begin{figure}
\centering
\includegraphics[width=\columnwidth]{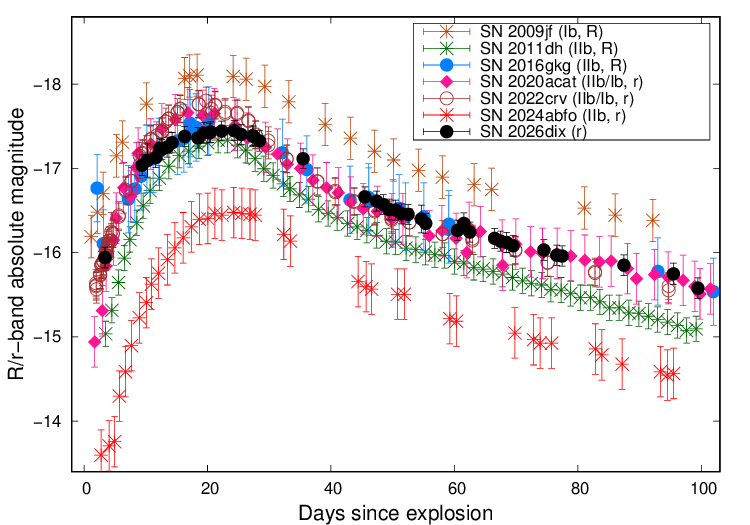}
\caption{Comparison of LCs of SN~2026dix to those of SNe IIb 2011dh, 2016gkg, and 2024abfo, as well as of SN Ib 2009jf, and of IIb/Ib transitional SNe 2020acat and 2022crv. Sources of data are given in Table \ref{tab:sne}.}
\label{fig:lc_comp}
\end{figure}

Both the peak absolute magnitudes and rise times of SN~2026dix are very similar to those of SNe IIb 2011dh \citep{Sahu_2013,Ergon_2014,Marion_2014} and 2016gkg \citep{Bersten2018}; nevertheless, its later-time LC evolution resembles much more that of known SN~IIb/Ib transitional objects (SNe 2009mg, 2020acat, 2022crv); see Figs. \ref{fig:lc_comp} and \ref{fig:22crv_lc_comp}. This is in agreement with the spectral evolution of SN~2026dix described in Sec. \ref{sec:anal_sp} and also suggests we found realistic values for both $E(B-V)_\textrm{tot}$ and $t_0$.
Note, however, that SN~2026dix 
was less luminous
in every filter than 
SNe 2020acat  \citep{Medler_2022,Ergon_2024} and 2022crv \citep{Gangopadhyay_2023,Dong_2024}; thus, it was likely a less energetic event (see also Sec. \ref{sec:anal_bol_model}). SN~2009mg \citep{Oates_2012}, our closest spectral analog of SN~2026dix, seems to be even fainter (however, it only has $B$ and $V$ photometry in the literature).

We also note that the observed LCs of SN~2026dix lack signs of the initial shock-cooling phase. This is not unexpected, since the compactness of the progenitor
causes the shock-cooling emission to decay very quickly. This is the general case for SNe Ib and was found also in the well-sampled LCs of the other transitional IIb/Ib SNe 2020acat and 2022crv, or in the case of Type cIIb 2024abfo \citep[where only the ultraviolet (UV) fluxes show the initial decrease; see][]{Reguitti2025}. 

\begin{figure}
\centering
\includegraphics[width=\columnwidth]{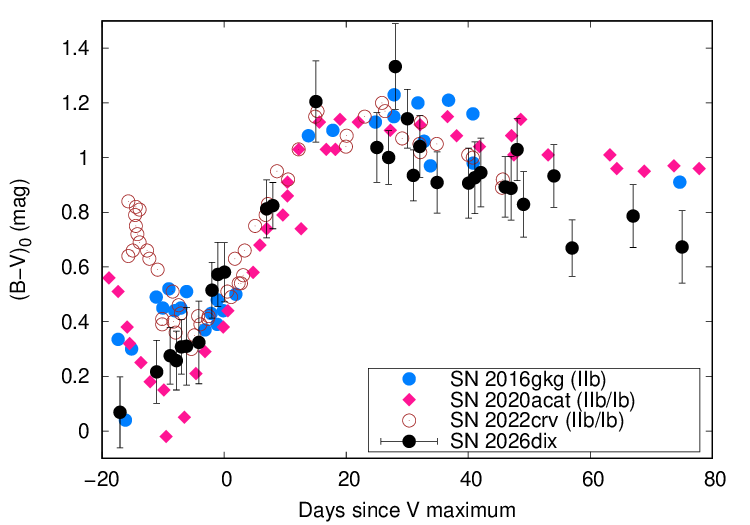}
\includegraphics[width=\columnwidth]{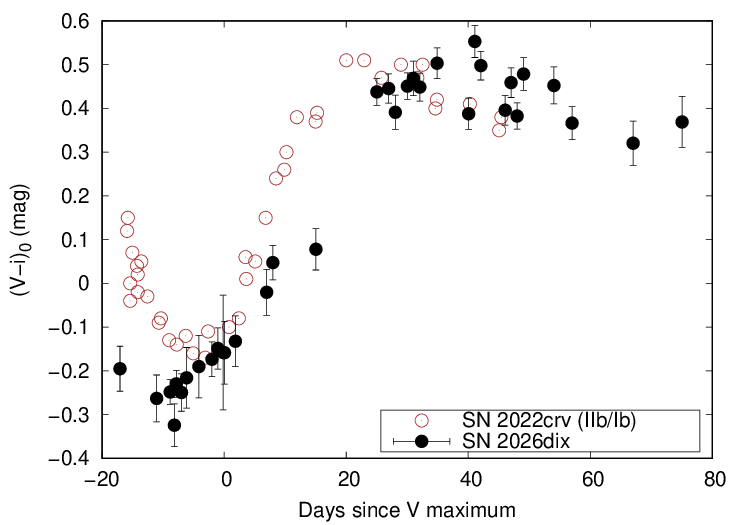}
\caption{Comparison of the reddening-corrected $B-V$ and $V-i$ color curves of SN~2026dix to those of IIb/Ib transitional SNe 2020acat and 2022crv and of SN IIb 2016gkg. Sources of data are given in Table \ref{tab:sne}.}
\label{fig:color_comp}
\end{figure}

While we already compared the color evolution of SN~2026dix to general SN IIb and SN~Ib templates, in order to estimate the total extinction (see Fig. \ref{fig:color_ebv} in Sec. \ref{sec:anal_d_ebv}), we also show a comparison of $(B-V)_0$ and $(V-i)_0$ curves of SN~2026dix to those of IIb/Ib transitional SNe 2020acat and 2022crv, as well as of SN~2016gkg (see Fig. \ref{fig:color_comp}); just as for their spectra and LCs, the color evolution of the three SNe looks very similar.

\subsection{Bolometric LC modeling}\label{sec:anal_bol}

\subsubsection{Constructing the bolometric LC of SN~2026dix}\label{sec:anal_bol_constr}

The bolometric LC of SN 2026dix was constructed using the \texttt{SuperBol} code \citep[developed by][]{Nicholl_2018}. First, we generated a pseudobolometric LC from  simple integration of the observed fluxes measured in the {\it BVr'i'z'} bands; we omitted $g'$-band data because of the large scatter at post-maximum epochs, and because of the overlap between the $BV$ and $g$ bands (note that \texttt{SuperBol} is prepared to handle this overlap).
The code corrects the magnitudes for extinction and converts them to flux densities. For the integration of the flux densities, the code uses a simple trapezoidal rule and assumes zero flux below and above the limit defined by the filter equivalent width of the bluest and reddest filters, respectively. Finally, the measured {\it BVr'i'z'} flux was converted into luminosity using the host distance given in Table \ref{tab:sne}.

Next, it is also necessary to estimate the flux contribution from the unobserved UV and infrared (IR) regimes. In the absence of such observed data, a basic step is to fit a blackbody (BB) function to the observed flux densities to estimate the UV/IR part of the spectral energy distributions (SEDs). 
While the IR contribution can be well estimated with the integration of the Rayleigh-Jeans tail of the fitted BB function, this is usually not the case in the UV, since the strong depletion caused by metal lines leads to depressed UV fluxes with respect to the BB function.
In \texttt{SuperBol}, there is an option to take into account the UV line blanketing, applying the formula $L_\textrm{UV}(\lambda < \lambda_\textrm{cutoff}) = L_\textrm{BB}(\lambda)*(\lambda/\lambda_\textrm{cutoff})^x$, where users can choose the values of the cutoff wavelength ($\lambda_\textrm{cutoff}$) and of the suppression index ($x$). We found that applying the value $x=4.0$ results in a reliable shape for the suppressed BB curve, assuming $\lambda_\textrm{cutoff}=4000$ \AA\ and 5000 \AA\ before and after {\it V}-band maximum brightness, respectively.
This ``final'' bolometric LC of SN~2026dix and the results of other approximations are shown in Fig. \ref{fig:lc_bol}. 

\subsubsection{Two-component semi-analytic modeling of the bolometric LC of SN~2026dix}\label{sec:anal_bol_model}

\begin{figure*}
\centering
\includegraphics[width=0.48\textwidth]{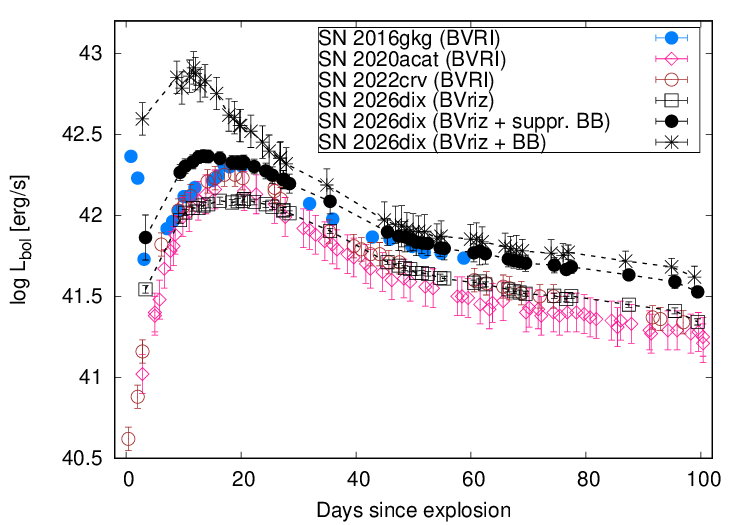}
\includegraphics[width=0.48\textwidth]{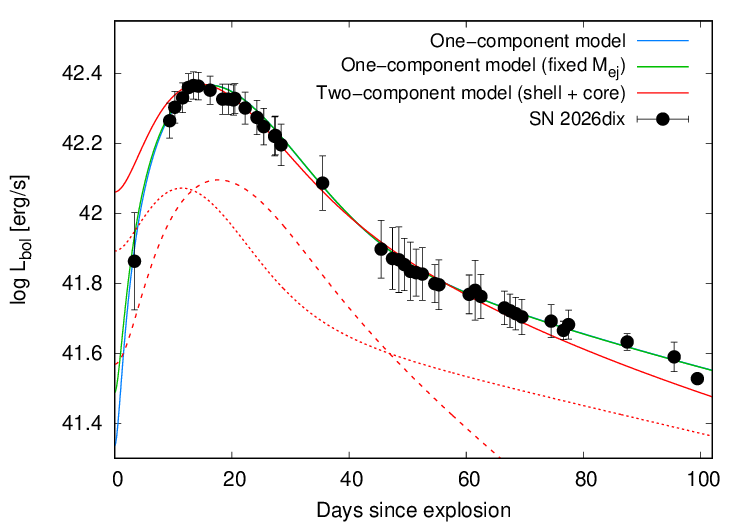}
\caption{{\it Left:} (Pseudo)bolometric LCs of SN~2026dix calculated with the {\texttt SuperBol} code \citep{Nicholl_2018}. Black open squares denote the ``pure'' pseudobolometric LC (i.e., integration of $BVriz$ fluxes), also compared to the $BVRI$ pseudobolometric LCs of SNe 2016gkg, 2020acat, and 2022crv (adopted from \citealp{Bersten2018}, \citealp{Medler_2022}, and \citealp{Gangopadhyay_2023}, respectively). Filled black circles and crosses denote the results of applying ordinary and suppressed BBs for taking into account the UV and IR contributions, respectively (see text for details). The bolometric LC evolution of SN~2026dix is also traced with dashed lines for better visibility.
{\it Right:} One- and two-component semi-analytic models compared to the bolometric LC of SN~2026dix; for the two-component model, dotted and dashed curves mark the ``shell'' and ``core'' components, respectively (see text for details.}
\label{fig:lc_bol}
\end{figure*}

Some of us \citep[][hereafter NV16]{NV16} previously developed a semi-analytical diffusion model for estimating the physical properties of SN explosions based on modeling their bolometric LCs. The model, based on the method of \citet{Arnett_Fu_1989}, assumes homologously expanding and spherically symmetric SN ejecta with a uniform density profile and also takes into account the effect of recombination. In general, the early shock cooling phase of SESN LCs cannot be fitted by the original Arnett-Fu model. Thus, our model, introduced in NV16, assumes a two-component structure for the SN ejecta: the first peak of the LC is dominated by the adiabatic cooling of the shock-heated, H-rich envelope (referred to as the 'shell'), while the second peak is powered by radioactive decay deposited in the denser inner region ('core'). 
The method was used by NV16 and also by other authors for analyzing the bolometric LCs of several SNe IIb (including e.g. SN~2022crv, see \citealp{Gangopadhyay_2023}).
During the analysis of the 'final' bolometric LC of SN~2026dix (i.e. integrated $BVr'i'z'$ fluxes + UV/IR contribution calculated by fitting suppressed BB functions, see Sec. \ref{sec:anal_bol_constr}), we started our modeling process with adopting the typical SN~IIb parameters from NV16.

As discussed above, the LC of SN~2026dix -- just as that of the other known IIb/Ib transitional SNe -- lacks the initial shock-cooling phase, which is basically described with the ``shell'' component in our modeling method. Thus, first, we applied single-component (i.e., pure ``core'') models,
finding them applicable for describing the bolometric LC of SN~2026dix if we take into account the rotational energy of a newborn magnetar. This is a long-supposed energy source in SESNe (see, e.g., \citealp{Kasen_Bildsten_2010}) and has been assumed during the LC modeling of several SNe IIb/Ib (see, e.g., \citealt{Bose_2021,Wang_2023,Dong_2024}). 
We initially fixed the average opacity of the single ``core'' component as $\kappa=0.2$ 
cm$^2$ g$^{-1}$
(Model IA), adopting the value found for H-poor, He-rich ejecta by NV16 and by \cite{Nagy_2018}. This model (see Table \ref{tab:bol_lc_models}) provides a good fit to the bolometric LC of the SN, as can be seen in the right panel of Fig. \ref{fig:lc_bol}, assuming ejecta and $^{56}$Ni masses of $M_{\rm ej}=2.2$ and $M_{\rm Ni}=0.053$ \msolar, respectively, thermal and kinetic energies of the ejecta of $E_{\rm th} = 2.3 \times 10^{51}$ and $E_{\rm kin} = 2.9  \times 10^{51}$ erg, respectively, and a progenitor radius of $R_0=2.0 \times 10^{11}$ cm.

As described by NV16 and \cite{Nagy_2018}, the average opacity of the ejecta ($\kappa$) is strongly correlated with several parameters in the Arnett-Fu model, especially with the ejecta mass ($M_{\rm ej}$). Thus, we repeated the single ``core''-component modeling with a fixed value of $M_{\rm ej}=3.1$ \msolar (being more consistent with the results of the hydrodynamical models and the direct progenitor analysis; see below). Applying this Model IB, we get a very similar result with a somewhat larger progenitor radius ($R_0 = 2.4 \times 10^{11}$ cm) and a slightly smaller kinetic energy ($E_{\rm kin} = 2.9 \times 10^{51}$ erg), and an average opacity of $\kappa=0.12$ 
cm$^2$ g$^{-1}$,
this latter value suggests ejecta that are (almost) H-free, but rather He- and metal-rich, consistent with the picture of an SN~IIb/Ib transitional object.

Moreover, we checked whether the original conception of a two-component (``core'' + ``shell'') model could work in the case of SN~2026dix. Here we worked again with fixed $\kappa$ values (0.18 and 0.34 
cm$^2$ g$^{-1}$
for the ``core'' and the ``shell'' components, respectively,  which mimics a H-poor core and a H-rich shell). Although this Model II curve cannot adequately fit the (poorly) observed early-phase LC evolution, it gives us some additional constraints on the ejecta.
Along with similar ``core'' parameters, we were required to assume a significant Ni mass ($M_{\rm Ni}=0.039$ \msolar) in the ``shell'' as well. This suggests strong mixing in the ejecta, which has been mentioned as a possible scenario in various hydrodinamical models of SESNe \citep[see, e.g.,][]{Bersten2012,Dessart_2016}.

We also compared our results to the LC-modeling outputs of similar SNe. \cite{Gangopadhyay_2023} applied the same NV16 model for SN~2022crv. Since this SN has a well-sampled early-phase LC, those authors were able to place stronger constraints on both the ``core'' and the ``shel'' components. Although their results differ in some regards from ours (i.e., obtaining somewhat larger kinetic energies and ejecta masses, and a very small progenitor radius of $R_0 \approx 0.3~R_{\odot}$), we note that they assumed a very low value for the ``core'' opacity ($\kappa = 0.06$ cm$^2$ g$^{-1}$),
which, based on our previous studies, assumes H and He-free ejecta (and, as mentioned above, strongly affects other model parameters).
Further ``Arnett-like'' LC modeling was carried out for SN~2022crv by \cite{Dong_2024} (also involving a magnetar as an extra energy source), as well as by \cite{Medler_2022} for SNe 2020acat and 2016gkg (and also for some other SNe~IIb). In general, they get similar parameters for these SNe (even though they also adopted the very low optical opacity values of $\kappa=0.06$--0.07 
cm$^2$ g$^{-1}$.

\begin{table*}
\caption{Parameters of the final semi-analytic bolometric LC models for SN~2026dix: Model IA (single ``core''-component model), Model IB (single ``core''-component model with fixed ejecta mass); Model II (two-component model consisting of a ``core'' and a ``shell" component).
}
\label{tab:bol_lc_models}
\centering
\renewcommand{\arraystretch}{2}
\begin{tabular}{l|cccccccc} 
\hline 
\hline
~ & $R_0$ & $M_{\rm ej}$ & $M_{\rm Ni}$ & $E_{\rm th}$ & $E_{\rm kin}$ & $\kappa$ & $E_{\rm p}$ & $t_{\rm p}$ \\
~ & (10$^{11}$ cm) & ($M_\odot$) & ($M_\odot$) & (10$^{51}$ erg) & (10$^{51}$ erg) & (cm$^2$ g$^{-1}$) & (10$^{51}$ erg) & (days) \\
\hline
Model IA & 2.0 & 2.2$^{+0.2}_{-0.3}$ & 0.053$^{+0.002}_{-0.003}$ & 2.3$^{+0.2}_{-0.3}$ & 2.9$^{+0.2}_{-0.1}$ & 0.20$^a$ & 0.012$^{+0.002}_{-0.003}$ & 3.0 $\pm$ 0.3 \\
Model IB & 2.4 & 3.1$^a$ & 0.053$^{+0.002}_{-0.003}$ & 2.3$^{+0.1}_{-0.3}$ & 2.8 $\pm$ 0.1 & 0.12$^{+0.01}_{-0.02}$ & 0.012$^{+0.001}_{-0.003}$ & 3.0$^{+0.3}_{-0.2}$ \\
~ & \multicolumn{8}{c}{Model II} \\
``core'' & 2.8 & 2.0$^{+0.1}_{-0.2}$ & 0.060$^{+0.10}_{-0.20}$ & 2.3$^{+0.2}_{-0.3}$ & 2.7$^{+0.6}_{-0.1}$ & 0.18 $\pm$ 0.1 & -- & -- \\
``shell'' & 5.2 & 0.3 & 0.039 & 2.1 & 0.6 & 0.34 & -- & -- \\
\hline
\end{tabular}
\tablefoot{$^a$Fixed value. Parameters are the progenitor radius ($R_0$), the ejecta mass ($M_{\rm ej}$), the initial mass of the radioactive $^{56}$Ni ($M_{\rm Ni}$), the thermal and kinetic energies ($E_{\rm th}$ and $E_{\rm kin}$), the average opacity of the ejecta ($\kappa$), and the spin-down energy of the neutron star ($E_{\rm p}$) and its characteristic timescale ($t_{\rm p}$).}
\end{table*}

\subsection{Comparison of spectra and LCs with Dessart+16 models}\label{sec:anal_D16}

We also compared the spectra, as well as the filtered and bolometric LCs of SN~2026dix, to the outputs of the radiative-transfer ({\tt CMFGEN}) models constructed for studying SNe IIb/Ib/Ic \cite[][hereafter D16]{Dessart_2016}.
Most of the $3p65$ and $4p64$ model configurations (both describe the assumed progenitors of SNe~IIb), as well as the $6p5$ models (SN~Ib progenitors), result in spectra near peak brightness that are reasonably similar to the observed (2026-03-13) spectrum of SN~2026dix (Fig. \ref{fig:D16_comp_sp_0313}. Similarity also persists at later times ($\sim$80 and $\sim$120 days after explosion; see Figs. \ref{fig:D16_comp_sp_0508} and \ref{fig:D16_comp_sp_0611}, respectively), but the flux levels of the $6p5$ models are significantly higher at these times than the observed ones. 

Comparison of the LC evolution of SN~2026dix with the D16 model output reveals a larger heterogeneity. In the {\it B} band, only $4p464D$ models are luminous enough to describe the peak of the LC; however, at longer wavelengths and in post-peak phases, the $4p64B$ and $4p64D$ models work better (note, however, that none of the models provides a perfect match during the whole observed phase range). The $4p64A$ models are subluminous, while the LC shapes of $6p5$ models are quite different from those of SN~2026dix.

Thus, we conclude that the $4p64B$ and $4p64D$ models are the most similar to our observations. These models represent the outcomes of the explosion of a single $M_{\rm ZAMS}$ = 18 $M_{\odot}$ progenitor (ZAMs is the zero age main sequence) with a final core mass of 4.6 \msolar, an ejecta mass of 3.2 \msolar, a $^{56}$Ni mass of 0.09--0.11 \msolar, a progenitor radius of $1.1 \times 10^{12}$ cm ($\sim 15~R_{\odot}$), and a kinetic energy of the ejecta of $\sim$ 2.5 and $5.1 \times 10^{51}$ erg (for $4p64B$ and $4p64D$, respectively). These progenitor and ejecta parameters
are consistent
with the results of our semi-analytical LC modeling (Sec. \ref{sec:anal_bol_model}). Note, however, that the M$_\textrm{Ni}$ and (for $4p64D$) the $E_{\rm kin}$ values are higher in the D16 models; the energy difference in our (one-component) semi-analytic models arises from the magnetic energy of the assumed neutron star.

\subsection{Progenitor identification and characterization}\label{sec:anal_prog}

\begin{figure*}[h!]
\centering
\includegraphics[width=\textwidth]{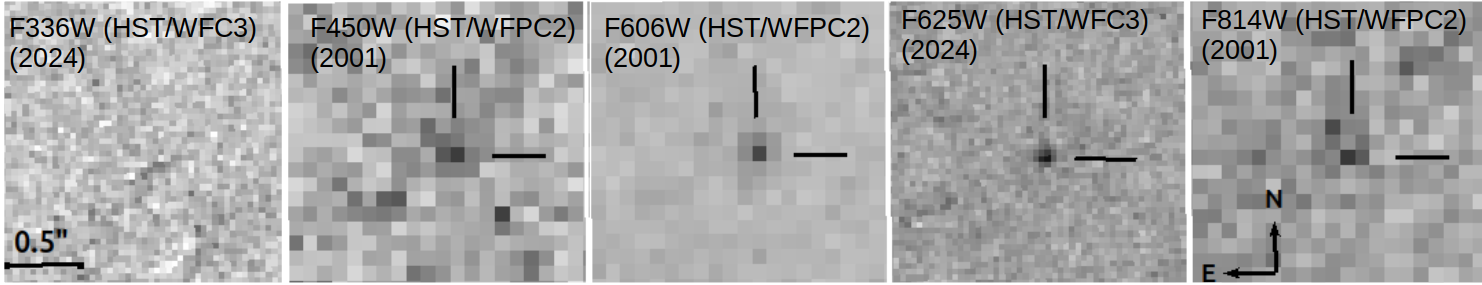}
\caption{Pre-explosion {\it HST} WFPC2 and WFC3 images (obtained in 2001 and 2024, respectively) of the site of SN~2026dix. Black lines mark the position of the assumed progenitor (see details in text).}
\label{fig:HST_obs}
\end{figure*}

The SN 2026dix site was serendipitously captured by {\it HST} with the Wide Field Planetary Camera 2 (WFPC2) on 2001 January 17 (GO-8559, PI T.~Boeker) in band F814W and on 2001 October 2 (GO-9042, PI S.~Smartt) in bands F450W and F606W, as well as with the Wide Field Camera 3 (WFC3) in the UVIS channel on 2024 January 25 (GO-17194, PI L.~Kelsey) in F336W and F625W (see Fig. \ref{fig:HST_obs}).


We used IRAF/PyRAF {\tt geomap} to establish the astrometric solutions based on 5 stars on the BRC80 2026-03-14 $r'$-band image, in common with the pre-explosion {\it HST} WFPC2 F814W HLA image (and also matching the two different HLA versions -- the 0.1\arcsec\ pixel$^{-1}$ WFPC2 mosaics and the 0.05\arcsec\ pixel$^{-1}$ PC mosaic -- to each other and then transforming the WFPC2 image position to the PC image). 
We found a plate solution with uncertainties of 1.67 WFPC2 pixels (= 0.167\arcsec\ or 167 milliarcsec). A progenitor candidate is within this uncertainty relative to the position of the SN.

After that, we employed the photometric methods outlined by \citet{VanDyk2024} to analyze the available {\sl HST\/} data. We first processed the  WFPC2 and WFC3 frames with {\tt Astrodrizzle} \citep{Avila2015} and then conducted the photometric Vega magnitude measurements with {\tt Dolphot} \citep{Dolphin2016}. The detection upper limit for F336W is at $5\sigma$, following \citet{VanDyk2023}. We also estimated an upper limit in this band using the Python routine {\tt space{\textunderscore}phot}\footnote{https://space-phot.readthedocs.io/en/latest/} \citep{Pierel2024} and received a similar value. The results are given in Table~\ref{tab:hstphot}.

\begin{table}
\begin{center}
\caption{HST photometry of the SN 2026dix progenitor candidate.}
\label{tab:hstphot}
\begin{tabular}{cccc}
\hline
\hline
Date & Instrument & Band & Magnitude \\
- & ~ & ~ & (Vega) \\
\hline
2024-01-25 & WFC3/UVIS & F336W & $>$24.43\\
2001-10-02 & WFPC2 & F450W & $25.63 \pm 0.26$ \\
2001-10-02 & WFPC2 & F606W & $24.68 \pm 0.09$ \\
2024-01-25 & WFC3/UVIS & F625W & $24.86 \pm 0.09$ \\
2001-01-17 & WFPC2 & F814W & $23.97 \pm 0.16$ \\
\hline
\end{tabular}
\end{center}
\end{table}

\begin{figure}[h!]
\centering
\includegraphics[width=\columnwidth]{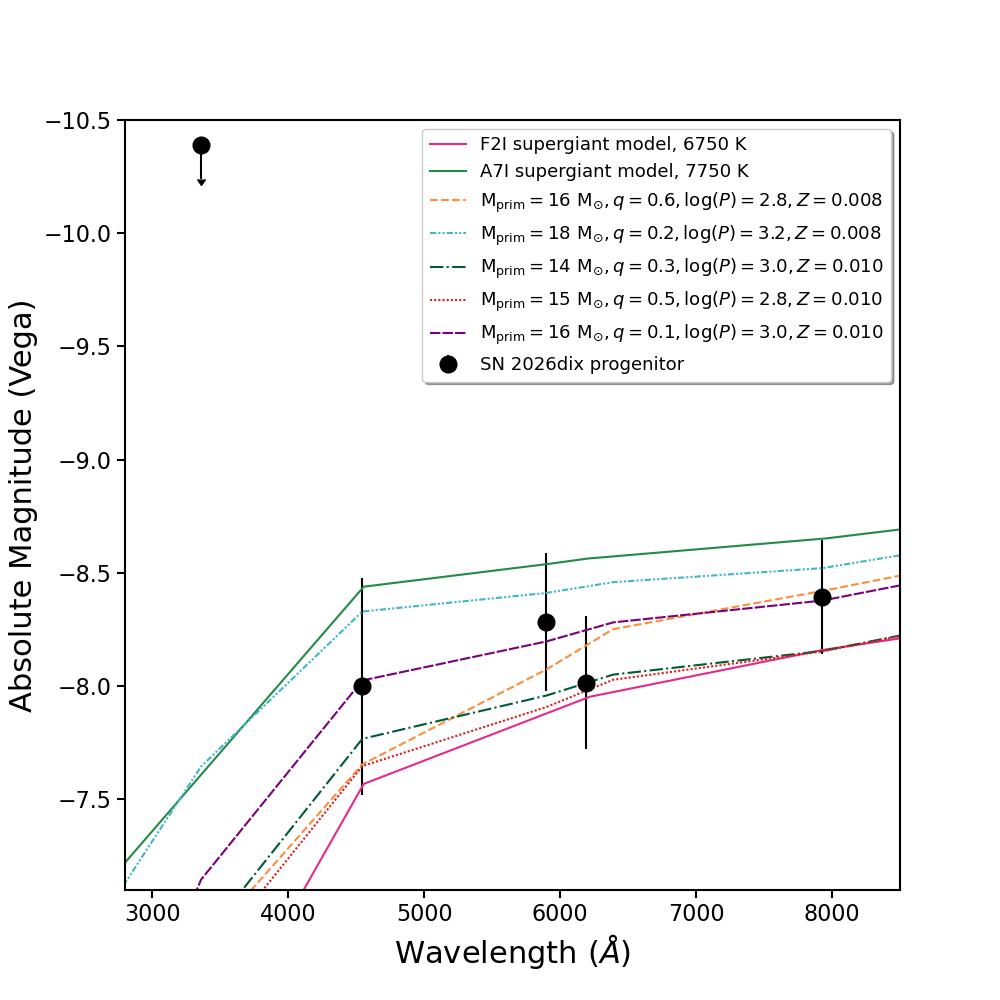}
\includegraphics[width=\columnwidth]{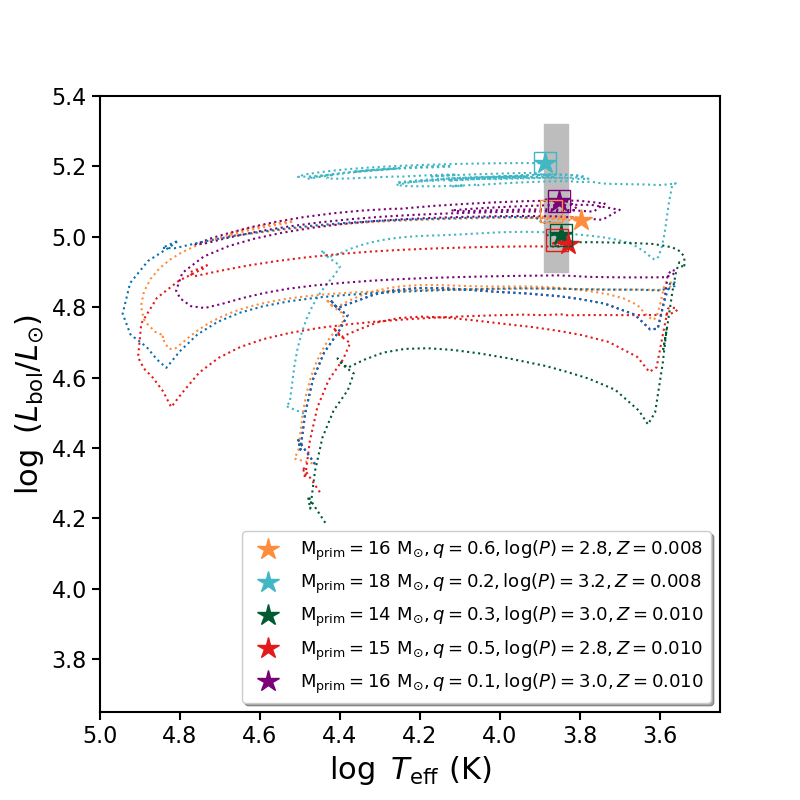}
\caption{{\it Upper panel:} Absolute brightness of the SN 2026dix progenitor candidate (solid circles). Shown for comparison are synthetic SEDs of 6750~K and 7750~K model supergiant stellar atmospheres \citep[approximate types F2\,I and A7\,I, respectively;][]{Castelli2003}, as well as synthetic SEDs at the termini of BPASS evolutionary tracks for model binary systems, with primary initial mass $M_{\rm prim}$, mass ratio $q$, and orbital period $(\log)P$ (in days), at metallicities $Z=0.008$ and $Z=0.010$ \citep{Stanway2018}. {\it Lower panel:} HRD showing the range in effective temperature ($T_{\rm eff}$) and bolometric luminosity ($L_{\rm bol}$) inferred from the SED fitting (gray shaded box). The same BPASS evolutionary tracks as in the left panel are shown for comparison. The total $T_{\rm eff}$ and $L_{\rm bol}$ of the binary systems are indicated as open squares, whereas the termini of the primary stellar component are  solid stars; the tracks of the primaries are  dotted curves.
}
\label{fig:HST_prog}
\end{figure}

The reddening to the progenitor candidate was assumed to be the same as to the SN itself. We therefore corrected the {\it HST} photometry for both the assumed total reddening and distance (see Section~\ref{sec:anal_d_ebv}). The total photometric uncertainties included the reddening uncertainty and the distance uncertainty added in quadrature. This corrected SED for the progenitor candidate is shown in Figure~\ref{fig:HST_prog} (left panel). We compared the SEDs derived from model stellar atmospheres for supergiants \citep{Castelli2003} to the candidate SED and found that, within the uncertainties, it could be ``fit'' by models with effective temperatures ($T_{\rm eff}$) in the range of 6750~K to 7750~K (see Fig.~\ref{fig:HST_prog}) -- that is, approximately by stars in the range of spectral type and luminosity class F2\,I to A7\,I. Integrating these spectral models over wavelength, the progenitor candidate would  have had a bolometric luminosity in the range  $\log(L_{\rm bol}/L_{\odot})=4.90$--5.32. The locus of the progenitor candidate is shown in a Hertzsprung-Russell diagram (HRD) in Figure~\ref{fig:HST_prog} (right panel). The progenitor candidate can be considered to be a yellow supergiant (YSG). 

We then selected BPASS binary-star models \citep{Stanway2018} for which the colors of the evolutionary-track endpoints matched those of the progenitor candidate to within the uncertainties. This comparison was made for tracks within a range of metallicities from $Z=0.008$ to 0.020. We also imposed the criterion that the models had to result in an SN IIb at the track terminus (the BPASS tracks terminate at the end of C burning), following standard prescriptions  \citep{Eldridge2013,Eldridge2017,Warwick2026}: for the primary component of the binary, the final CO core had to be $>1.38\ M_{\odot}$, the final ONe core mass $>0.1\ M_{\odot}$, the total core mass $>1.5\ M_{\odot}$, the total H mass in the range $M(H) \ge 10^{-3}$ and $<1.5\ M_{\odot}$, and the ratio of the total H to He mass $<1.05$. We then further compared the termini of the evolutionary tracks to the ($T_{\rm eff}$, $L_{\rm bol}$) locus of the progenitor candidate in the HRD and eliminated models which did not match the locus to within the uncertainties. Interestingly, the systems that survived these criteria were all of subsolar metallicity, $Z=0.008$ and $Z=0.010$. The resulting tracks are shown in Figure~\ref{fig:HST_prog}.

We emphasize here that none of the BPASS single-star models matched the  colors and luminosity of the progenitor candidate, nor did any of the binary models that should strictly terminate as SNe~Ib, i.e., with $M(H) < 10^{-3}$ \citep{Eldridge2017}.

 Besides the metallicity, the three properties that define the BPASS model tracks are the initial mass of the primary ($M_{\rm prim}$), the initial mass ratio of the binary ($q$), and the initial orbital period $P$ (in days). The models that best compared with the candidate locus are all long-period binaries, with massive primaries and low-to-moderate mass ratios. These models, specifically for $Z=0.008$, are $M_{\rm prim}=16\ M_{\odot}$, $q=0.6$, $\log(P)=2.8$; and $M_{\rm prim}=18\ M_{\odot}$, $q=0.2$, $\log(P)=3.2$. For $Z=0.010$ these are $M_{\rm prim}=14\ M_{\odot}$, $q=0.3$, $\log(P)=3.0$; $M_{\rm prim}=15\ M_{\odot}$, $q=0.5$, $\log(P)=2.8$; and $M_{\rm prim}=16\ M_{\odot}$, $q=0.1$, $\log(P)=3.0$.

In all of these systems, the light from the primary dominates over that from the secondary. For the systems with the lowest value of $q$ (i.e., with a comparatively low-mass, low-luminosity companion), the contribution from the secondary ($\lesssim 330\ L_{\odot}$) at primary terminus is nearly negligible, at $<1\%$. For the system with $q=0.6$ the luminosity ratio is $\sim 6\%$, and for the system with $q=0.5$ it is $\sim 3$\%. For these last two systems, the primary actually terminates at a cooler temperature than the ensemble, $T_{\rm eff}\approx6290$~K (F7\,I) and $\approx 6770$~K (F2\,I), respectively, and at a somewhat lower luminosity than the overall system, since the contribution from the blue dwarf (main-sequence) companion in both cases is relatively high in comparison; see Figure~\ref{fig:HST_prog}.

The systems all appear to experience Case B mass transfer during He burning via Roche-lobe overflow (RLOF), as models predict (e.g., \citealp{Claeys2011}; \citealp{Yoon2017} and references therein; \citealp{Dessart2024}). As expected, the primaries all lose substantial mass at that point, and the system angular momentum, orbital periods, and orbital separations drastically decrease, with the binary effectively within the remaining envelope of the primary; the systems experience common-envelope evolution \citep{Eldridge2017}. 

As mentioned, all of the primaries in the model systems experience significant mass loss, from $\sim 9$ to $11\ M_{\odot}$.  
Only the systems with comparatively high-mass secondaries (higher $q$) accrete significant mass, from $\sim 0.1$ to $0.4\ M_{\odot}$; for the rest of the systems, the accretion is $\lesssim 0.04\ M_{\odot}$. 
Nevertheless, the inferred mass transfer in all of these model systems is  nonconservative.

The primary-star radii at terminus of the five models, with initial $M_{\rm prim}$ from 14 to $18\ M_{\odot}$, are in the range $\sim 215$--$281\ R_{\odot}$. The final He core mass, predicted explosion remnant mass, and ejecta mass range from 4.4 to $6.9\ M_{\odot}$, 1.44 to $1.51\ M_{\odot}$ (i.e., these are all neutron stars), and 3.1 to $5.7\ M_{\odot}$, respectively. More importantly, the remaining H mass at the model termini is low, ranging from 0.02 to $0.12\ M_{\odot}$. 
These parameter ranges overlap with the values found during either our semi-analytical modeling of the bolometric LC of SN~2026dix, and the comparison of LCs and spectra to the D16 radiative-transfer models (see Secs. \ref{sec:anal_bol_model} and \ref{sec:anal_D16}, respectively). The only significant difference is in the progenitor radius, for which both modeling methods give much smaller values ($\sim$ 3$-$7 and $\sim$ 15 $R_{\odot}$, respectively) than the direct progenitor analysis. Differences, however, do exist for all the SNe IIb with an identified progenitor in NV16 and D16, and may arise from the effect of binarity and/or intense mass-loss processes very shortly before explosion.

Finally, we show the synthetic SEDs associated with the five model binary systems above, relative to the progenitor candidate SED in Figure~\ref{fig:HST_prog}. As one can see, the overall SEDs of these models provide a reasonable representation of the candidate SED, which is why these models were selected in the first place. The implication from the comparison with the models is that the SN arose from the explosion in an interacting binary system, consistent with the properties of a warm (F2 to A7), luminous ($\log(L_{\rm bol}/L_{\odot})\approx 4.9$--5.3) supergiant primary and a less luminous, less massive hot dwarf companion.

Is the metallicity of the evolutionary tracks which best account for the SN 2026dix locus consistent with expectations for the host galaxy?
The global metallicity (oxygen abundance $12+\log({\rm O/H})$ as proxy) of NGC 3913 ranges from 8.42 to 8.94, depending on the metallicity indicator \citep{Sanchez2017}. We can also estimate the local metallicity from the abundance gradient given by \citet{Galbany2016}. SN 2026dix is offset from the host nucleus by $\sim 12{\farcs}8$ nearly due west. Neglecting inclination (NGC 3913 is nearly face-on) and assuming 17.5 Mpc for the host distance, the offset is equivalent to 1.08 kpc and therefore the oxygen abundance is $\sim 8.6$. This, then, corresponds to a metallicity $Z\approx 0.01$, which is consistent with one set of tracks, but is somewhat higher than (although not entirely inconsistent with) $Z=0.008$ for the other group of tracks \citep[$Z=0.008$ corresponds to $12+\log({\rm O/H})\approx 8.44$;][]{Asplund2009}.

\begin{figure}[h!]
\centering
\includegraphics[width=\columnwidth]{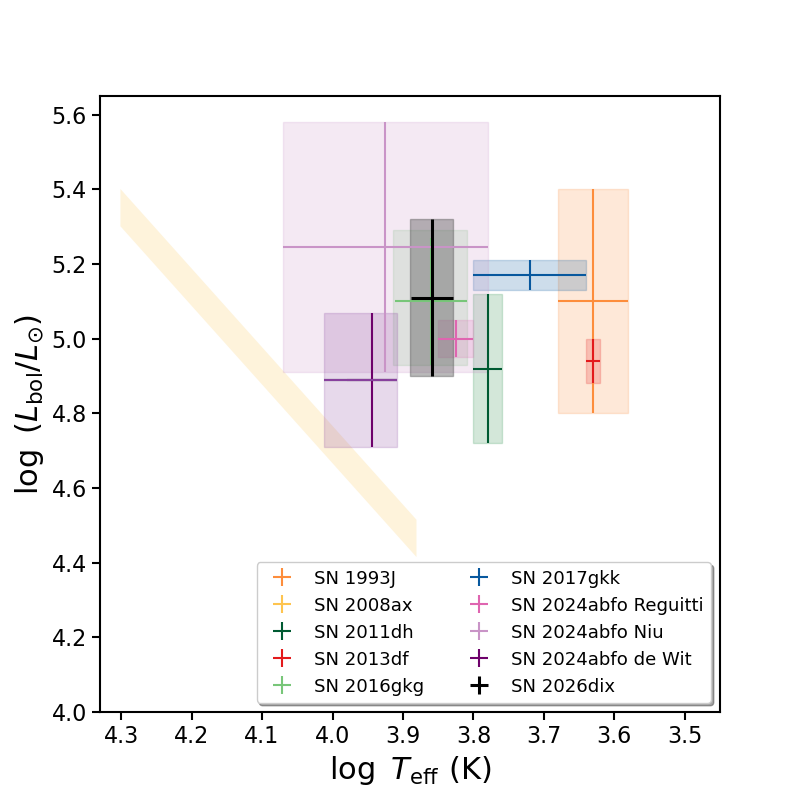}
\caption{HRD showing the locus of the SN 2026dix progenitor candidate, in comparison with the loci for SN 1993J \citep{Aldering1994,VanDyk2002,Maund2004,Stancliffe2009}, SN 2008ax \citep{Folatelli2015}, SN 2011dh \citep{Maund2011,VanDyk2011}, SN 2013df \citep{VanDyk2014}, SN 2016gkg \citep{Bersten2018}, SN 2017gkk \citep{Niu2024}, and SN 2024abfo \citep{Reguitti2025,Niu2025,deWet2025}.}
\label{fig:prog_comp}
\end{figure}

We note that the progenitor candidate locus in the HRD is also roughly consistent with (although somewhat warmer than) the endpoint of the binary model {\em Lm13p500} at Large Magellanic Cloud metallicity from \citet{Yoon2017}, which has an initial $13\ M_{\odot}$ primary in an initial 500 day orbit, experiences ``early'' Case B mass transfer (while the primary crosses the HRD Hertzsprung gap), and terminates as an ``SN~IIb-Y'' (a YSG leading to an SN~IIb). The locus is also consistent with the endpoint of the solar-metallicity model from that same study, {\em Sm16p1900}, which experiences ``late'' Case B mass transfer (after the H envelope of the primary has become fully convective), and also would terminate as an ''SN~IIb-Y.''

We can compare our assessment of the SN 2026dix progenitor candidate with all of the SN IIb progenitors that have been directly identified to date. The most famous example is SN 1993J: an object coincident with the SN position was reported by J.-M. Perelmuter \citep{Perelmuter_1993_IAUC5736} and independently by A. V. Filippenko \citep{Filippenko_1993_IAUC5737}, while R. M. Humphreys et al. \citep{Humphreys_93_IAUC5739} subsequently identified it explicitly as a candidate progenitor and noted that its colors indicated a composite source containing a late-type supergiant and a hotter star. Later, homogeneous photometry favored the red component as the progenitor, with a spectral type of approximately K0 I and a bolometric luminosity of ($M_{\rm bol} \approx -7.8$, \citealp{Aldering1994}).
\citet{Cohen1995} found that the star was not variable prior to explosion. \citet{VanDyk2002}, correcting the ground-based photometry with that from {\it HST}, determined that the star was an early K-type supergiant with $M_V = -7.0 \pm 0.4$ mag and $M_{\rm ini}=13$--$22\ M_{\odot}$. \citet{Stancliffe2009}, based on the analysis by \citet{Maund2004} of the presumed binary, consisting of a  primary with $\log(L_{\rm bol}/L_{\odot})=5.1 \pm 0.3$ and $\log(T_{\rm eff}\ [{\rm K}])=3.63 \pm 0.05$ and secondary with $\log(L_{\rm bol}/L_{\odot})=5.0 \pm 0.3$ and $\log(T_{\rm eff}\ [{\rm K}])=4.3 \pm 0.1$, modeled the progenitor as a 
$15\ M_{\odot}+14\ M_{\odot}$ system. \citet{Fox2014} attempted to further constrain the surviving companion as a hot B2 star.

Unlike the extended progenitor of SN 1993J, \citet{Folatelli2015} determined that the star \citet{Crockett2008} first detected as the progenitor of SN~2008ax was actually multiple, based on higher-resolution {\it HST} data, and concluded that the progenitor was a more compact supergiant of B to mid-A type, somewhere in the range of $L_{\rm bol}= (2.6-20)\times 10^4\ L_{\odot}$ and $T=7600$--20,000~K, modeling it as a $18\ M_{\odot}+12\ M_{\odot}$ close binary. \citet{Maund2011} and \citet{VanDyk2011} both identified the SN 2011dh progenitor in archival {\it HST} data, the former assigning to the star $\log(L_{\rm bol}/L_{\odot})=4.92 \pm 0.2$ and $\log(T_{\rm eff}\ [{\rm K}])=3.78 \pm 0.02$; \citet{Bersten2012} convincingly argued that the star was a supergiant, and \citet{Benvenuto2013} subsequently modeled it as a $16\ M_{\odot}+10\ M_{\odot}$ binary with initial $P=125$ days and final $M(H) \approx (3-5) \times 10^{-3}\ M_{\odot}$.

\citet{VanDyk2014} identified the SN 2013df progenitor in {\it HST} data and concluded that $\log(L_{\rm bol}/L_{\odot})=4.94 \pm 0.06$, $T_{\rm eff}=4250\pm 100$~K, and $M_{\rm ini}=13$--$17\ M_{\odot}$, making it strikingly similar to the SN 1993J progenitor, which would be consistent with the  properties of the SN itself  \citep[e.g.,][]{Maeda2015}. Based on the progenitor-candidate identification for SN 2016gkg, \citet[][also \citealp{Tartaglia2017,Kilpatrick2017}]{Bersten2018} concluded that the primary had $\log(L_{\rm bol}/L_{\odot})=5.10^{+0.17}_{-0.19}$ and $T_{\rm eff}=7250^{+900}_{-850}$~K, and modeled it as a $19.5\ M_{\odot}+13.5\ M_{\odot}$ system with an initial period 70 days, resulting in a final period of 631 days and $M(H)\approx6\times10^{-3}\ M_{\odot}$.

More recently, \citet{Niu2024} identified a progenitor candidate for SN 2017gkk, with $\log(L_{\rm bol}/L_{\odot})=5.17 \pm 0.04$ and $\log(T_{\rm eff}\ [{\rm K}])=3.72 \pm 0.08$, leading to an inference of $M_{\rm ini}\approx16\ M_{\odot}$. A possibly variable progenitor candidate has also been identified, based on a combination of {\it HST} and ground-based data, for SN 2024abfo; the analysis results differ among studies, from $\log(L_{\rm bol}/L_{\odot})=5.00 \pm 0.05$ and $\log(T_{\rm eff}\ [{\rm K}])=3.80$--3.85 \citep{Reguitti2025}, to $\log(L_{\rm bol}/L_{\odot})=4.91$--5.58 and $\log(T_{\rm eff}\ [{\rm K}])=3.78$--4.07 \citep[][a $M_{\rm ini}=12$--$18\ M_{\odot}$ YSG]{Niu2025}, to $\log(L_{\rm bol}/L_{\odot})=4.89 \pm 0.18$ and $\log(T_{\rm eff}\ [{\rm K}])=3.94^{+0.07}_{-0.04}$ \citep[][an A3 supergiant]{deWet2025}.

We show the comparison in Figure~\ref{fig:prog_comp}. The SN 2026dix progenitor candidate appears to be remarkably similar to the SN 2016gkg progenitor, as well as showing similarities with the SN 2024abfo progenitor candidate, at least based on the \citet{Reguitti2025} and \citet{Niu2025} studies (considerably less so for the results of \citealp{deWet2025}). Although still within roughly the same luminosity range, the progenitor candidate bears far less resemblance to that of SN 2017gkk and the progenitors of SN 2011dh, the much cooler SN 1993J and SN 2013df, and the hotter SN 2008ax. The SN 2026dix progenitor candidate truly appears to be a warm supergiant. 

The comparison with SN 2016gkg is particularly interesting, since the \citet{Bersten2018} modeling for that progenitor requires a more massive primary in a far shorter initial orbital period than what we infer from our comparison with the BPASS binary tracks (somewhat lower masses and long initial periods). We speculate that these discrepancies arise from differences in modeling initial conditions, assumptions, and procedures (the \citealp{Bersten2018} approach focused on hydrodynamical modeling after explosion of H-poor He stars and specific reproduction of the SN 2016gkg light curves; these models were also of solar metallicity). Our comparison with the BPASS tracks is predominantly meant to be suggestive, based purely on our SED fitting of four optical {\it HST} photometric points.


\section{Conclusions} \label{sec:concl}

We described above our findings regarding the nearby SN~2026dix. After determining the total extinction of the object (which turned out to be quite high, $E(B-V)_\textrm{tot}=0.6\pm 0.1$ mag, strongly dominated by the host extinction), we carried out a detailed comparative LC and spectral analysis of the SN. We found that SN~2026dix closely resembles SNe 2009mg, 2020acat, and 2022crv; thus, it is another member of the recently identified group of transitional SNe~IIb/Ib.  This finding strengthens the picture of the continuous distribution of SESNe.

We also identified the presumed progenitor star on multiple pre-explosion ({\it HST}) images, finding a point source within a very small (167 mas) uncertainty radius relative to the position of the SN.
After performing photometry, we constructed the SED of this object and compared the SED to that of model stellar atmospheres for supergiant stars.
The comparison with these models and BPASS binary stellar evolution tracks implies that SN~2026dix arose from the explosion in an interacting binary system, consistent with the properties of a warm ($T_\textrm{eff} \approx 6750-7750 K$, i.e., F2 to A7), luminous ($\log(L_{\rm bol}/L_{\odot})\approx 4.9$--5.3) supergiant primary and a less luminous, less massive hot dwarf companion.
The nature of the identified progenitor is similar to that of other known ones of SNe IIb, especially SNe 2016gkg and 2024abfo; the former of these two SNe also has LC evolution very similar  to that of SN~2026dix (while 2024abfo is significantly less luminous).
Characteristics of such a progenitor (including the final core mass, the ejecta mass, and the remaining H-mass) also harmonize well with both the results of our semi-analytical modeling of the bolometric LC of SN~2026dix, and the comparison of its LCs and spectra to the output of {\tt CMFGEN} radiative-transfer models of an exploding star with $M_\textrm{ini}$ = 18 \msolar.
More detailed modeling of our results for the SN 2026dix progenitor candidate is beyond the scope of this paper; however, it is highly encouraged.

\begin{acknowledgements}
This project has received funding from the HUN-REN Hun-
garian Research Network. 
The operation of the BRC80 (Baja Observatory) and RC80 telescopes (Konkoly Observatory) has been supported by the GINOP 2.3.2-15-2016-00033 grant of the National Research, Development and Innovation Office (NKFIH), Hungary, funded by the European Union.
This research was supported by NKFIH OTKA grants
K-142534.
{\'A}.S. acknowledges support by the KKP-137523 ``SeismoLab'' \'Elvonal grant of NKFIH.
A.V.F.’s research group at U.C. Berkeley acknowledges financial assistance from Gary and Cynthia Bengier, Clark and Sharon Winslow, Alan Eustace and Kathy Kwan          (W.Z. is a Bengier-Winslow-Eustace Specialist in Astronomy), Timothy and Melissa Draper, Briggs and Kathleen Wood, Ellyn and Alan 
Seelenfreund (T.G.B. is Draper-Wood-Seelenfreund Specialist in Astronomy), and numerous other donors. 

This research is based in part on observations made with the NASA/ESA {\it Hubble Space Telescope} obtained from the Space Telescope Science Institute (STScI), which is operated by the Association of Universities for Research in Astronomy, Inc., under NASA contract NAS5–26555. This work is based in part on observations made with the NASA/ESA/CSA {\it James Webb Space Telescope}; the data were obtained from the Mikulski Archive for Space Telescopes at STScI.  
Based in part on data obtained from the Hubble Legacy Archive, which is a collaboration between 
STScI/NASA, the Space Telescope European Coordinating Facility (ST-ECF/ESA), and the Canadian Astronomy Data Centre (CADC/NRC/CSA).

A major upgrade of the Kast spectrograph on the Shane 3\,m telescope at Lick Observatory, led by Brad Holden, was made possible through gifts from the Heising-Simons Foundation, 
William and Marina Kast, and the University of California Observatories. We appreciate the expert assistance of the staff at Lick  Observatory. Research at Lick Observatory 
is partially supported by a gift from Google.

\end{acknowledgements}

\bibliographystyle{aa}
\bibliography{main} 

@INPROCEEDINGS{1993ASPC...52..173T,
       author = {{Tody}, Doug},
        title = "{IRAF in the Nineties}",
    booktitle = {Astronomical Data Analysis Software and Systems II},
         year = 1993,
       editor = {{Hanisch}, R.~J. and {Brissenden}, R.~J.~V. and {Barnes}, J.},
       series = {Astronomical Society of the Pacific Conference Series},
       volume = {52},
        month = jan,
        pages = {173},
       adsurl = {https://ui.adsabs.harvard.edu/abs/1993ASPC...52..173T}
}

@ARTICLE{Yaron_2012,
       author = {{Yaron}, Ofer and {Gal-Yam}, Avishay},
        title = "{WISeREP{\textemdash}An Interactive Supernova Data Repository}",
      journal = {\pasp},
         year = 2012,
        month = jul,
       volume = {124},
       number = {917},
        pages = {668},
          doi = {10.1086/666656},
archivePrefix = {arXiv},
       eprint = {1204.1891},
 primaryClass = {astro-ph.IM},
       adsurl = {https://ui.adsabs.harvard.edu/abs/2012PASP..124..668Y}
}

@ARTICLE{Tonry12,
       author = {{Tonry}, J.~L. and {Stubbs}, C.~W. and {Kilic}, M. and {Flewelling}, H.~A. and {Deacon}, N.~R. and {Chornock}, R. and {Berger}, E. and {Burgett}, W.~S. and {Chambers}, K.~C. and {Kaiser}, N. and {Kudritzki}, R. -P. and {Hodapp}, K.~W. and {Magnier}, E.~A. and {Morgan}, J.~S. and {Price}, P.~A. and {Wainscoat}, R.~J.},
        title = "{First Results from Pan-STARRS1: Faint, High Proper Motion White Dwarfs in the Medium-Deep Fields}",
      journal = {\apj},
         year = 2012,
        month = jan,
       volume = {745},
       number = {1},
          eid = {42},
        pages = {42},
          doi = {10.1088/0004-637X/745/1/42},
archivePrefix = {arXiv},
       eprint = {1110.0060},
 primaryClass = {astro-ph.GA},
       adsurl = {https://ui.adsabs.harvard.edu/abs/2012ApJ...745...42T}
}

@ARTICLE{MASTER_disc,
       author = {{Kuvshinov}, D. and {Lipunov}, V. and {Panchenko}, I. and {Balanutsa}, P. and {Kuznetsov}, A. and {Antipov}, G. and {Gorbovskoy}, E. and {Tiurina}, N. and {Sankovich}, A. and {Topolev}, V. and {Zhirkov}, K. and {Vlasenko}, D. and {Tselik}, Y. and {Kechin}, Y. and {Senik}, V. and {Cheryasov}, D. and {Chasovnikov}, A. and {Gress}, O. and {Budnev}, N. and {Francile}, C. and {Podesta}, F. and {Podesta}, R. and {Gonzalez}, E. and {Tlatov}, A. and {Dormidontov}, D. and {Sosnovskij}, A. and {Gabovich}, A. and {Yurkov}, V. and {Buckley}, D. and {Rebolo}, R. and {Carrasco}, L. and {Valdes}, J.~R. and {Chavushyan}, V. and {Alvarez}, V.~M.~P. and {Martinez}, J. and {Tanori}, J. and {Corella}, A.~R. and {Rodriguez}, L.~H.},
        title = "{MASTER Transient Discovery Report for 2026-02-16}",
      journal = {Transient Name Server Discovery Report},
         year = 2026,
        month = feb,
       volume = {2026-660},
        pages = {1},
       adsurl = {https://ui.adsabs.harvard.edu/abs/2026TNSTR.660....1K}
}

@ARTICLE{TNS_classific,
       author = {{Pursiainen}, M. and {O'Neill}, D. and {Killestein}, T. and {Kotak}, R. and {Lyman}, J. and {Godson}, B. and {Ackley}, K. and {Dyer}, M.~J. and {Ulaczyk}, K. and {Belkin}, S. and {Steeghs}, D. and {Galloway}, D.~K. and {Dhillon}, V. and {O'Brien}, P. and {Ramsay}, G. and {Noysena}, K. and {Breton}, R.~P. and {Nuttall}, L.~K. and {Pollacco}, D. and {Vel'azquez}, J.~C. and {Kumar}, A. and {O'Neill}, D.},
        title = "{GOTO Transient Classification Report for 2026-02-18}",
      journal = {Transient Name Server Classification Report},
         year = 2026,
        month = feb,
       volume = {2026-701},
        pages = {1},
       adsurl = {https://ui.adsabs.harvard.edu/abs/2026TNSCR.701....1P}
}

@ARTICLE{Barbon_1982,
       author = {{Barbon}, R. and {Ciatti}, F. and {Rosino}, L. and {Ortolani}, S. and {Rafanelli}, P.},
        title = "{Spectra and light curves of three recent supernovae}",
      journal = {\aap},
         year = 1982,
        month = dec,
       volume = {116},
       number = {1},
        pages = {43-53},
       adsurl = {https://ui.adsabs.harvard.edu/abs/1982A&A...116...43B}
}

@ARTICLE{Silverman2012,
       author = {{Silverman}, Jeffrey M. and {Foley}, Ryan J. and {Filippenko}, Alexei V. and {Ganeshalingam}, Mohan and {Barth}, Aaron J. and {Chornock}, Ryan and {Griffith}, Christopher V. and {Kong}, Jason J. and {Lee}, Nicholas and {Leonard}, Douglas C. and {Matheson}, Thomas and {Miller}, Emily G. and {Steele}, Thea N. and {Barris}, Brian J. and {Bloom}, Joshua S. and {Cobb}, Bethany E. and {Coil}, Alison L. and {Desroches}, Louis-Benoit and {Gates}, Elinor L. and {Ho}, Luis C. and {Jha}, Saurabh W. and {Kandrashoff}, Michael T. and {Li}, Weidong and {Mandel}, Kaisey S. and {Modjaz}, Maryam and {Moore}, Matthew R. and {Mostardi}, Robin E. and {Papenkova}, Marina S. and {Park}, Sung and {Perley}, Daniel A. and {Poznanski}, Dovi and {Reuter}, Cassie A. and {Scala}, James and {Serduke}, Franklin J.~D. and {Shields}, Joseph C. and {Swift}, Brandon J. and {Tonry}, John L. and {Van Dyk}, Schuyler D. and {Wang}, Xiaofeng and {Wong}, Diane S.},
        title = "{Berkeley Supernova Ia Program - I. Observations, data reduction and spectroscopic sample of 582 low-redshift Type Ia supernovae}",
      journal = {\mnras},
         year = 2012,
        month = sep,
       volume = {425},
       number = {3},
        pages = {1789-1818},
          doi = {10.1111/j.1365-2966.2012.21270.x},
archivePrefix = {arXiv},
       eprint = {1202.2128},
 primaryClass = {astro-ph.CO},
       adsurl = {https://ui.adsabs.harvard.edu/abs/2012MNRAS.425.1789S}
}

@ARTICLE{Filippenko1982,
       author = {{Filippenko}, A.~V.},
        title = "{The importance of atmospheric differential refraction in spectrophotometry.}",
      journal = {\pasp},
         year = 1982,
        month = aug,
       volume = {94},
        pages = {715-721},
          doi = {10.1086/131052},
       adsurl = {https://ui.adsabs.harvard.edu/abs/1982PASP...94..715F}
}

@INPROCEEDINGS{Tody1986,
       author = {{Tody}, Doug},
        title = "{The IRAF Data Reduction and Analysis System}",
    booktitle = {Instrumentation in astronomy VI},
         year = 1986,
       editor = {{Crawford}, David L.},
       series = {Society of Photo-Optical Instrumentation Engineers (SPIE) Conference Series},
       volume = {627},
        month = jan,
        pages = {733},
          doi = {10.1117/12.968154},
       adsurl = {https://ui.adsabs.harvard.edu/abs/1986SPIE..627..733T}
}

@ARTICLE{SF11,
       author = {{Schlafly}, Edward F. and {Finkbeiner}, Douglas P.},
        title = "{Measuring Reddening with Sloan Digital Sky Survey Stellar Spectra and Recalibrating SFD}",
      journal = {\apj},
         year = 2011,
        month = aug,
       volume = {737},
       number = {2},
          eid = {103},
        pages = {103},
          doi = {10.1088/0004-637X/737/2/103},
archivePrefix = {arXiv},
       eprint = {1012.4804},
 primaryClass = {astro-ph.GA},
       adsurl = {https://ui.adsabs.harvard.edu/abs/2011ApJ...737..103S}
}

@ARTICLE{Stritzinger_2018,
       author = {{Stritzinger}, M.~D. and {Taddia}, F. and {Burns}, C.~R. and {Phillips}, M.~M. and {Bersten}, M. and {Contreras}, C. and {Folatelli}, G. and {Holmbo}, S. and {Hsiao}, E.~Y. and {Hoeflich}, P. and {Leloudas}, G. and {Morrell}, N. and {Sollerman}, J. and {Suntzeff}, N.~B.},
        title = "{The Carnegie Supernova Project I. Methods to estimate host-galaxy reddening of stripped-envelope supernovae}",
      journal = {\aap},
         year = 2018,
        month = feb,
       volume = {609},
          eid = {A135},
        pages = {A135},
          doi = {10.1051/0004-6361/201730843},
archivePrefix = {arXiv},
       eprint = {1707.07615},
 primaryClass = {astro-ph.HE},
       adsurl = {https://ui.adsabs.harvard.edu/abs/2018A&A...609A.135S}
}

@ARTICLE{Abreu_2022,
       author = {{Abreu}, P. and {Aglietta}, M. and {Albury}, J.~M. and {Allekotte}, I. and {Almeida Cheminant}, K. and {Almela}, A. and {Alvarez-Mu{\~n}iz}, J. and {Alves Batista}, R. and {Ammerman Yebra}, J. and {Anastasi}, G.~A. and {Anchordoqui}, L. and {Andrada}, B. and {Andringa}, S. and {Aramo}, C. and {Ara{\'u}jo Ferreira}, P.~R. and {Arnone}, E. and {Arteaga Vel{\'a}zquez}, J.~C. and {Asorey}, H. and {Assis}, P. and {Avila}, G. and {Avocone}, E. and {Badescu}, A.~M. and {Bakalova}, A. and {Balaceanu}, A. and {Barbato}, F. and {Bellido}, J.~A. and {Berat}, C. and {Bertaina}, M.~E. and {Bhatta}, G. and {Biermann}, P.~L. and {Binet}, V. and {Bismark}, K. and {Bister}, T. and {Biteau}, J. and {Blazek}, J. and {Bleve}, C. and {Bl{\"u}mer}, J. and {Boh{\'a}{\v{c}}ov{\'a}}, M. and {Boncioli}, D. and {Bonifazi}, C. and {Bonneau Arbeletche}, L. and {Borodai}, N. and {Botti}, A.~M. and {Brack}, J. and {Bretz}, T. and {Brichetto Orchera}, P.~G. and {Briechle}, F.~L. and {Buchholz}, P. and {Bueno}, A. and {Buitink}, S. and {Buscemi}, M. and {B{\"u}sken}, M. and {Caballero-Mora}, K.~S. and {Caccianiga}, L. and {Canfora}, F. and {Caracas}, I. and {Caruso}, R. and {Castellina}, A. and {Catalani}, F. and {Cataldi}, G. and {Cazon}, L. and {Cerda}, M. and {Chinellato}, J.~A. and {Chudoba}, J. and {Chytka}, L. and {Clay}, R.~W. and {Cobos Cerutti}, A.~C. and {Colalillo}, R. and {Coleman}, A. and {Coluccia}, M.~R. and {Concei{\c{c}}{\~a}o}, R. and {Condorelli}, A. and {Consolati}, G. and {Contreras}, F. and {Convenga}, F. and {Correia dos Santos}, D. and {Covault}, C.~E. and {Dasso}, S. and {Daumiller}, K. and {Dawson}, B.~R. and {Day}, J.~A. and {de Almeida}, R.~M. and {de Jes{\'u}s}, J. and {de Jong}, S.~J. and {de Mello Neto}, J.~R.~T. and {De Mitri}, I. and {de Oliveira}, J. and {de Oliveira Franco}, D. and {de Palma}, F. and {de Souza}, V. and {De Vito}, E. and {Del Popolo}, A. and {del R{\'\i}o}, M. and {Deligny}, O. and {Deval}, L. and {di Matteo}, A. and {Dobre}, M. and {Dobrigkeit}, C. and {D'Olivo}, J.~C. and {Domingues Mendes}, L.~M. and {dos Anjos}, R.~C. and {Dova}, M.~T. and {Ebr}, J. and {Engel}, R. and {Epicoco}, I. and {Erdmann}, M. and {Escobar}, C.~O. and {Etchegoyen}, A. and {Falcke}, H. and {Farmer}, J. and {Farrar}, G. and {Fauth}, A.~C. and {Fazzini}, N. and {Feldbusch}, F. and {Fenu}, F. and {Fick}, B. and {Figueira}, J.~M. and {Filip{\v{c}}i{\v{c}}}, A. and {Fitoussi}, T. and {Fodran}, T. and {Fujii}, T. and {Fuster}, A. and {Galea}, C. and {Galelli}, C. and {Garc{\'\i}a}, B. and {Gemmeke}, H. and {Gesualdi}, F. and {Gherghel-Lascu}, A. and {Ghia}, P.~L. and {Giaccari}, U. and {Giammarchi}, M. and {Glombitza}, J. and {Gobbi}, F. and {Gollan}, F. and {Golup}, G. and {G{\'o}mez Berisso}, M. and {G{\'o}mez Vitale}, P.~F. and {Gongora}, J.~P. and {Gonz{\'a}lez}, J.~M. and {Gonz{\'a}lez}, N. and {Goos}, I. and {G{\'o}ra}, D. and {Gorgi}, A. and {Gottowik}, M. and {Grubb}, T.~D. and {Guarino}, F. and {Guedes}, G.~P. and {Guido}, E. and {Hahn}, S. and {Hamal}, P. and {Hampel}, M.~R. and {Hansen}, P. and {Harari}, D. and {Harvey}, V.~M. and {Haungs}, A. and {Hebbeker}, T. and {Heck}, D. and {Hill}, G.~C. and {Hojvat}, C. and {H{\"o}randel}, J.~R. and {Horvath}, P. and {Hrabovsk{\'y}}, M. and {Huege}, T. and {Insolia}, A. and {Isar}, P.~G. and {Janecek}, P. and {Johnsen}, J.~A. and {Jurysek}, J. and {K{\"a}{\"a}p{\"a}}, A. and {Kampert}, K.~H. and {Keilhauer}, B. and {Khakurdikar}, A. and {Kizakke Covilakam}, V.~V. and {Klages}, H.~O. and {Kleifges}, M. and {Kleinfeller}, J. and {Knapp}, F. and {Kunka}, N. and {Lago}, B.~L. and {Langner}, N. and {Leigui de Oliveira}, M.~A. and {Lenok}, V. and {Letessier-Selvon}, A. and {Lhenry-Yvon}, I. and {Lo Presti}, D. and {Lopes}, L. and {L{\'o}pez}, R. and {Lu}, L. and {Luce}, Q. and {Lundquist}, J.~P. and {Machado Payeras}, A. and {Mancarella}, G. and {Mandat}, D. and {Manning}, B.~C. and {Manshanden}, J. and {Mantsch}, P. and {Marafico}, S. and {Mariani}, F.~M. and {Mariazzi}, A.~G. and {Mari{\textcommabelow s}}, I.~C.},
        title = "{Arrival Directions of Cosmic Rays above 32 EeV from Phase One of the Pierre Auger Observatory}",
      journal = {\apj},
         year = 2022,
        month = aug,
       volume = {935},
       number = {2},
          eid = {170},
        pages = {170},
          doi = {10.3847/1538-4357/ac7d4e},
archivePrefix = {arXiv},
       eprint = {2206.13492},
 primaryClass = {astro-ph.HE},
       adsurl = {https://ui.adsabs.harvard.edu/abs/2022ApJ...935..170A}
}

@ARTICLE{Leroy_2019,
       author = {{Leroy}, Adam K. and {Sandstrom}, Karin M. and {Lang}, Dustin and {Lewis}, Alexia and {Salim}, Samir and {Behrens}, Erica A. and {Chastenet}, J{\'e}r{\'e}my and {Chiang}, I-Da and {Gallagher}, Molly J. and {Kessler}, Sarah and {Utomo}, Dyas},
        title = "{A z = 0 Multiwavelength Galaxy Synthesis. I. A WISE and GALEX Atlas of Local Galaxies}",
      journal = {\apjs},
         year = 2019,
        month = oct,
       volume = {244},
       number = {2},
          eid = {24},
        pages = {24},
          doi = {10.3847/1538-4365/ab3925},
archivePrefix = {arXiv},
       eprint = {1910.13470},
 primaryClass = {astro-ph.GA},
       adsurl = {https://ui.adsabs.harvard.edu/abs/2019ApJS..244...24L}
}

@ARTICLE{Tully_2009,
       author = {{Tully}, R. Brent and {Rizzi}, Luca and {Shaya}, Edward J. and {Courtois}, H{\'e}l{\`e}ne M. and {Makarov}, Dmitry I. and {Jacobs}, Bradley A.},
        title = "{The Extragalactic Distance Database}",
      journal = {\aj},
         year = 2009,
        month = aug,
       volume = {138},
       number = {2},
        pages = {323-331},
          doi = {10.1088/0004-6256/138/2/323},
       adsurl = {https://ui.adsabs.harvard.edu/abs/2009AJ....138..323T}
}

@ARTICLE{Tully_2016,
       author = {{Tully}, R. Brent and {Courtois}, H{\'e}l{\`e}ne M. and {Sorce}, Jenny G.},
        title = "{Cosmicflows-3}",
      journal = {\aj},
         year = 2016,
        month = aug,
       volume = {152},
       number = {2},
          eid = {50},
        pages = {50},
          doi = {10.3847/0004-6256/152/2/50},
archivePrefix = {arXiv},
       eprint = {1605.01765},
 primaryClass = {astro-ph.CO},
       adsurl = {https://ui.adsabs.harvard.edu/abs/2016AJ....152...50T}
}

@ARTICLE{Courtois_2012,
       author = {{Courtois}, H{\'e}l{\`e}ne M. and {Hoffman}, Yehuda and {Tully}, R. Brent and {Gottl{\"o}ber}, Stefan},
        title = "{Three-dimensional Velocity and Density Reconstructions of the Local Universe with Cosmicflows-1}",
      journal = {\apj},
         year = 2012,
        month = jan,
       volume = {744},
       number = {1},
          eid = {43},
        pages = {43},
          doi = {10.1088/0004-637X/744/1/43},
archivePrefix = {arXiv},
       eprint = {1109.3856},
 primaryClass = {astro-ph.CO},
       adsurl = {https://ui.adsabs.harvard.edu/abs/2012ApJ...744...43C}
}

@ARTICLE{Paturel_2003a,
       author = {{Paturel}, G. and {Petit}, C. and {Prugniel}, Ph. and {Theureau}, G. and {Rousseau}, J. and {Brouty}, M. and {Dubois}, P. and {Cambr{\'e}sy}, L.},
        title = "{HYPERLEDA.  I. Identification and designation of galaxies}",
      journal = {\aap},
         year = 2003,
        month = dec,
       volume = {412},
        pages = {45-55},
          doi = {10.1051/0004-6361:20031411},
       adsurl = {https://ui.adsabs.harvard.edu/abs/2003A&A...412...45P}
}

@ARTICLE{Paturel_2003ab,
       author = {{Paturel}, G. and {Theureau}, G. and {Bottinelli}, L. and {Gouguenheim}, L. and {Coudreau-Durand}, N. and {Hallet}, N. and {Petit}, C.},
        title = "{HYPERLEDA. II. The homogenized HI data}",
      journal = {\aap},
         year = 2003,
        month = dec,
       volume = {412},
        pages = {57-67},
          doi = {10.1051/0004-6361:20031412},
       adsurl = {https://ui.adsabs.harvard.edu/abs/2003A&A...412...57P}
}

@ARTICLE{Makarov_2014,
       author = {{Makarov}, Dmitry and {Prugniel}, Philippe and {Terekhova}, Nataliya and {Courtois}, H{\'e}l{\`e}ne and {Vauglin}, Isabelle},
        title = "{HyperLEDA. III. The catalogue of extragalactic distances}",
      journal = {\aap},
         year = 2014,
        month = oct,
       volume = {570},
          eid = {A13},
        pages = {A13},
          doi = {10.1051/0004-6361/201423496},
archivePrefix = {arXiv},
       eprint = {1408.3476},
 primaryClass = {astro-ph.GA},
       adsurl = {https://ui.adsabs.harvard.edu/abs/2014A&A...570A..13M}
}

@ARTICLE{Davis_2024,
       author = {{Davis}, Benjamin L. and {Graham}, Alister W. and {Soria}, Roberto and {Jin}, Zehao and {Karachentsev}, Igor D. and {Karachentseva}, Valentina E. and {D'Onghia}, Elena},
        title = "{Identification of Intermediate-mass Black Hole Candidates among a Sample of Sd Galaxies}",
      journal = {\apj},
         year = 2024,
        month = aug,
       volume = {971},
       number = {2},
          eid = {123},
        pages = {123},
          doi = {10.3847/1538-4357/ad55eb},
archivePrefix = {arXiv},
       eprint = {2406.05778},
 primaryClass = {astro-ph.GA},
       adsurl = {https://ui.adsabs.harvard.edu/abs/2024ApJ...971..123D}
}

@ARTICLE{SNID-SAGE_2026,
       author = {{Stoppa}, Fiorenzo and {Smartt}, Stephen J.},
        title = "{SNID─SAGE: a modern framework for interactive supernova classification and spectral analysis}",
      journal = {\mnras},
         year = 2026,
        month = jul,
       volume = {549},
       number = {4},
          eid = {stag1066},
        pages = {stag1066},
          doi = {10.1093/mnras/stag1066},
archivePrefix = {arXiv},
       eprint = {2603.28741},
 primaryClass = {astro-ph.IM},
       adsurl = {https://ui.adsabs.harvard.edu/abs/2026MNRAS.549g1066S}
}

@ARTICLE{Gangopadhyay_2026,
       author = {{Gangopadhyay}, Anjasha and {Pessi}, Priscila J.},
        title = "{Hydrogen-rich to stripped-envelope: observational continuity and biases in CCSNe}",
      journal = {Frontiers in Astronomy and Space Sciences},
         year = 2026,
        month = jan,
       volume = {12},
          eid = {1708372},
        pages = {1708372},
          doi = {10.3389/fspas.2025.1708372},
archivePrefix = {arXiv},
       eprint = {2512.04010},
 primaryClass = {astro-ph.HE},
       adsurl = {https://ui.adsabs.harvard.edu/abs/2026FrASS..1208372G}
}

@ARTICLE{Gangopadhyay_2023,
       author = {{Gangopadhyay}, Anjasha and {Maeda}, Keiichi and {Singh}, Avinash and {Nayana}, A.~J. and {Nakaoka}, Tatsuya and {Kawabata}, Koji S. and {Taguchi}, Kenta and {Singh}, Mridweeka and {Chandra}, Poonam and {Ryder}, Stuart D. and {Dastidar}, Raya and {Yamanaka}, Masayuki and {Kawabata}, Miho and {Alsaberi}, Rami Z.~E. and {Dukiya}, Naveen and {Teja}, Rishabh Singh and {Ailawadhi}, Bhavya and {Dutta}, Anirban and {Sahu}, D.~K. and {Moriya}, Takashi J. and {Misra}, Kuntal and {Tanaka}, Masaomi and {Chevalier}, Roger and {Tominaga}, Nozomu and {Uno}, Kohki and {Imazawa}, Ryo and {Hamada}, Taisei and {Hori}, Tomoya and {Isogai}, Keisuke},
        title = "{Bridging between Type IIb and Ib Supernovae: SN IIb 2022crv with a Very Thin Hydrogen Envelope}",
      journal = {\apj},
         year = 2023,
        month = nov,
       volume = {957},
       number = {2},
          eid = {100},
        pages = {100},
          doi = {10.3847/1538-4357/acfa94},
archivePrefix = {arXiv},
       eprint = {2309.07463},
 primaryClass = {astro-ph.HE},
       adsurl = {https://ui.adsabs.harvard.edu/abs/2023ApJ...957..100G}
}

@ARTICLE{Medler_2022,
       author = {{Medler}, K. and {Mazzali}, P.~A. and {Teffs}, J. and {Ashall}, C. and {Anderson}, J.~P. and {Arcavi}, I. and {Benetti}, S. and {Bostroem}, K.~A. and {Burke}, J. and {Cai}, Y.-Z. and {Charalampopoulos}, P. and {Elias-Rosa}, N. and {Ergon}, M. and {Galbany}, L. and {Gromadzki}, M. and {Hiramatsu}, D. and {Howell}, D.~A. and {Inserra}, C. and {Lundqvist}, P. and {McCully}, C. and {M{\"u}ller-Bravo}, T. and {Newsome}, M. and {Nicholl}, M. and {Padilla Gonzalez}, E. and {Paraskeva}, E. and {Pastorello}, A. and {Pellegrino}, C. and {Pessi}, P.~J. and {Reguitti}, A. and {Reynolds}, T.~M. and {Roy}, R. and {Terreran}, G. and {Tomasella}, L. and {Young}, D.~R.},
        title = "{SN 2020acat: an energetic fast rising Type IIb supernova}",
      journal = {\mnras},
         year = 2022,
        month = jul,
       volume = {513},
       number = {4},
        pages = {5540-5558},
          doi = {10.1093/mnras/stac1192},
archivePrefix = {arXiv},
       eprint = {2201.06991},
 primaryClass = {astro-ph.HE},
       adsurl = {https://ui.adsabs.harvard.edu/abs/2022MNRAS.513.5540M}
}

@ARTICLE{Ergon_2024,
       author = {{Ergon}, Mattias and {Lundqvist}, Peter and {Fransson}, Claes and {Kuncarayakti}, Hanindyo and {Das}, Kaustav K. and {De}, Kishalay and {Ferrari}, Lucia and {Fremling}, Christoffer and {Medler}, Kyle and {Maeda}, Keiichi and {Pastorello}, Andrea and {Sollerman}, Jesper and {Stritzinger}, Maximilian D.},
        title = "{Light curve and spectral modelling of the type IIb SN 2020acat. Evidence for a strong Ni bubble effect on the diffusion time}",
      journal = {\aap},
         year = 2024,
        month = mar,
       volume = {683},
          eid = {A241},
        pages = {A241},
          doi = {10.1051/0004-6361/202346718},
archivePrefix = {arXiv},
       eprint = {2308.07158},
 primaryClass = {astro-ph.HE},
       adsurl = {https://ui.adsabs.harvard.edu/abs/2024A&A...683A.241E}
}

@ARTICLE{Dong_2024,
       author = {{Dong}, Yize and {Valenti}, Stefano and {Ashall}, Chris and {Williamson}, Marc and {Sand}, David J. and {Van Dyk}, Schuyler D. and {Filippenko}, Alexei V. and {Jha}, Saurabh W. and {Lundquist}, Michael and {Modjaz}, Maryam and {Andrews}, Jennifer E. and {Jencson}, Jacob E. and {Hosseinzadeh}, Griffin and {Pearson}, Jeniveve and {Kwok}, Lindsey A. and {Boland}, Teresa and {Hsiao}, Eric Y. and {Smith}, Nathan and {Elias-Rosa}, Nancy and {Srivastav}, Shubham and {Smartt}, Stephen and {Fulton}, Michael and {Zheng}, WeiKang and {Brink}, Thomas G. and {Shahbandeh}, Melissa and {Bostroem}, K. Azalee and {Hoang}, Emily and {Janzen}, Daryl and {Mehta}, Darshana and {Meza}, Nicolas and {Shrestha}, Manisha and {Wyatt}, Samuel and {Auchettl}, Katie and {Burns}, Christopher R. and {Farah}, Joseph and {Galbany}, Llu{\'\i}s and {Padilla Gonzalez}, Estefania and {Haislip}, Joshua and {Hinkle}, Jason T. and {Howell}, D. Andrew and {De Jaeger}, Thomas and {Kouprianov}, Vladimir and {Kumar}, Sahana and {Lu}, Jing and {McCully}, Curtis and {Moran}, Shane and {Morrell}, Nidia and {Newsome}, Megan and {Pellegrino}, Craig and {Polin}, Abigail and {Reichart}, Daniel E. and {Shappee}, B.~J. and {Stritzinger}, Maximilian D. and {Terreran}, Giacomo and {Tucker}, M.~A.},
        title = "{Characterizing the Rapid Hydrogen Disappearance in SN 2022crv: Evidence of a Continuum between Type Ib and IIb Supernova Properties}",
      journal = {\apj},
         year = 2024,
        month = oct,
       volume = {974},
       number = {2},
          eid = {316},
        pages = {316},
          doi = {10.3847/1538-4357/ad710e},
archivePrefix = {arXiv},
       eprint = {2309.09433},
 primaryClass = {astro-ph.HE},
       adsurl = {https://ui.adsabs.harvard.edu/abs/2024ApJ...974..316D}
}

@ARTICLE{Oates_2012,
       author = {{Oates}, S.~R. and {Bayless}, A.~J. and {Stritzinger}, M.~D. and {Prichard}, T. and {Prieto}, J.~L. and {Immler}, S. and {Brown}, P.~J. and {Breeveld}, A.~A. and {De Pasquale}, M. and {Kuin}, N.~P.~M. and {Hamuy}, M. and {Holland}, S.~T. and {Taddia}, F. and {Roming}, P.~W.~A.},
        title = "{Multiwavelength observations of the Type IIb supernova 2009mg}",
      journal = {\mnras},
         year = 2012,
        month = aug,
       volume = {424},
       number = {2},
        pages = {1297-1306},
          doi = {10.1111/j.1365-2966.2012.21311.x},
archivePrefix = {arXiv},
       eprint = {1205.3738},
 primaryClass = {astro-ph.HE},
       adsurl = {https://ui.adsabs.harvard.edu/abs/2012MNRAS.424.1297O}
}

@ARTICLE{Sahu_2011,
       author = {{Sahu}, D.~K. and {Gurugubelli}, U.~K. and {Anupama}, G.~C. and {Nomoto}, K.},
        title = "{Optical studies of SN 2009jf: a Type Ib supernova with an extremely slow decline and aspherical signature}",
      journal = {\mnras},
         year = 2011,
        month = jun,
       volume = {413},
       number = {4},
        pages = {2583-2594},
          doi = {10.1111/j.1365-2966.2011.18326.x},
archivePrefix = {arXiv},
       eprint = {1101.2068},
 primaryClass = {astro-ph.SR},
       adsurl = {https://ui.adsabs.harvard.edu/abs/2011MNRAS.413.2583S}
}

@ARTICLE{Sahu_2013,
       author = {{Sahu}, D.~K. and {Anupama}, G.~C. and {Chakradhari}, N.~K.},
        title = "{One year of monitoring of the Type IIb supernova SN 2011dh}",
      journal = {\mnras},
         year = 2013,
        month = jul,
       volume = {433},
       number = {1},
        pages = {2-22},
          doi = {10.1093/mnras/stt647},
archivePrefix = {arXiv},
       eprint = {1305.7067},
 primaryClass = {astro-ph.SR},
       adsurl = {https://ui.adsabs.harvard.edu/abs/2013MNRAS.433....2S}
}

@ARTICLE{Marion_2014,
       author = {{Marion}, G.~H. and {Vinko}, Jozsef and {Kirshner}, Robert P. and {Foley}, Ryan J. and {Berlind}, Perry and {Bieryla}, Allyson and {Bloom}, Joshua S. and {Calkins}, Michael L. and {Challis}, Peter and {Chevalier}, Roger A. and {Chornock}, Ryan and {Culliton}, Chris and {Curtis}, Jason L. and {Esquerdo}, Gilbert A. and {Everett}, Mark E. and {Falco}, Emilio E. and {France}, Kevin and {Fransson}, Claes and {Friedman}, Andrew S. and {Garnavich}, Peter and {Leibundgut}, Bruno and {Meyer}, Samuel and {Smith}, Nathan and {Soderberg}, Alicia M. and {Sollerman}, Jesper and {Starr}, Dan L. and {Szklenar}, Tamas and {Takats}, Katalin and {Wheeler}, J. Craig},
        title = "{Type IIb Supernova SN 2011dh: Spectra and Photometry from the Ultraviolet to the Near-infrared}",
      journal = {\apj},
         year = 2014,
        month = feb,
       volume = {781},
       number = {2},
          eid = {69},
        pages = {69},
          doi = {10.1088/0004-637X/781/2/69},
archivePrefix = {arXiv},
       eprint = {1303.5482},
 primaryClass = {astro-ph.SR},
       adsurl = {https://ui.adsabs.harvard.edu/abs/2014ApJ...781...69M}
}

@ARTICLE{Ergon_2014,
       author = {{Ergon}, M. and {Sollerman}, J. and {Fraser}, M. and {Pastorello}, A. and {Taubenberger}, S. and {Elias-Rosa}, N. and {Bersten}, M. and {Jerkstrand}, A. and {Benetti}, S. and {Botticella}, M.~T. and {Fransson}, C. and {Harutyunyan}, A. and {Kotak}, R. and {Smartt}, S. and {Valenti}, S. and {Bufano}, F. and {Cappellaro}, E. and {Fiaschi}, M. and {Howell}, A. and {Kankare}, E. and {Magill}, L. and {Mattila}, S. and {Maund}, J. and {Naves}, R. and {Ochner}, P. and {Ruiz}, J. and {Smith}, K. and {Tomasella}, L. and {Turatto}, M.},
        title = "{Optical and near-infrared observations of SN 2011dh - The first 100 days}",
      journal = {\aap},
         year = 2014,
        month = feb,
       volume = {562},
          eid = {A17},
        pages = {A17},
          doi = {10.1051/0004-6361/201321850},
archivePrefix = {arXiv},
       eprint = {1305.1851},
 primaryClass = {astro-ph.SR},
       adsurl = {https://ui.adsabs.harvard.edu/abs/2014A&A...562A..17E}
}

@ARTICLE{Nicholl_2018,
       author = {{Nicholl}, Matt},
        title = "{SuperBol: A User-friendly Python Routine for Bolometric Light Curves}",
      journal = {Research Notes of the American Astronomical Society},
         year = 2018,
        month = dec,
       volume = {2},
       number = {4},
          eid = {230},
        pages = {230},
          doi = {10.3847/2515-5172/aaf799},
       adsurl = {https://ui.adsabs.harvard.edu/abs/2018RNAAS...2..230N}
}

@ARTICLE{thomas11,
       author = {{Thomas}, R.~C. and {Nugent}, P.~E. and {Meza}, J.~C.},
        title = "{SYNAPPS: Data-Driven Analysis for Supernova Spectroscopy}",
      journal = {\pasp},
         year = 2011,
        month = feb,
       volume = {123},
       number = {900},
        pages = {237},
          doi = {10.1086/658673},
       adsurl = {https://ui.adsabs.harvard.edu/abs/2011PASP..123..237T}
}

@INPROCEEDINGS{Castelli2003,
       author = {{Castelli}, F. and {Kurucz}, R.~L.},
        title = "{New Grids of ATLAS9 Model Atmospheres}",
    booktitle = {Modelling of Stellar Atmospheres},
         year = 2003,
       editor = {{Piskunov}, N. and {Weiss}, W.~W. and {Gray}, D.~F.},
       series = {IAU Symposium},
       volume = {210},
        month = jan,
        pages = {A20},
          doi = {10.48550/arXiv.astro-ph/0405087},
archivePrefix = {arXiv},
       eprint = {astro-ph/0405087},
 primaryClass = {astro-ph},
       adsurl = {https://ui.adsabs.harvard.edu/abs/2003IAUS..210P.A20C}
}

@software{Pierel2024,
  author       = {Pierel, Justin},
  title        = {{Space-Phot: Simple Python-Based Photometry for Space Telescopes, Zenodo 12100100}},
  month        = jun,
  year         = 2024,
  publisher    = {Zenodo},
  doi          = {10.5281/zenodo.12100100},
  url          = {https://doi.org/10.5281/zenodo.12100100}
}

@software{Dolphin2016,
       author = {{Dolphin}, Andrew},
        title = "{DOLPHOT: Stellar photometry}",
 howpublished = {Astrophysics Source Code Library, record ascl:1608.013},
         year = 2016,
        month = aug,
          eid = {ascl:1608.013},
archivePrefix = {ascl},
       eprint = {1608.013},
       adsurl = {https://ui.adsabs.harvard.edu/abs/2016ascl.soft08013D}
}

@ARTICLE{VanDyk2023,
       author = {{Van Dyk}, Schuyler D. and {de Graw}, Asia and {Baer-Way}, Raphael and {Zheng}, WeiKang and {Filippenko}, Alexei V. and {Fox}, Ori D. and {Smith}, Nathan and {Brink}, Thomas G. and {de Jaeger}, Thomas and {Kelly}, Patrick L. and {Vasylyev}, Sergiy S.},
        title = "{The disappearances of six supernova progenitors}",
      journal = {\mnras},
         year = 2023,
        month = feb,
       volume = {519},
       number = {1},
        pages = {471-482},
          doi = {10.1093/mnras/stac3549},
archivePrefix = {arXiv},
       eprint = {2212.00179},
 primaryClass = {astro-ph.HE},
       adsurl = {https://ui.adsabs.harvard.edu/abs/2023MNRAS.519..471V}
}

@ARTICLE{VanDyk2024,
       author = {{Van Dyk}, Schuyler D. and {Srinivasan}, Sundar and {Andrews}, Jennifer E. and {Soraisam}, Monika and {Szalai}, Tam{\'a}s and {Howell}, Steve B. and {Isaacson}, Howard and {Matheson}, Thomas and {Petigura}, Erik and {Scicluna}, Peter and {Stephens}, Andrew W. and {Van Zandt}, Judah and {Zheng}, WeiKang and {Chun}, Sang-Hyun and {Fillippenko}, Alexei V.},
        title = "{The SN 2023ixf Progenitor in M101. II. Properties}",
      journal = {\apj},
         year = 2024,
        month = jun,
       volume = {968},
       number = {1},
          eid = {27},
        pages = {27},
          doi = {10.3847/1538-4357/ad414b},
archivePrefix = {arXiv},
       eprint = {2308.14844},
 primaryClass = {astro-ph.SR},
       adsurl = {https://ui.adsabs.harvard.edu/abs/2024ApJ...968...27V}
}

@INPROCEEDINGS{Avila2015,
       author = {{Avila}, R.~J. and {Hack}, W. and {Cara}, M. and {Borncamp}, D. and {Mack}, J. and {Smith}, L. and {Ubeda}, L.},
        title = "{DrizzlePac 2.0 - Introducing New Features}",
    booktitle = {Astronomical Data Analysis Software and Systems XXIV (ADASS XXIV)},
         year = 2015,
       editor = {{Taylor}, A.~R. and {Rosolowsky}, E.},
       series = {Astronomical Society of the Pacific Conference Series},
       volume = {495},
        month = sep,
        pages = {281},
          doi = {10.48550/arXiv.1411.5605},
archivePrefix = {arXiv},
       eprint = {1411.5605},
 primaryClass = {astro-ph.IM},
       adsurl = {https://ui.adsabs.harvard.edu/abs/2015ASPC..495..281A}
}

@ARTICLE{Stanway2018,
       author = {{Stanway}, E.~R. and {Eldridge}, J.~J.},
        title = "{Re-evaluating old stellar populations}",
      journal = {\mnras},
         year = 2018,
        month = sep,
       volume = {479},
       number = {1},
        pages = {75-93},
          doi = {10.1093/mnras/sty1353},
archivePrefix = {arXiv},
       eprint = {1805.08784},
 primaryClass = {astro-ph.GA},
       adsurl = {https://ui.adsabs.harvard.edu/abs/2018MNRAS.479...75S}
}

@ARTICLE{Eldridge2013,
       author = {{Eldridge}, John J. and {Fraser}, Morgan and {Smartt}, Stephen J. and {Maund}, Justyn R. and {Crockett}, R. Mark},
        title = "{The death of massive stars - II. Observational constraints on the progenitors of Type Ibc supernovae}",
      journal = {\mnras},
         year = 2013,
        month = nov,
       volume = {436},
       number = {1},
        pages = {774-795},
          doi = {10.1093/mnras/stt1612},
archivePrefix = {arXiv},
       eprint = {1301.1975},
 primaryClass = {astro-ph.SR},
       adsurl = {https://ui.adsabs.harvard.edu/abs/2013MNRAS.436..774E}
}

@ARTICLE{Eldridge2017,
       author = {{Eldridge}, J.~J. and {Stanway}, E.~R. and {Xiao}, L. and {McClelland}, L.~A.~S. and {Taylor}, G. and {Ng}, M. and {Greis}, S.~M.~L. and {Bray}, J.~C.},
        title = "{Binary Population and Spectral Synthesis Version 2.1: Construction, Observational Verification, and New Results}",
      journal = {\pasa},
         year = 2017,
        month = nov,
       volume = {34},
          eid = {e058},
        pages = {e058},
          doi = {10.1017/pasa.2017.51},
archivePrefix = {arXiv},
       eprint = {1710.02154},
 primaryClass = {astro-ph.SR},
       adsurl = {https://ui.adsabs.harvard.edu/abs/2017PASA...34...58E}
}

@ARTICLE{Sanchez2017,
       author = {{S{\'a}nchez}, S.~F. and {Barrera-Ballesteros}, J.~K. and {S{\'a}nchez-Menguiano}, L. and {Walcher}, C.~J. and {Marino}, R.~A. and {Galbany}, L. and {Bland-Hawthorn}, J. and {Cano-D{\'\i}az}, M. and {Garc{\'\i}a-Benito}, R. and {L{\'o}pez-Cob{\'a}}, C. and {Zibetti}, S. and {Vilchez}, J.~M. and {Igl{\'e}sias-P{\'a}ramo}, J. and {Kehrig}, C. and {L{\'o}pez S{\'a}nchez}, A.~R. and {Duarte Puertas}, S. and {Ziegler}, B.},
        title = "{The mass-metallicity relation revisited with CALIFA}",
      journal = {\mnras},
         year = 2017,
        month = aug,
       volume = {469},
       number = {2},
        pages = {2121-2140},
          doi = {10.1093/mnras/stx808},
archivePrefix = {arXiv},
       eprint = {1703.09769},
 primaryClass = {astro-ph.GA},
       adsurl = {https://ui.adsabs.harvard.edu/abs/2017MNRAS.469.2121S}
}

@ARTICLE{Galbany2016,
       author = {{Galbany}, L. and {Stanishev}, V. and {Mour{\~a}o}, A.~M. and {Rodrigues}, M. and {Flores}, H. and {Walcher}, C.~J. and {S{\'a}nchez}, S.~F. and {Garc{\'\i}a-Benito}, R. and {Mast}, D. and {Badenes}, C. and {Gonz{\'a}lez Delgado}, R.~M. and {Kehrig}, C. and {Lyubenova}, M. and {Marino}, R.~A. and {Moll{\'a}}, M. and {Meidt}, S. and {P{\'e}rez}, E. and {van de Ven}, G. and {V{\'\i}lchez}, J.~M.},
        title = "{Nearby supernova host galaxies from the CALIFA survey. II. Supernova environmental metallicity}",
      journal = {\aap},
         year = 2016,
        month = jun,
       volume = {591},
          eid = {A48},
        pages = {A48},
          doi = {10.1051/0004-6361/201528045},
archivePrefix = {arXiv},
       eprint = {1603.07808},
 primaryClass = {astro-ph.GA},
       adsurl = {https://ui.adsabs.harvard.edu/abs/2016A&A...591A..48G}
}

@ARTICLE{Asplund2009,
       author = {{Asplund}, Martin and {Grevesse}, Nicolas and {Sauval}, A. Jacques and {Scott}, Pat},
        title = "{The Chemical Composition of the Sun}",
      journal = {\araa},
         year = 2009,
        month = sep,
       volume = {47},
       number = {1},
        pages = {481-522},
          doi = {10.1146/annurev.astro.46.060407.145222},
archivePrefix = {arXiv},
       eprint = {0909.0948},
 primaryClass = {astro-ph.SR},
       adsurl = {https://ui.adsabs.harvard.edu/abs/2009ARA&A..47..481A}
}

@ARTICLE{Warwick2026,
       author = {{Warwick}, B. and {Lyman}, J. and {Coppejans}, D.~L. and {Byrne}, C.~M. and {Eldridge}, J.~J. and {Pursiainen}, M. and {Stanway}, E.~R.},
        title = "{The progenitors and circumstellar environments of stripped-envelope interacting supernovae from BPASS}",
      journal = {arXiv e-prints},
         year = 2026,
        month = aug,
          eid = {arXiv:2608.02022},
        pages = {arXiv:2608.02022},
archivePrefix = {arXiv},
       eprint = {2608.02022},
 primaryClass = {astro-ph.HE},
       adsurl = {https://ui.adsabs.harvard.edu/abs/2026arXiv260802022W}
}

@ARTICLE{Claeys2011,
       author = {{Claeys}, J.~S.~W. and {de Mink}, S.~E. and {Pols}, O.~R. and {Eldridge}, J.~J. and {Baes}, M.},
        title = "{Binary progenitor models of type IIb supernovae}",
      journal = {\aap},
         year = 2011,
        month = apr,
       volume = {528},
          eid = {A131},
        pages = {A131},
          doi = {10.1051/0004-6361/201015410},
archivePrefix = {arXiv},
       eprint = {1102.1732},
 primaryClass = {astro-ph.SR},
       adsurl = {https://ui.adsabs.harvard.edu/abs/2011A&A...528A.131C}
}

@ARTICLE{Yoon2017,
       author = {{Yoon}, Sung-Chul and {Dessart}, Luc and {Clocchiatti}, Alejandro},
        title = "{Type Ib and IIb Supernova Progenitors in Interacting Binary Systems}",
      journal = {\apj},
         year = 2017,
        month = may,
       volume = {840},
       number = {1},
          eid = {10},
        pages = {10},
          doi = {10.3847/1538-4357/aa6afe},
archivePrefix = {arXiv},
       eprint = {1701.02089},
 primaryClass = {astro-ph.SR},
       adsurl = {https://ui.adsabs.harvard.edu/abs/2017ApJ...840...10Y}
}

@ARTICLE{Dessart2024,
       author = {{Dessart}, Luc and {Guti{\'e}rrez}, Claudia P. and {Ercolino}, Andrea and {Jin}, Harim and {Langer}, Norbert},
        title = "{A sequence of Type Ib, IIb, II-L, and II-P supernovae from binary-star progenitors with varying initial separations}",
      journal = {\aap},
         year = 2024,
        month = may,
       volume = {685},
          eid = {A169},
        pages = {A169},
          doi = {10.1051/0004-6361/202349066},
archivePrefix = {arXiv},
       eprint = {2402.12977},
 primaryClass = {astro-ph.SR},
       adsurl = {https://ui.adsabs.harvard.edu/abs/2024A&A...685A.169D}
}

@ARTICLE{Folatelli2015,
       author = {{Folatelli}, Gast{\'o}n and {Bersten}, Melina C. and {Kuncarayakti}, Hanindyo and {Benvenuto}, Omar G. and {Maeda}, Keiichi and {Nomoto}, Ken'ichi},
        title = "{The Progenitor of the Type IIb SN 2008ax Revisited}",
      journal = {\apj},
         year = 2015,
        month = oct,
       volume = {811},
       number = {2},
          eid = {147},
        pages = {147},
          doi = {10.1088/0004-637X/811/2/147},
archivePrefix = {arXiv},
       eprint = {1509.01588},
 primaryClass = {astro-ph.SR},
       adsurl = {https://ui.adsabs.harvard.edu/abs/2015ApJ...811..147F}
}

@ARTICLE{Crockett2008,
       author = {{Crockett}, R.~M. and {Eldridge}, J.~J. and {Smartt}, S.~J. and {Pastorello}, A. and {Gal-Yam}, A. and {Fox}, D.~B. and {Leonard}, D.~C. and {Kasliwal}, M.~M. and {Mattila}, S. and {Maund}, J.~R. and {Stephens}, A.~W. and {Danziger}, I.~J.},
        title = "{The type IIb SN 2008ax: the nature of the progenitor}",
      journal = {\mnras},
         year = 2008,
        month = nov,
       volume = {391},
       number = {1},
        pages = {L5-L9},
          doi = {10.1111/j.1745-3933.2008.00540.x},
archivePrefix = {arXiv},
       eprint = {0805.1913},
 primaryClass = {astro-ph},
       adsurl = {https://ui.adsabs.harvard.edu/abs/2008MNRAS.391L...5C}
}

@ARTICLE{Aldering1994,
       author = {{Aldering}, G. and {Humphreys}, R.~M. and {Richmond}, M.},
        title = "{SN 1993J: The Optical Properties of its Progenitor}",
      journal = {\aj},
         year = 1994,
        month = feb,
       volume = {107},
        pages = {662},
          doi = {10.1086/116886},
       adsurl = {https://ui.adsabs.harvard.edu/abs/1994AJ....107..662A}
}

@ARTICLE{Cohen1995,
       author = {{Cohen}, Judith G. and {Darling}, Jeremy and {Porter}, Alain},
        title = "{The Nonvariability of the Progenitor of Supernova 1993J in M81}",
      journal = {\aj},
         year = 1995,
        month = jul,
       volume = {110},
        pages = {308},
          doi = {10.1086/117520},
       adsurl = {https://ui.adsabs.harvard.edu/abs/1995AJ....110..308C}
}

@ARTICLE{VanDyk2002,
       author = {{Van Dyk}, Schuyler D. and {Garnavich}, Peter M. and {Filippenko}, Alexei V. and {H{\"o}flich}, Peter and {Kirshner}, Robert P. and {Kurucz}, Robert L. and {Challis}, Peter},
        title = "{The Progenitor of Supernova 1993J Revisited}",
      journal = {\pasp},
         year = 2002,
        month = dec,
       volume = {114},
       number = {802},
        pages = {1322-1332},
          doi = {10.1086/344382},
archivePrefix = {arXiv},
       eprint = {astro-ph/0208382},
 primaryClass = {astro-ph},
       adsurl = {https://ui.adsabs.harvard.edu/abs/2002PASP..114.1322V}
}

@ARTICLE{Maund2004,
       author = {{Maund}, Justyn R. and {Smartt}, Stephen J. and {Kudritzki}, Rolf P. and {Podsiadlowski}, Philipp and {Gilmore}, Gerard F.},
        title = "{The massive binary companion star to the progenitor of supernova 1993J}",
      journal = {\nat},
         year = 2004,
        month = jan,
       volume = {427},
       number = {6970},
        pages = {129-131},
          doi = {10.1038/nature02161},
archivePrefix = {arXiv},
       eprint = {astro-ph/0401090},
 primaryClass = {astro-ph},
       adsurl = {https://ui.adsabs.harvard.edu/abs/2004Natur.427..129M}
}

@ARTICLE{Stancliffe2009,
       author = {{Stancliffe}, Richard J. and {Eldridge}, John J.},
        title = "{Modelling the binary progenitor of Supernova 1993J}",
      journal = {\mnras},
         year = 2009,
        month = jul,
       volume = {396},
       number = {3},
        pages = {1699-1708},
          doi = {10.1111/j.1365-2966.2009.14849.x},
archivePrefix = {arXiv},
       eprint = {0904.0282},
 primaryClass = {astro-ph.SR},
       adsurl = {https://ui.adsabs.harvard.edu/abs/2009MNRAS.396.1699S}
}

@ARTICLE{Fox2014,
       author = {{Fox}, Ori D. and {Azalee Bostroem}, K. and {Van Dyk}, Schuyler D. and {Filippenko}, Alexei V. and {Fransson}, Claes and {Matheson}, Thomas and {Cenko}, S. Bradley and {Chandra}, Poonam and {Dwarkadas}, Vikram and {Li}, Weidong and {Parker}, Alex H. and {Smith}, Nathan},
        title = "{Uncovering the Putative B-star Binary Companion of the SN 1993J Progenitor}",
      journal = {\apj},
         year = 2014,
        month = jul,
       volume = {790},
       number = {1},
          eid = {17},
        pages = {17},
          doi = {10.1088/0004-637X/790/1/17},
archivePrefix = {arXiv},
       eprint = {1405.4863},
 primaryClass = {astro-ph.HE},
       adsurl = {https://ui.adsabs.harvard.edu/abs/2014ApJ...790...17F}
}

@ARTICLE{Maund2011,
       author = {{Maund}, J.~R. and {Fraser}, M. and {Ergon}, M. and {Pastorello}, A. and {Smartt}, S.~J. and {Sollerman}, J. and {Benetti}, S. and {Botticella}, M.-T. and {Bufano}, F. and {Danziger}, I.~J. and {Kotak}, R. and {Magill}, L. and {Stephens}, A.~W. and {Valenti}, S.},
        title = "{The Yellow Supergiant Progenitor of the Type II Supernova 2011dh in M51}",
      journal = {\apjl},
         year = 2011,
        month = oct,
       volume = {739},
       number = {2},
          eid = {L37},
        pages = {L37},
          doi = {10.1088/2041-8205/739/2/L37},
archivePrefix = {arXiv},
       eprint = {1106.2565},
 primaryClass = {astro-ph.SR},
       adsurl = {https://ui.adsabs.harvard.edu/abs/2011ApJ...739L..37M}
}

@ARTICLE{VanDyk2011,
       author = {{Van Dyk}, Schuyler D. and {Li}, Weidong and {Cenko}, S. Bradley and {Kasliwal}, Mansi M. and {Horesh}, Assaf and {Ofek}, Eran O. and {Kraus}, Adam L. and {Silverman}, Jeffrey M. and {Arcavi}, Iair and {Filippenko}, Alexei V. and {Gal-Yam}, Avishay and {Quimby}, Robert M. and {Kulkarni}, Shrinivas R. and {Yaron}, Ofer and {Polishook}, David},
        title = "{The Progenitor of Supernova 2011dh/PTF11eon in Messier 51}",
      journal = {\apjl},
         year = 2011,
        month = nov,
       volume = {741},
       number = {2},
          eid = {L28},
        pages = {L28},
          doi = {10.1088/2041-8205/741/2/L28},
archivePrefix = {arXiv},
       eprint = {1106.2897},
 primaryClass = {astro-ph.CO},
       adsurl = {https://ui.adsabs.harvard.edu/abs/2011ApJ...741L..28V}
}

@ARTICLE{Bersten2012,
       author = {{Bersten}, Melina C. and {Benvenuto}, Omar G. and {Nomoto}, Ken'ichi and {Ergon}, Mattias and {Folatelli}, Gast{\'o}n and {Sollerman}, Jesper and {Benetti}, Stefano and {Botticella}, Maria Teresa and {Fraser}, Morgan and {Kotak}, Rubina and {Maeda}, Keiichi and {Ochner}, Paolo and {Tomasella}, Lina},
        title = "{The Type IIb Supernova 2011dh from a Supergiant Progenitor}",
      journal = {\apj},
         year = 2012,
        month = sep,
       volume = {757},
       number = {1},
          eid = {31},
        pages = {31},
          doi = {10.1088/0004-637X/757/1/31},
archivePrefix = {arXiv},
       eprint = {1207.5975},
 primaryClass = {astro-ph.HE},
       adsurl = {https://ui.adsabs.harvard.edu/abs/2012ApJ...757...31B}
}

@ARTICLE{Benvenuto2013,
       author = {{Benvenuto}, Omar G. and {Bersten}, Melina C. and {Nomoto}, Ken'ichi},
        title = "{A Binary Progenitor for the Type IIb Supernova 2011dh in M51}",
      journal = {\apj},
         year = 2013,
        month = jan,
       volume = {762},
       number = {2},
          eid = {74},
        pages = {74},
          doi = {10.1088/0004-637X/762/2/74},
archivePrefix = {arXiv},
       eprint = {1207.5807},
 primaryClass = {astro-ph.SR},
       adsurl = {https://ui.adsabs.harvard.edu/abs/2013ApJ...762...74B}
}

@ARTICLE{VanDyk2014,
       author = {{Van Dyk}, Schuyler D. and {Zheng}, WeiKang and {Fox}, Ori D. and {Cenko}, S. Bradley and {Clubb}, Kelsey I. and {Filippenko}, Alexei V. and {Foley}, Ryan J. and {Miller}, Adam A. and {Smith}, Nathan and {Kelly}, Patrick L. and {Lee}, William H. and {Ben-Ami}, Sagi and {Gal-Yam}, Avishay},
        title = "{The Type IIb Supernova 2013df and its Cool Supergiant Progenitor}",
      journal = {\aj},
         year = 2014,
        month = feb,
       volume = {147},
       number = {2},
          eid = {37},
        pages = {37},
          doi = {10.1088/0004-6256/147/2/37},
archivePrefix = {arXiv},
       eprint = {1312.3984},
 primaryClass = {astro-ph.SR},
       adsurl = {https://ui.adsabs.harvard.edu/abs/2014AJ....147...37V}
}

@ARTICLE{Maeda2015,
       author = {{Maeda}, K. and {Hattori}, T. and {Milisavljevic}, D. and {Folatelli}, G. and {Drout}, M.~R. and {Kuncarayakti}, H. and {Margutti}, R. and {Kamble}, A. and {Soderberg}, A. and {Tanaka}, M. and {Kawabata}, M. and {Kawabata}, K.~S. and {Yamanaka}, M. and {Nomoto}, K. and {Kim}, J.~H. and {Simon}, J.~D. and {Phillips}, M.~M. and {Parrent}, J. and {Nakaoka}, T. and {Moriya}, T.~J. and {Suzuki}, A. and {Takaki}, K. and {Ishigaki}, M. and {Sakon}, I. and {Tajitsu}, A. and {Iye}, M.},
        title = "{Type IIb Supernova 2013df Entering into an Interaction Phase: A Link between the Progenitor and the Mass Loss}",
      journal = {\apj},
         year = 2015,
        month = jul,
       volume = {807},
       number = {1},
          eid = {35},
        pages = {35},
          doi = {10.1088/0004-637X/807/1/35},
archivePrefix = {arXiv},
       eprint = {1504.06668},
 primaryClass = {astro-ph.SR},
       adsurl = {https://ui.adsabs.harvard.edu/abs/2015ApJ...807...35M}
}

@ARTICLE{Bersten2018,
       author = {{Bersten}, M.~C. and {Folatelli}, G. and {Garc{\'\i}a}, F. and {van Dyk}, S.~D. and {Benvenuto}, O.~G. and {Orellana}, M. and {Buso}, V. and {S{\'a}nchez}, J.~L. and {Tanaka}, M. and {Maeda}, K. and {Filippenko}, A.~V. and {Zheng}, W. and {Brink}, T.~G. and {Cenko}, S.~B. and {de Jaeger}, T. and {Kumar}, S. and {Moriya}, T.~J. and {Nomoto}, K. and {Perley}, D.~A. and {Shivvers}, I. and {Smith}, N.},
        title = "{A surge of light at the birth of a supernova}",
      journal = {\nat},
         year = 2018,
        month = feb,
       volume = {554},
       number = {7693},
        pages = {497-499},
          doi = {10.1038/nature25151},
archivePrefix = {arXiv},
       eprint = {1802.09360},
 primaryClass = {astro-ph.HE},
       adsurl = {https://ui.adsabs.harvard.edu/abs/2018Natur.554..497B}
}

@ARTICLE{Tartaglia2017,
       author = {{Tartaglia}, L. and {Fraser}, M. and {Sand}, D.~J. and {Valenti}, S. and {Smartt}, S.~J. and {McCully}, C. and {Anderson}, J.~P. and {Arcavi}, I. and {Elias-Rosa}, N. and {Galbany}, L. and {Gal-Yam}, A. and {Haislip}, J.~B. and {Hosseinzadeh}, G. and {Howell}, D.~A. and {Inserra}, C. and {Jha}, S.~W. and {Kankare}, E. and {Lundqvist}, P. and {Maguire}, K. and {Mattila}, S. and {Reichart}, D. and {Smith}, K.~W. and {Smith}, M. and {Stritzinger}, M. and {Sullivan}, M. and {Taddia}, F. and {Tomasella}, L.},
        title = "{The Progenitor and Early Evolution of the Type IIb SN 2016gkg}",
      journal = {\apjl},
         year = 2017,
        month = feb,
       volume = {836},
       number = {1},
          eid = {L12},
        pages = {L12},
          doi = {10.3847/2041-8213/aa5c7f},
archivePrefix = {arXiv},
       eprint = {1611.00419},
 primaryClass = {astro-ph.HE},
       adsurl = {https://ui.adsabs.harvard.edu/abs/2017ApJ...836L..12T}
}

@ARTICLE{Kilpatrick2017,
       author = {{Kilpatrick}, Charles D. and {Foley}, Ryan J. and {Abramson}, Louis E. and {Pan}, Yen-Chen and {Lu}, Cicero-Xinyu and {Williams}, Peter and {Treu}, Tommaso and {Siebert}, Matthew R. and {Fassnacht}, Christopher D. and {Max}, Claire E.},
        title = "{On the progenitor of the Type IIb supernova 2016gkg}",
      journal = {\mnras},
         year = 2017,
        month = mar,
       volume = {465},
       number = {4},
        pages = {4650-4657},
          doi = {10.1093/mnras/stw3082},
archivePrefix = {arXiv},
       eprint = {1610.04587},
 primaryClass = {astro-ph.SR},
       adsurl = {https://ui.adsabs.harvard.edu/abs/2017MNRAS.465.4650K}
}

@ARTICLE{Niu2024,
       author = {{Niu}, Zexi and {Sun}, Ning-Chen and {Liu}, Jifeng},
        title = "{Discovery of a Dusty Yellow Supergiant Progenitor for the Type IIb SN 2017gkk}",
      journal = {\apjl},
         year = 2024,
        month = jul,
       volume = {970},
       number = {1},
          eid = {L9},
        pages = {L9},
          doi = {10.3847/2041-8213/ad5f20},
archivePrefix = {arXiv},
       eprint = {2407.03721},
 primaryClass = {astro-ph.HE},
       adsurl = {https://ui.adsabs.harvard.edu/abs/2024ApJ...970L...9N}
}

@ARTICLE{Reguitti2025,
       author = {{Reguitti}, A. and {Pastorello}, A. and {Smartt}, S.~J. and {Valerin}, G. and {Pignata}, G. and {Campana}, S. and {Chen}, T.-W. and {Sankar}, A.~K. and {Moran}, S. and {Mazzali}, P.~A. and {Duarte}, J. and {Salmaso}, I. and {Anderson}, J.~P. and {Ashall}, C. and {Benetti}, S. and {Gromadzki}, M. and {Guti{\'e}rrez}, C.~P. and {Humina}, C. and {Inserra}, C. and {Kankare}, E. and {Kravtsov}, T. and {Muller-Bravo}, T.~E. and {Pessi}, P.~J. and {Sollerman}, J. and {Young}, D.~R. and {Chambers}, K. and {de Boer}, T. and {Gao}, H. and {Huber}, M. and {Lin}, C.-C. and {Lowe}, T. and {Magnier}, E. and {Minguez}, P. and {Smith}, I.~A. and {Smith}, K.~W. and {Srivastav}, S. and {Wainscoat}, R. and {Benedet}, M.},
        title = "{SN 2024abfo: A partially stripped type II supernova from a yellow supergiant}",
      journal = {\aap},
         year = 2025,
        month = jun,
       volume = {698},
          eid = {A129},
        pages = {A129},
          doi = {10.1051/0004-6361/202554388},
archivePrefix = {arXiv},
       eprint = {2503.03851},
 primaryClass = {astro-ph.HE},
       adsurl = {https://ui.adsabs.harvard.edu/abs/2025A&A...698A.129R}
}

@ARTICLE{Niu2025,
       author = {{Niu}, Zexi and {Sun}, Ning-Chen and {Maund}, Justyn R. and {Guo}, Zhen and {Li}, Wenxiong and {Sun}, Meng and {Liu}, Jifeng},
        title = "{Discovery of a Variable Yellow Supergiant Progenitor for the Type IIb SN 2024abfo}",
      journal = {\apjl},
         year = 2025,
        month = jul,
       volume = {987},
       number = {1},
          eid = {L10},
        pages = {L10},
          doi = {10.3847/2041-8213/ade4cd},
archivePrefix = {arXiv},
       eprint = {2504.20407},
 primaryClass = {astro-ph.HE},
       adsurl = {https://ui.adsabs.harvard.edu/abs/2025ApJ...987L..10N}
}

@ARTICLE{deWet2025,
       author = {{de Wet}, S. and {Leloudas}, G. and {Buckley}, D.~A.~H. and {Erasmus}, N. and {Groot}, P.~J. and {Zimmerman}, E.~A.},
        title = "{A low mass, binary-stripped envelope for the Type IIb SN 2024abfo}",
      journal = {\aap},
         year = 2025,
        month = dec,
       volume = {704},
          eid = {A89},
        pages = {A89},
          doi = {10.1051/0004-6361/202556356},
archivePrefix = {arXiv},
       eprint = {2507.11131},
 primaryClass = {astro-ph.HE},
       adsurl = {https://ui.adsabs.harvard.edu/abs/2025A&A...704A..89D}
}

@ARTICLE{Majumdar_2026_BHTOM,
       author = {{Majumdar}, J. and {Pessi}, P. and {Wyrzykowski}, L. and {Mikolajczyk}, P.~J. and {Kotysz}, K. and {Qvam}, J.~K.~M. and {Dubois}, F. and {Kurowski}, S. and {Skrobacz}, W. and {Siwak}, M. and {Zola}, S. and {Nikolajuk}, M. and {Burzy{\'n}ski}, W. and {Glowacki}, E. and {Zdanavicius}, J. and {Ogloza}, W.},
        title = "{Multi-band Photometric Follow-up of SN 2026dix with BHTOM.space Global Telescope Network, along with the 50-mm Seestar S50 camera}",
      journal = {Transient Name Server AstroNote},
         year = 2026,
        month = apr,
       volume = {105},
        pages = {1},
       adsurl = {https://ui.adsabs.harvard.edu/abs/2026TNSAN.105....1M}
}

@ARTICLE{NV16,
       author = {{Nagy}, A.~P. and {Vink{\'o}}, J.},
        title = "{A two-component model for fitting light curves of core-collapse supernovae}",
      journal = {\aap},
         year = 2016,
        month = may,
       volume = {589},
          eid = {A53},
        pages = {A53},
          doi = {10.1051/0004-6361/201527931},
archivePrefix = {arXiv},
       eprint = {1602.04001},
 primaryClass = {astro-ph.IM},
       adsurl = {https://ui.adsabs.harvard.edu/abs/2016A&A...589A..53N}
}

@ARTICLE{Arnett_Fu_1989,
       author = {{Arnett}, W. David and {Fu}, Albert},
        title = "{The Late Behavior of Supernova 1987A. I. The Light Curve}",
      journal = {\apj},
         year = 1989,
        month = may,
       volume = {340},
        pages = {396},
          doi = {10.1086/167402},
       adsurl = {https://ui.adsabs.harvard.edu/abs/1989ApJ...340..396A}
}

@ARTICLE{Dessart_2016,
       author = {{Dessart}, Luc and {Hillier}, D. John and {Woosley}, Stan and {Livne}, Eli and {Waldman}, Roni and {Yoon}, Sung-Chul and {Langer}, Norbert},
        title = "{Inferring supernova IIb/Ib/Ic ejecta properties from light curves and spectra: correlations from radiative-transfer models}",
      journal = {\mnras},
         year = 2016,
        month = may,
       volume = {458},
       number = {2},
        pages = {1618-1635},
          doi = {10.1093/mnras/stw418},
archivePrefix = {arXiv},
       eprint = {1602.06280},
 primaryClass = {astro-ph.SR},
       adsurl = {https://ui.adsabs.harvard.edu/abs/2016MNRAS.458.1618D}
}

@ARTICLE{Wang_2023,
       author = {{Wang}, Tao and {Wang}, Shan-Qin and {Gan}, Wen-Pei and {Li}, Long},
        title = "{SN 2018gk Revisited: the Photosphere, the Central Engine, and the Putative Dust}",
      journal = {\apj},
         year = 2023,
        month = may,
       volume = {948},
       number = {2},
          eid = {138},
        pages = {138},
          doi = {10.3847/1538-4357/acc24d},
archivePrefix = {arXiv},
       eprint = {2211.15966},
 primaryClass = {astro-ph.HE},
       adsurl = {https://ui.adsabs.harvard.edu/abs/2023ApJ...948..138W}
}

@ARTICLE{Bose_2021,
       author = {{Bose}, Subhash and {Dong}, Subo and {Kochanek}, C.~S. and {Stritzinger}, M.~D. and {Ashall}, Chris and {Benetti}, Stefano and {Falco}, E. and {Filippenko}, Alexei V. and {Pastorello}, Andrea and {Prieto}, Jose L. and {Somero}, Auni and {Sukhbold}, Tuguldur and {Zhang}, Junbo and {Auchettl}, Katie and {Brink}, Thomas G. and {Brown}, J.~S. and {Chen}, Ping and {Fiore}, A. and {Grupe}, Dirk and {Holoien}, T.~W.-S. and {Lundqvist}, Peter and {Mattila}, Seppo and {Mutel}, Robert and {Pooley}, David and {Post}, R.~S. and {Reddy}, Naveen and {Reynolds}, Thomas M. and {Shappee}, Benjamin J. and {Stanek}, K.~Z. and {Thompson}, Todd A. and {Villanueva}, Jr., S. and {Zheng}, WeiKang},
        title = "{ASASSN-18am/SN 2018gk: an overluminous Type IIb supernova from a massive progenitor}",
      journal = {\mnras},
         year = 2021,
        month = may,
       volume = {503},
       number = {3},
        pages = {3472-3491},
          doi = {10.1093/mnras/stab629},
archivePrefix = {arXiv},
       eprint = {2007.00008},
 primaryClass = {astro-ph.HE},
       adsurl = {https://ui.adsabs.harvard.edu/abs/2021MNRAS.503.3472B}
}

@ARTICLE{Nagy_2018,
       author = {{Nagy}, Andrea P.},
        title = "{Average Opacity Calculation for Core-collapse Supernovae}",
      journal = {\apj},
         year = 2018,
        month = aug,
       volume = {862},
       number = {2},
          eid = {143},
        pages = {143},
          doi = {10.3847/1538-4357/aace56},
archivePrefix = {arXiv},
       eprint = {1806.07188},
 primaryClass = {astro-ph.HE},
       adsurl = {https://ui.adsabs.harvard.edu/abs/2018ApJ...862..143N}
}

@ARTICLE{Kasen_Bildsten_2010,
       author = {{Kasen}, Daniel and {Bildsten}, Lars},
        title = "{Supernova Light Curves Powered by Young Magnetars}",
      journal = {\apj},
         year = 2010,
        month = jul,
       volume = {717},
       number = {1},
        pages = {245-249},
          doi = {10.1088/0004-637X/717/1/245},
archivePrefix = {arXiv},
       eprint = {0911.0680},
 primaryClass = {astro-ph.HE},
       adsurl = {https://ui.adsabs.harvard.edu/abs/2010ApJ...717..245K}
}

@ARTICLE{Laplace_2021,
       author = {{Laplace}, E. and {Justham}, S. and {Renzo}, M. and {G{\"o}tberg}, Y. and {Farmer}, R. and {Vartanyan}, D. and {de Mink}, S.~E.},
        title = "{Different to the core: The pre-supernova structures of massive single and binary-stripped stars}",
      journal = {\aap},
         year = 2021,
        month = dec,
       volume = {656},
          eid = {A58},
        pages = {A58},
          doi = {10.1051/0004-6361/202140506},
archivePrefix = {arXiv},
       eprint = {2102.05036},
 primaryClass = {astro-ph.SR},
       adsurl = {https://ui.adsabs.harvard.edu/abs/2021A&A...656A..58L}
}

@ARTICLE{Smartt_2009,
       author = {{Smartt}, Stephen J.},
        title = "{Progenitors of Core-Collapse Supernovae}",
      journal = {\araa},
         year = 2009,
        month = sep,
       volume = {47},
       number = {1},
        pages = {63-106},
          doi = {10.1146/annurev-astro-082708-101737},
archivePrefix = {arXiv},
       eprint = {0908.0700},
 primaryClass = {astro-ph.SR},
       adsurl = {https://ui.adsabs.harvard.edu/abs/2009ARA&A..47...63S}
}

@INCOLLECTION{Limongi_2017_HSN,
       author = {{Limongi}, Marco},
        title = "{Supernovae from Massive Stars}",
    booktitle = {Handbook of Supernovae},
         year = 2017,
       editor = {{Alsabti}, Athem W. and {Murdin}, Paul},
        pages = {513},
          doi = {10.1007/978-3-319-21846-5_119},
       adsurl = {https://ui.adsabs.harvard.edu/abs/2017hsn..book..513L}
}

@ARTICLE{Smith_2014_rev,
       author = {{Smith}, Nathan},
        title = "{Mass Loss: Its Effect on the Evolution and Fate of High-Mass Stars}",
      journal = {\araa},
         year = 2014,
        month = aug,
       volume = {52},
        pages = {487-528},
          doi = {10.1146/annurev-astro-081913-040025},
archivePrefix = {arXiv},
       eprint = {1402.1237},
 primaryClass = {astro-ph.SR},
       adsurl = {https://ui.adsabs.harvard.edu/abs/2014ARA&A..52..487S}
}

@ARTICLE{Modjaz_2014,
       author = {{Modjaz}, M. and {Blondin}, S. and {Kirshner}, R.~P. and {Matheson}, T. and {Berlind}, P. and {Bianco}, F.~B. and {Calkins}, M.~L. and {Challis}, P. and {Garnavich}, P. and {Hicken}, M. and {Jha}, S. and {Liu}, Y.~Q. and {Marion}, G.~H.},
        title = "{Optical Spectra of 73 Stripped-envelope Core-collapse Supernovae}",
      journal = {\aj},
         year = 2014,
        month = may,
       volume = {147},
       number = {5},
          eid = {99},
        pages = {99},
          doi = {10.1088/0004-6256/147/5/99},
archivePrefix = {arXiv},
       eprint = {1405.1910},
 primaryClass = {astro-ph.HE},
       adsurl = {https://ui.adsabs.harvard.edu/abs/2014AJ....147...99M}
}

@ARTICLE{Bianco_2014,
       author = {{Bianco}, F.~B. and {Modjaz}, M. and {Hicken}, M. and {Friedman}, A. and {Kirshner}, R.~P. and {Bloom}, J.~S. and {Challis}, P. and {Marion}, G.~H. and {Wood-Vasey}, W.~M. and {Rest}, A.},
        title = "{Multi-color Optical and Near-infrared Light Curves of 64 Stripped-envelope Core-Collapse Supernovae}",
      journal = {\apjs},
         year = 2014,
        month = aug,
       volume = {213},
       number = {2},
          eid = {19},
        pages = {19},
          doi = {10.1088/0067-0049/213/2/19},
archivePrefix = {arXiv},
       eprint = {1405.1428},
 primaryClass = {astro-ph.SR},
       adsurl = {https://ui.adsabs.harvard.edu/abs/2014ApJS..213...19B}
}

@ARTICLE{Taddia_2018,
       author = {{Taddia}, F. and {Stritzinger}, M.~D. and {Bersten}, M. and {Baron}, E. and {Burns}, C. and {Contreras}, C. and {Holmbo}, S. and {Hsiao}, E.~Y. and {Morrell}, N. and {Phillips}, M.~M. and {Sollerman}, J. and {Suntzeff}, N.~B.},
        title = "{The Carnegie Supernova Project I. Analysis of stripped-envelope supernova light curves}",
      journal = {\aap},
         year = 2018,
        month = feb,
       volume = {609},
          eid = {A136},
        pages = {A136},
          doi = {10.1051/0004-6361/201730844},
archivePrefix = {arXiv},
       eprint = {1707.07614},
 primaryClass = {astro-ph.HE},
       adsurl = {https://ui.adsabs.harvard.edu/abs/2018A&A...609A.136T}
}

@ARTICLE{Liu_2016,
       author = {{Liu}, Yu-Qian and {Modjaz}, Maryam and {Bianco}, Federica B. and {Graur}, Or},
        title = "{Analyzing the Largest Spectroscopic Data Set of Stripped Supernovae to Improve Their Identifications and Constrain Their Progenitors}",
      journal = {\apj},
         year = 2016,
        month = aug,
       volume = {827},
       number = {2},
          eid = {90},
        pages = {90},
          doi = {10.3847/0004-637X/827/2/90},
archivePrefix = {arXiv},
       eprint = {1510.08049},
 primaryClass = {astro-ph.HE},
       adsurl = {https://ui.adsabs.harvard.edu/abs/2016ApJ...827...90L}
}

@INCOLLECTION{Gal-Yam_2017_HSN,
       author = {{Gal-Yam}, Avishay},
        title = "{Observational and Physical Classification of Supernovae}",
    booktitle = {Handbook of Supernovae},
         year = 2017,
       editor = {{Alsabti}, Athem W. and {Murdin}, Paul},
        pages = {195},
          doi = {10.1007/978-3-319-21846-5_35},
       adsurl = {https://ui.adsabs.harvard.edu/abs/2017hsn..book..195G}
}

@ARTICLE{Prentice_2019,
       author = {{Prentice}, S.~J. and {Ashall}, C. and {James}, P.~A. and {Short}, L. and {Mazzali}, P.~A. and {Bersier}, D. and {Crowther}, P.~A. and {Barbarino}, C. and {Chen}, T.-W. and {Copperwheat}, C.~M. and {Darnley}, M.~J. and {Denneau}, L. and {Elias-Rosa}, N. and {Fraser}, M. and {Galbany}, L. and {Gal-Yam}, A. and {Harmanen}, J. and {Howell}, D.~A. and {Hosseinzadeh}, G. and {Inserra}, C. and {Kankare}, E. and {Karamehmetoglu}, E. and {Lamb}, G.~P. and {Limongi}, M. and {Maguire}, K. and {McCully}, C. and {Olivares E}, F. and {Piascik}, A.~S. and {Pignata}, G. and {Reichart}, D.~E. and {Rest}, A. and {Reynolds}, T. and {Rodr{\'\i}guez}, {\'O}. and {Saario}, J.~L.~O. and {Schulze}, S. and {Smartt}, S.~J. and {Smith}, K.~W. and {Sollerman}, J. and {Stalder}, B. and {Sullivan}, M. and {Taddia}, F. and {Valenti}, S. and {Vergani}, S.~D. and {Williams}, S.~C. and {Young}, D.~R.},
        title = "{Investigating the properties of stripped-envelope supernovae; what are the implications for their progenitors?}",
      journal = {\mnras},
         year = 2019,
        month = may,
       volume = {485},
       number = {2},
        pages = {1559-1578},
          doi = {10.1093/mnras/sty3399},
archivePrefix = {arXiv},
       eprint = {1812.03716},
 primaryClass = {astro-ph.HE},
       adsurl = {https://ui.adsabs.harvard.edu/abs/2019MNRAS.485.1559P}
}

@ARTICLE{Sukhbold_2016,
       author = {{Sukhbold}, Tuguldur and {Ertl}, T. and {Woosley}, S.~E. and {Brown}, Justin M. and {Janka}, H.-T.},
        title = "{Core-collapse Supernovae from 9 to 120 Solar Masses Based on Neutrino-powered Explosions}",
      journal = {\apj},
         year = 2016,
        month = apr,
       volume = {821},
       number = {1},
          eid = {38},
        pages = {38},
          doi = {10.3847/0004-637X/821/1/38},
archivePrefix = {arXiv},
       eprint = {1510.04643},
 primaryClass = {astro-ph.HE},
       adsurl = {https://ui.adsabs.harvard.edu/abs/2016ApJ...821...38S}
}

@ARTICLE{Dessart_2015,
       author = {{Dessart}, Luc and {Hillier}, D. John and {Woosley}, Stan and {Livne}, Eli and {Waldman}, Roni and {Yoon}, Sung-Chul and {Langer}, Norbert},
        title = "{Radiative-transfer models for supernovae IIb/Ib/Ic from binary-star progenitors}",
      journal = {\mnras},
         year = 2015,
        month = oct,
       volume = {453},
       number = {2},
        pages = {2189-2213},
          doi = {10.1093/mnras/stv1747},
archivePrefix = {arXiv},
       eprint = {1507.07783},
 primaryClass = {astro-ph.SR},
       adsurl = {https://ui.adsabs.harvard.edu/abs/2015MNRAS.453.2189D}
}

@ARTICLE{Chevalier_Soderberg_2010,
       author = {{Chevalier}, Roger A. and {Soderberg}, Alicia M.},
        title = "{Type IIb Supernovae with Compact and Extended Progenitors}",
      journal = {\apjl},
         year = 2010,
        month = mar,
       volume = {711},
       number = {1},
        pages = {L40-L43},
          doi = {10.1088/2041-8205/711/1/L40},
archivePrefix = {arXiv},
       eprint = {0911.3408},
 primaryClass = {astro-ph.HE},
       adsurl = {https://ui.adsabs.harvard.edu/abs/2010ApJ...711L..40C}
}

@ARTICLE{Szalai_2016,
       author = {{Szalai}, Tam{\'a}s and {Vink{\'o}}, J{\'o}zsef and {Nagy}, Andrea P. and {Silverman}, Jeffrey M. and {Wheeler}, J. Craig and {Dhungana}, Govinda and {Marion}, G. Howie and {Kehoe}, Robert and {Fox}, Ori D. and {S{\'a}rneczky}, Kriszti{\'a}n and {Marschalk{\'o}}, G{\'a}bor and {B{\'\i}r{\'o}}, Barna I. and {Borkovits}, Tam{\'a}s and {Heged{\"u}s}, Tibor and {Szak{\'a}ts}, R{\'o}bert and {Ferrante}, Farley V. and {B{\'a}nyai}, Evelin and {Hodos{\'a}n}, Gabriella and {Kelemen}, J{\'a}nos and {P{\'a}l}, Andr{\'a}s},
        title = "{The continuing story of SN IIb 2013df: new optical and IR observations and analysis}",
      journal = {\mnras},
         year = 2016,
        month = aug,
       volume = {460},
       number = {2},
        pages = {1500-1518},
          doi = {10.1093/mnras/stw1031},
archivePrefix = {arXiv},
       eprint = {1604.08046},
 primaryClass = {astro-ph.SR},
       adsurl = {https://ui.adsabs.harvard.edu/abs/2016MNRAS.460.1500S}
}

@ARTICLE{Arcavi_2011,
       author = {{Arcavi}, Iair and {Gal-Yam}, Avishay and {Yaron}, Ofer and {Sternberg}, Assaf and {Rabinak}, Itay and {Waxman}, Eli and {Kasliwal}, Mansi M. and {Quimby}, Robert M. and {Ofek}, Eran O. and {Horesh}, Assaf and {Kulkarni}, Shrinivas R. and {Filippenko}, Alexei V. and {Silverman}, Jeffrey M. and {Cenko}, S. Bradley and {Li}, Weidong and {Bloom}, Joshua S. and {Sullivan}, Mark and {Nugent}, Peter E. and {Poznanski}, Dovi and {Gorbikov}, Evgeny and {Fulton}, Benjamin J. and {Howell}, D. Andrew and {Bersier}, David and {Riou}, Amedee and {Lamotte-Bailey}, Stephane and {Griga}, Thomas and {Cohen}, Judith G. and {Hachinger}, Stephan and {Polishook}, David and {Xu}, Dong and {Ben-Ami}, Sagi and {Manulis}, Ilan and {Walker}, Emma S. and {Maguire}, Kate and {Pan}, Yen-Chen and {Matheson}, Thomas and {Mazzali}, Paolo A. and {Pian}, Elena and {Fox}, Derek B. and {Gehrels}, Neil and {Law}, Nicholas and {James}, Philip and {Marchant}, Jonathan M. and {Smith}, Robert J. and {Mottram}, Chris J. and {Barnsley}, Robert M. and {Kandrashoff}, Michael T. and {Clubb}, Kelsey I.},
        title = "{SN 2011dh: Discovery of a Type IIb Supernova from a Compact Progenitor in the Nearby Galaxy M51}",
      journal = {\apjl},
         year = 2011,
        month = dec,
       volume = {742},
       number = {2},
          eid = {L18},
        pages = {L18},
          doi = {10.1088/2041-8205/742/2/L18},
archivePrefix = {arXiv},
       eprint = {1106.3551},
 primaryClass = {astro-ph.CO},
       adsurl = {https://ui.adsabs.harvard.edu/abs/2011ApJ...742L..18A}
}

@ARTICLE{Banhidi_2025,
       author = {{B{\'a}nhidi}, D. and {Barna}, B. and {Szalai}, T. and {Vink{\'o}}, J. and {B{\'\i}r{\'o}}, I.~B. and {Bostroem}, K.~A. and {Cs{\'a}nyi}, I. and {Davis}, K.~W. and {Foley}, R.~J. and {Galbany}, L. and {Jha}, S.~W. and {Howell}, D.~A. and {Kwok}, L.~A. and {P{\'a}l}, A. and {Pellegrino}, C. and {Rojas-Bravo}, C. and {Sz{\'e}kely}, P. and {Taggart}, K. and {Terreran}, G. and {Tinyanont}, S.},
        title = "{SN 2022xlp: The second-known well-observed, intermediate-luminosity Iax supernova}",
      journal = {\aap},
         year = 2025,
        month = nov,
       volume = {703},
          eid = {A64},
        pages = {A64},
          doi = {10.1051/0004-6361/202553922},
archivePrefix = {arXiv},
       eprint = {2509.07717},
 primaryClass = {astro-ph.SR},
       adsurl = {https://ui.adsabs.harvard.edu/abs/2025A&A...703A..64B}
}

@ARTICLE{Boccioli_2024_rev,
       author = {{Boccioli}, Luca and {Roberti}, Lorenzo},
        title = "{The Physics of Core-Collapse Supernovae: Explosion Mechanism and Explosive Nucleosynthesis}",
      journal = {Universe},
         year = 2024,
        month = mar,
       volume = {10},
       number = {3},
          eid = {148},
        pages = {148},
          doi = {10.3390/universe10030148},
archivePrefix = {arXiv},
       eprint = {2403.12942},
 primaryClass = {astro-ph.SR},
       adsurl = {https://ui.adsabs.harvard.edu/abs/2024Univ...10..148B}
}

@ARTICLE{Shivvers_2019,
       author = {{Shivvers}, Isaac and {Filippenko}, Alexei V. and {Silverman}, Jeffrey M. and {Zheng}, WeiKang and {Foley}, Ryan J. and {Chornock}, Ryan and {Barth}, Aaron J. and {Cenko}, S. Bradley and {Clubb}, Kelsey I. and {Fox}, Ori D. and {Ganeshalingam}, Mohan and {Graham}, Melissa L. and {Kelly}, Patrick L. and {Kleiser}, Io K.~W. and {Leonard}, Douglas C. and {Li}, Weidong and {Matheson}, Thomas and {Mauerhan}, Jon C. and {Modjaz}, Maryam and {Serduke}, Franklin J.~D. and {Shields}, Joseph C. and {Steele}, Thea N. and {Swift}, Brandon J. and {Wong}, Diane S. and {Yuk}, Heechan},
        title = "{The Berkeley sample of stripped-envelope supernovae}",
      journal = {\mnras},
         year = 2019,
        month = jan,
       volume = {482},
       number = {2},
        pages = {1545-1556},
          doi = {10.1093/mnras/sty2719},
archivePrefix = {arXiv},
       eprint = {1810.03650},
 primaryClass = {astro-ph.SR},
       adsurl = {https://ui.adsabs.harvard.edu/abs/2019MNRAS.482.1545S}
}

@ARTICLE{Fang_2022,
       author = {{Fang}, Qiliang and {Maeda}, Keiichi and {Kuncarayakti}, Hanindyo and {Tanaka}, Masaomi and {Kawabata}, Koji S. and {Hattori}, Takashi and {Aoki}, Kentaro and {Moriya}, Takashi J. and {Yamanaka}, Masayuki},
        title = "{Statistical Properties of the Nebular Spectra of 103 Stripped-envelope Core-collapse Supernovae}",
      journal = {\apj},
         year = 2022,
        month = apr,
       volume = {928},
       number = {2},
          eid = {151},
        pages = {151},
          doi = {10.3847/1538-4357/ac4f60},
archivePrefix = {arXiv},
       eprint = {2201.11467},
 primaryClass = {astro-ph.HE},
       adsurl = {https://ui.adsabs.harvard.edu/abs/2022ApJ...928..151F}
}

@ARTICLE{Woosley_2021,
       author = {{Woosley}, S.~E. and {Sukhbold}, Tuguldur and {Kasen}, D.~N.},
        title = "{Model Light Curves for Type Ib and Ic Supernovae}",
      journal = {\apj},
         year = 2021,
        month = jun,
       volume = {913},
       number = {2},
          eid = {145},
        pages = {145},
          doi = {10.3847/1538-4357/abf3be},
archivePrefix = {arXiv},
       eprint = {2009.06868},
 primaryClass = {astro-ph.HE},
       adsurl = {https://ui.adsabs.harvard.edu/abs/2021ApJ...913..145W}
}

@ARTICLE{Stritzinger_2023,
       author = {{Stritzinger}, M.~D. and {Holmbo}, S. and {Morrell}, N. and {Phillips}, M.~M. and {Burns}, C.~R. and {Castell{\'o}n}, S. and {Folatelli}, G. and {Hamuy}, M. and {Leloudas}, G. and {Suntzeff}, N.~B. and {Anderson}, J.~P. and {Ashall}, C. and {Baron}, E. and {Boissier}, S. and {Hsiao}, E.~Y. and {Karamehmetoglu}, E. and {Olivares}, F.},
        title = "{The Carnegie Supernova Project I. Optical spectroscopy of stripped-envelope supernovae}",
      journal = {\aap},
         year = 2023,
        month = jul,
       volume = {675},
          eid = {A82},
        pages = {A82},
          doi = {10.1051/0004-6361/202243376},
archivePrefix = {arXiv},
       eprint = {2302.11303},
 primaryClass = {astro-ph.HE},
       adsurl = {https://ui.adsabs.harvard.edu/abs/2023A&A...675A..82S}
}

@ARTICLE{Holmbo_2023,
       author = {{Holmbo}, S. and {Stritzinger}, M.~D. and {Karamehmetoglu}, E. and {Burns}, C.~R. and {Morrell}, N. and {Ashall}, C. and {Hsiao}, E.~Y. and {Galbany}, L. and {Folatelli}, G. and {Phillips}, M.~M. and {Baron}, E. and {Guti{\'e}rrez}, C.~P. and {Leloudas}, G. and {M{\"u}ller-Bravo}, T.~E. and {Hoeflich}, P. and {Taddia}, F. and {Suntzeff}, N.~B.},
        title = "{The Carnegie Supernova Project I. Spectroscopic analysis of stripped-envelope supernovae}",
      journal = {\aap},
         year = 2023,
        month = jul,
       volume = {675},
          eid = {A83},
        pages = {A83},
          doi = {10.1051/0004-6361/202245334},
archivePrefix = {arXiv},
       eprint = {2302.11304},
 primaryClass = {astro-ph.HE},
       adsurl = {https://ui.adsabs.harvard.edu/abs/2023A&A...675A..83H}
}

@ARTICLE{Filippenko_97_rev,
       author = {{Filippenko}, Alexei V.},
        title = "{Optical Spectra of Supernovae}",
      journal = {\araa},
         year = 1997,
        month = jan,
       volume = {35},
        pages = {309-355},
          doi = {10.1146/annurev.astro.35.1.309},
       adsurl = {https://ui.adsabs.harvard.edu/abs/1997ARA&A..35..309F}
}

@ARTICLE{Filippenko_1988,
       author = {{Filippenko}, Alexei V.},
        title = "{Supernova 1987K: Type II in Youth, Type Ib in Old Age}",
      journal = {\aj},
         year = 1988,
        month = dec,
       volume = {96},
        pages = {1941},
          doi = {10.1086/114940},
       adsurl = {https://ui.adsabs.harvard.edu/abs/1988AJ.....96.1941F}
}

\newpage
\begin{appendix}
\nolinenumbers

\onecolumn
\section{Photometry}


\begin{table*}[!ht]
\begin{center}
\caption{$BVgriz$ photometry of SN~2026dix obtained from the Baja Observatory}
\label{tab:phot_data_Baja}
\begin{tabular}{ccccccc}
\hline
\hline
MJD & B (mag) & V (mag) & g (mag) & r (mag) & i (mag) & z (mag) \\ 
\hline
61089.8 & 17.80 (0.13) & 17.13 (0.05) & 17.30 (0.06) & 16.87 (0.04) & 16.83 (0.05) & 16.83 (0.07) \\
61095.8 & 16.81 (0.12) & 16.00 (0.05) & 16.25 (0.05) & 15.78 (0.04) & 15.76 (0.03) & 15.68 (0.05) \\
61098.1 & 16.73 (0.10) & 15.86 (0.03) & 16.04 (0.04) & 15.69 (0.02) & 15.60 (0.02) & 15.53 (0.03) \\
61099.1 & 16.66 (0.11) & 15.80 (0.03) & 16.00 (0.05) & 15.59 (0.02) & 15.53 (0.03) & 15.46 (0.04) \\
61099.9 & 16.65 (0.10) & 15.74 (0.04) & 16.02 (0.06) & 15.56 (0.02) & 15.49 (0.04) & 15.40 (0.05) \\
61100.8 & 16.69 (0.11) & 15.75 (0.04) & 15.98 (0.05) & 15.55 (0.02) & 15.47 (0.02) & 15.35 (0.04) \\
61103.0 & 16.64 (0.14) & 15.65 (0.06) & 15.98 (0.05) & 15.46 (0.05) & 15.32 (0.06) & 15.24 (0.05) \\
61104.9 & 16.76 (0.10) & 15.65 (0.04) & 15.98 (0.07) & 15.45 (0.03) & 15.32 (0.03) & 15.21 (0.04) \\
61105.9 & 16.81 (0.12) & 15.63 (0.05) & 15.99 (0.06) & 15.39 (0.04) & 15.28 (0.04) & 15.18 (0.03) \\
61107.0 & 16.77 (0.11) & 15.59 (0.07) & 15.95 (0.07) & 15.38 (0.05) & 15.25 (0.06) & 15.13 (0.06) \\
61113.9 & 17.18 (0.11) & 15.77 (0.05) & 16.28 (0.04) & 15.44 (0.03) & 15.29 (0.03) & ... \\
61114.9 & 17.27 (0.08) & 15.85 (0.04) & 16.35 (0.03) & 15.49 (0.03) & 15.30 (0.02) & 15.14 (0.04) \\
61121.9 & 17.88 (0.15) & 16.08 (0.05) & 16.80 (0.02) & 15.70 (0.01) & 15.50 (0.02) & 15.29 (0.02) \\
61131.9 & ... & 16.70 (0.09) & 17.35 (0.06) & 16.12 (0.02) & 15.79 (0.02) & 15.50 (0.03) \\
61133.9 & ... & 16.81 (0.03) & 17.34 (0.04) & 16.16 (0.02) & 15.77 (0.03) & 15.52 (0.03) \\
61134.9 & 18.71 (0.16) & 16.78 (0.04) & 17.51 (0.02) & 16.24 (0.01) & 15.89 (0.01) & 15.56 (0.02) \\
61135.9 & ... & ... & 17.43 (0.03) & 16.24 (0.03) & 15.86 (0.01) & 15.59 (0.02) \\
61136.9 & 18.82 (0.22) & 16.79 (0.06) & ... & ... & 15.96 (0.11) & ... \\
61137.9 & ... & 16.95 (0.04) & ... & 16.28 (0.13) & 15.97 (0.08) & ... \\
61139.0 & ... & 16.93 (0.03) & 17.33 (0.03) & 16.37 (0.01) & 15.98 (0.01) & 15.66 (0.02) \\
61141.8 & ... & 17.04 (0.03) & 17.61 (0.03) & 16.46 (0.01) & 16.04 (0.02) & 15.73 (0.02) \\
61147.0 & ... & 17.05 (0.04) & 17.31 (0.04) & 16.55 (0.01) & 16.15 (0.01) & 15.78 (0.02) \\
61148.0 & 18.99 (0.22) & 17.12 (0.04) & 17.64 (0.04) & ... & ... & 15.80 (0.03) \\
61149.0 & ... & 17.12 (0.03) & ... & 16.60 (0.02) & 16.17 (0.02) & 15.78 (0.02) \\
61153.0 & ... & 17.17 (0.03) & 17.62 (0.04) & 16.68 (0.02) & 16.27 (0.01) & 15.92 (0.03) \\
61153.9 & ... & 17.17 (0.03) & 17.70 (0.03) & 16.68 (0.02) & 16.29 (0.01) & 15.87 (0.03) \\
61154.9 & ... & 17.19 (0.03) & 17.70 (0.04) & 16.73 (0.02) & 16.32 (0.01) & 15.91 (0.02) \\
61156.0 & ... & 17.28 (0.04) & 17.78 (0.03) & 16.70 (0.02) & 16.30 (0.02) & 15.95 (0.02) \\
61160.9 & ... & 17.26 (0.04) & 17.88 (0.05) & 16.86 (0.03) & 16.36 (0.02) & 15.94 (0.03) \\
61163.9 & ... & 17.28 (0.04) & 17.84 (0.07) & 16.88 (0.02) & 16.50 (0.02) & 15.99 (0.04) \\
61173.9 & 19.08 (0.22) & 17.37 (0.05) & 17.89 (0.04) & 17.05 (0.02) & 16.64 (0.02) & 16.09 (0.04) \\
61181.9 & ... & 17.51 (0.06) & 18.06 (0.05) & 17.13 (0.02) & 16.80 (0.03) & 16.21 (0.03) \\
61183.9 & ... & ... & ... & 17.15 (0.03) & ... & ... \\
61185.9 & 19.07 (0.18) & ... & ... & 17.24 (0.04) & 16.85 (0.04) & 16.31 (0.03) \\
61187.9 & ... & ... & ... & ... & ... & ... \\
61189.9 & ... & ... & 18.16 (0.07) & ... & 16.93 (0.03) & ... \\
\hline
\end{tabular}
\end{center}
\end{table*}

\clearpage
\begin{table*}[!ht]
\begin{center}
\caption{$BVgriz$ photometry of SN~2026dix obtained from the Konkoly Observatory}
\label{tab:phot_data_Konkoly}
\begin{tabular}{ccccccc}
\hline
\hline
MJD & B (mag) & V (mag) & g (mag) & r (mag) & i (mag) & z (mag) \\ 
\hline
61096.7 & ... & ... & ... & 15.72 (0.16) & ... & ... \\
61097.7 & ... & 15.87 (0.16) & ... & ... & ... & ... \\
61098.7 & ... & 15.75 (0.05) & ... & 15.58 (0.02) & 15.57 (0.02) & 15.43 (0.03) \\
61100.1 & ... & ... & 16.00 (0.02) & 15.53 (0.01) & 15.45 (0.01) & 15.28 (0.03) \\
61100.7 & 16.65 (0.14) & 15.74 (0.07) & 15.99 (0.05) & 15.51 (0.03) & 15.46 (0.02) & 15.33 (0.04) \\
61102.1 & 16.75 (0.14) & 15.64 (0.06) & ... & ... & ... & ... \\
61102.8 & 16.62 (0.15) & 15.70 (0.07) & 16.00 (0.05) & 15.44 (0.02) & 15.39 (0.02) & 15.20 (0.06) \\
61104.0 & ... & 15.61 (0.09) & ... & ... & ... & ... \\
61104.7 & ... & 15.62 (0.04) & ... & ... & ... & ... \\
61105.7 & ... & ... & 15.99 (0.03) & 15.39 (0.02) & 15.29 (0.01) & 15.14 (0.02) \\
61106.7 & ... & 15.64 (0.13) & 16.02 (0.03) & 15.39 (0.01) & 15.30 (0.01) & 15.12 (0.03) \\
61108.7 & ... & 15.64 (0.06) & 15.97 (0.03) & 15.37 (0.02) & 15.27 (0.01) & 15.10 (0.03) \\
61109.7 & ... & 15.61 (0.11) & ... & ... & ... & ... \\
61110.8 & ... & ... & 16.00 (0.03) & 15.37 (0.01) & 15.25 (0.01) & 15.14 (0.04) \\
61111.9 & ... & ... & 16.34 (0.02) & 15.41 (0.01) & 15.26 (0.01) & 15.04 (0.03) \\
61113.8 & ... & ... & 16.25 (0.02) & 15.43 (0.01) & 15.27 (0.01) & 15.07 (0.03) \\
61114.8 & ... & 15.73 (0.11) & ... & ... & ... & ... \\
61135.0 & ... & ... & ... & 16.23 (0.12) & ... & ... \\
61137.9 & ... & ... & 17.23 (0.05) & 16.34 (0.02) & 15.96 (0.02) & 15.59 (0.05) \\
61141.1 & ... & ... & 17.73 (0.26) & 16.41 (0.12) & 16.06 (0.08) & ... \\
61147.1 & ... & ... & 17.28 (0.05) & 16.61 (0.02) & 16.20 (0.02) & 15.76 (0.05) \\
61156.1 & ... & ... & 17.82 (0.05) & 16.75 (0.02) & 16.35 (0.02) & 15.88 (0.05) \\
61163.0 & ... & ... & 17.87 (0.09) & 16.85 (0.03) & 16.52 (0.02) & 16.03 (0.06) \\
\hline
\end{tabular}
\end{center}
\end{table*}

\begin{table*}[!ht]
\begin{center}
\caption{Peak times, rise times, and peak magnitudes in the LCs of SN~2026dix}
\label{tab:LC_peaks}
\begin{tabular}{c|cccc}
\hline
\hline
~ & $t_\textrm{peak}$ (MJD) & $t_\textrm{rise}$ (days) & Peak mag & Peak abs. mag \\
\hline
$B$ & 61102.8 $\pm$ 0.3 & 15.8 $\pm$ 0.3 & 16.63 $\pm$ 0.12 & $-$17.19 $\pm$ 0.16 \\
$V$ & 61106.9 $\pm$ 0.2 & 19.9 $\pm$ 0.2 & 15.61 $\pm$ 0.06 & $-$17.51 $\pm$ 0.08 \\
$g$ & 61102.5 $\pm$ 0.6 & 15.2 $\pm$ 0.6 & 15.97 $\pm$ 0.15 & $-$17.45 $\pm$ 0.16 \\
$r$ & 61109.3 $\pm$ 0.2 & 22.3 $\pm$ 0.2 & 15.36 $\pm$ 0.04 & $-$17.46 $\pm$ 0.07 \\
$i$ & 61110.4 $\pm$ 0.1 & 23.4 $\pm$ 0.1 & 15.24 $\pm$ 0.04 & $-$17.23 $\pm$ 0.07 \\
$z$ & 61109.8 $\pm$ 0.3 & 22.8 $\pm$ 0.3 & 15.08 $\pm$ 0.06 & $-$17.14 $\pm$ 0.08 \\
\hline
\end{tabular}
\end{center}
\end{table*}

\section{Reddening estimation}

\begin{figure*}[!ht]
\centering
\includegraphics[width=0.32\textwidth]{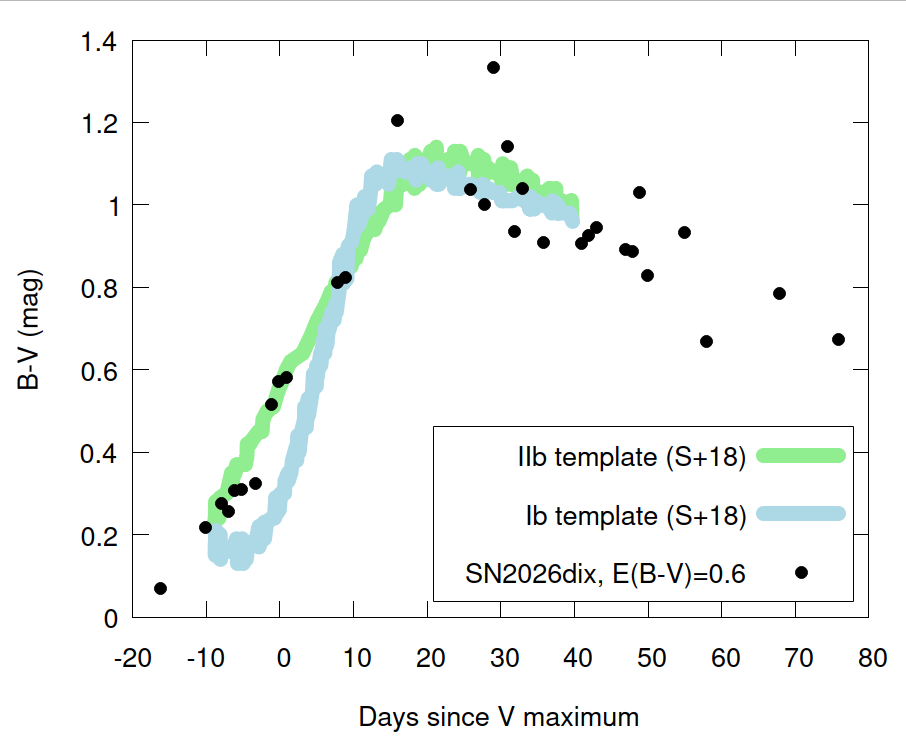}
\includegraphics[width=0.32\textwidth]{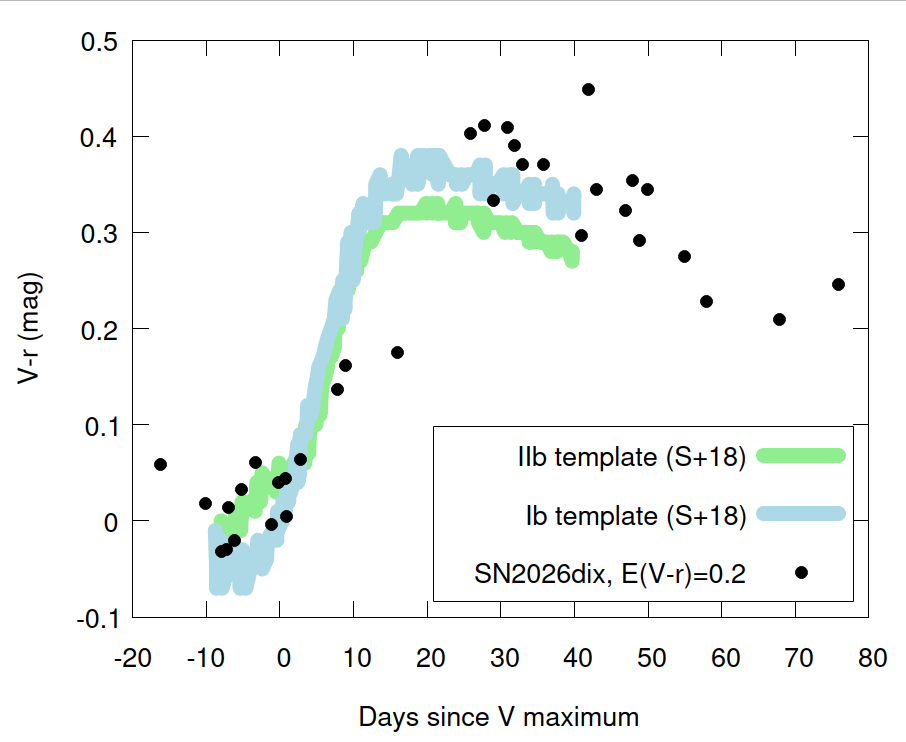}
\includegraphics[width=0.32\textwidth]{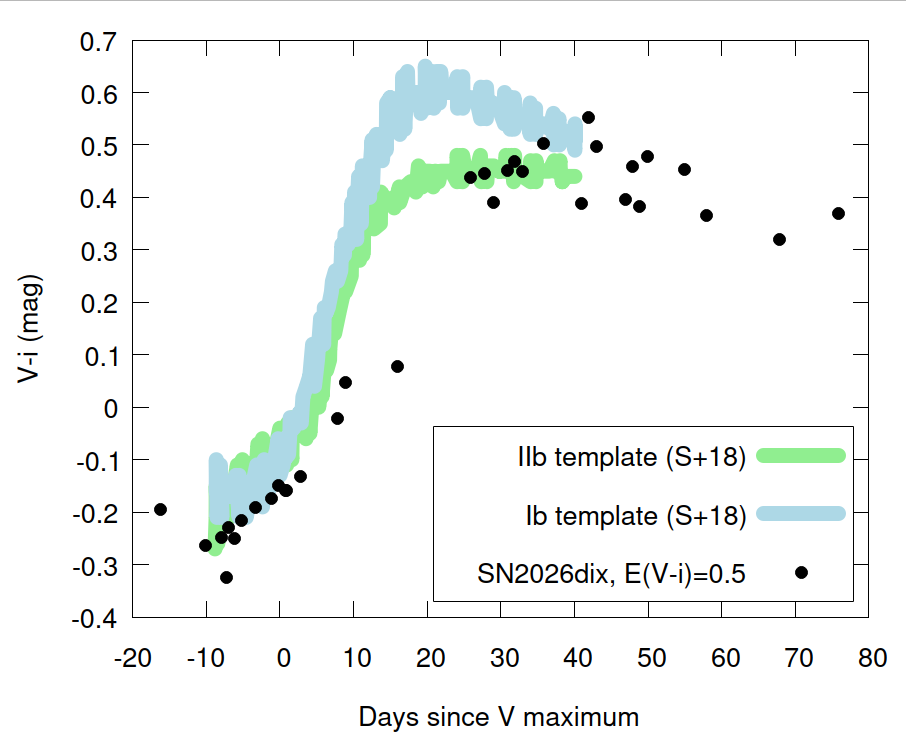}
\caption{Color evolution of SN~2026dix compared to $B-V$, $V-r$, and $V-i$ SESN color-curve templates from \cite{Stritzinger_2018}.}
\label{fig:color_ebv}
\end{figure*}

\clearpage
\section{Results of SYN++ modeling}

\begin{figure*}[!ht]
\centering
\includegraphics[width=5.5cm]{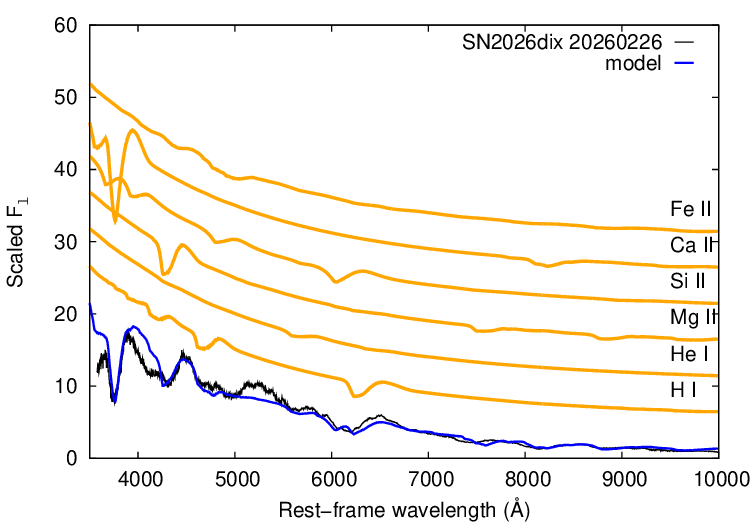}
\includegraphics[width=5.5cm]{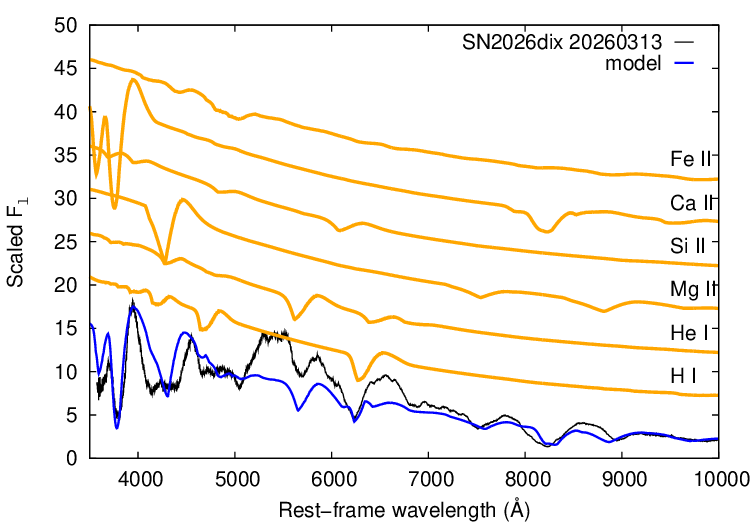}
\includegraphics[width=5.5cm]{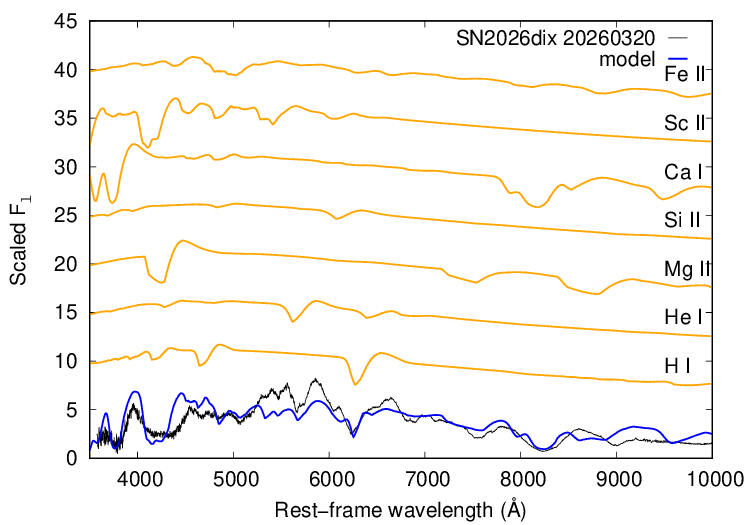}
\includegraphics[width=5.5cm]{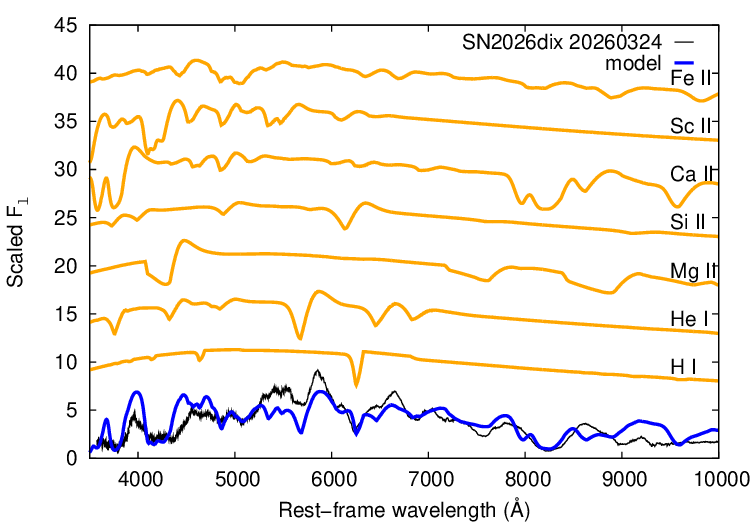}
\includegraphics[width=5.5cm]{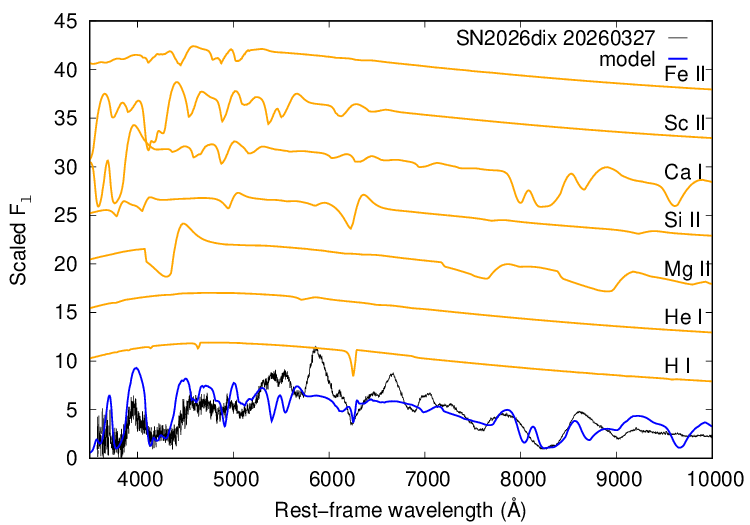}
\includegraphics[width=5.5cm]{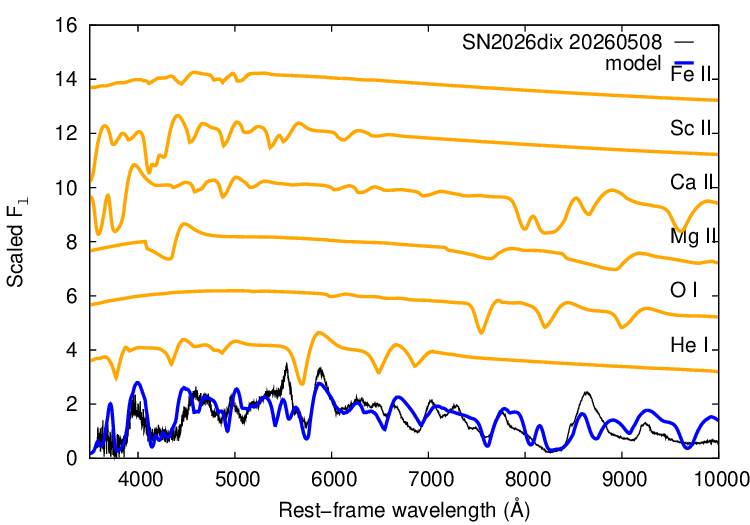}
\includegraphics[width=5.5cm]{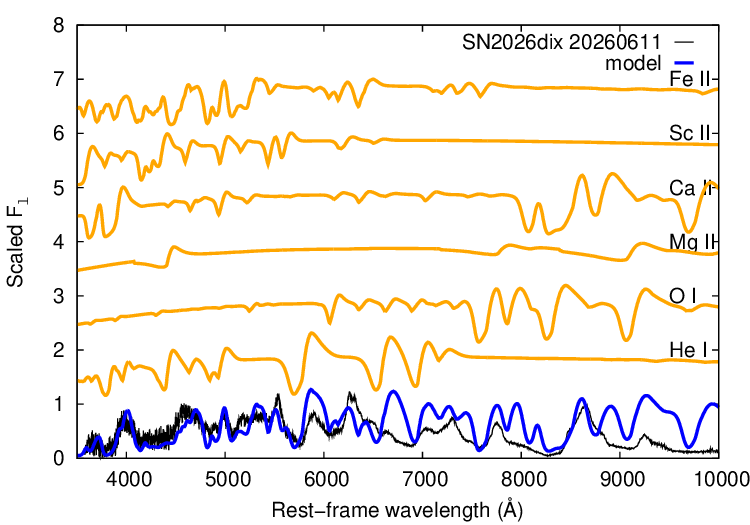} \hspace{3mm}
\includegraphics[width=0.4\columnwidth]{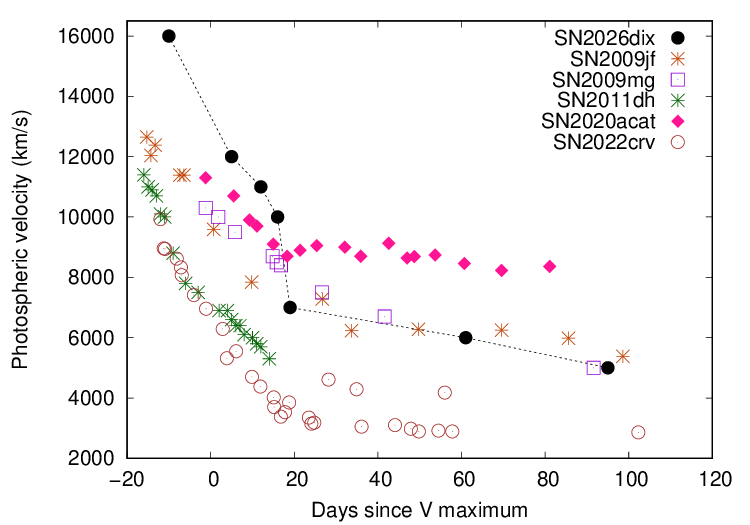}
\caption{Modeling  the spectra of SN~2026dix using the {\tt SYN++} code. The observed spectra (black) are corrected for redshift and interstellar reddening before plotting. The best-matched models are indicated in blue, while the vertically shifted orange curves represent the single-ion contributions to the spectra. {\it Bottom right:} Photospheric velocity evolution of SN~2026dix compared to that of Type Ib SN~2009jf, Type IIb SN~2011dh, and transitional Type IIb/Ib SNe 2009mg, 2020acat, and 2022crv.}
\label{fig:syn++}
\end{figure*}


\clearpage
\section{Comparative analysis of light curves}

\begin{figure*}[!h]
\centering
\includegraphics[width=0.45\textwidth]{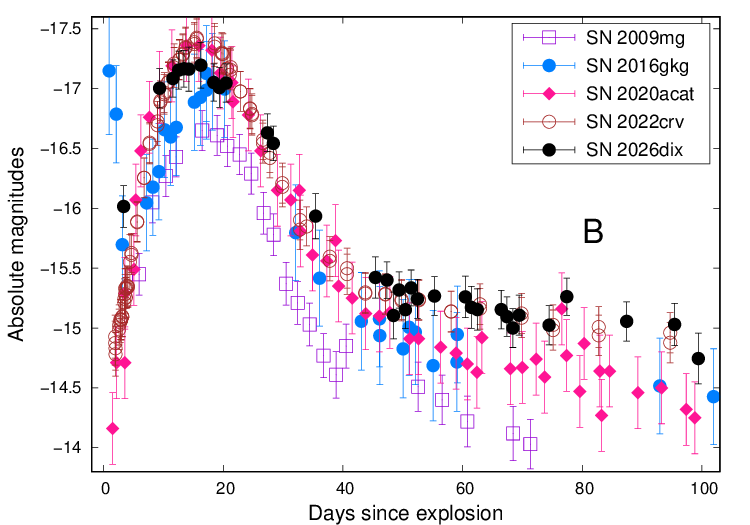}
\includegraphics[width=0.45\textwidth]{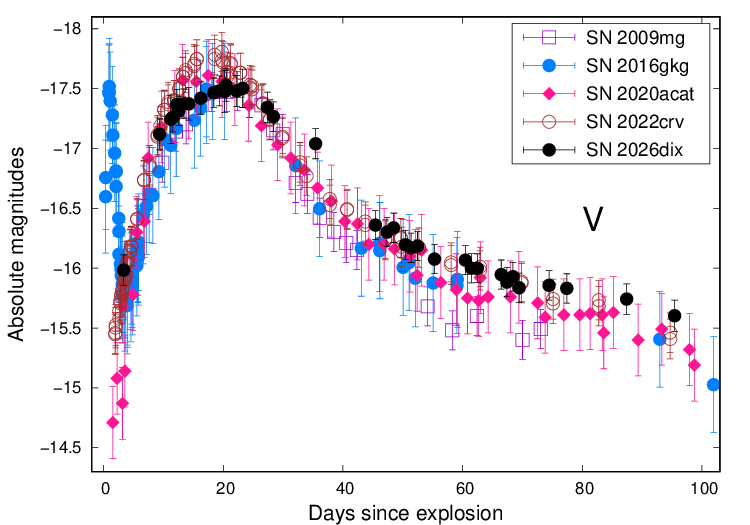}
\includegraphics[width=0.45\textwidth]{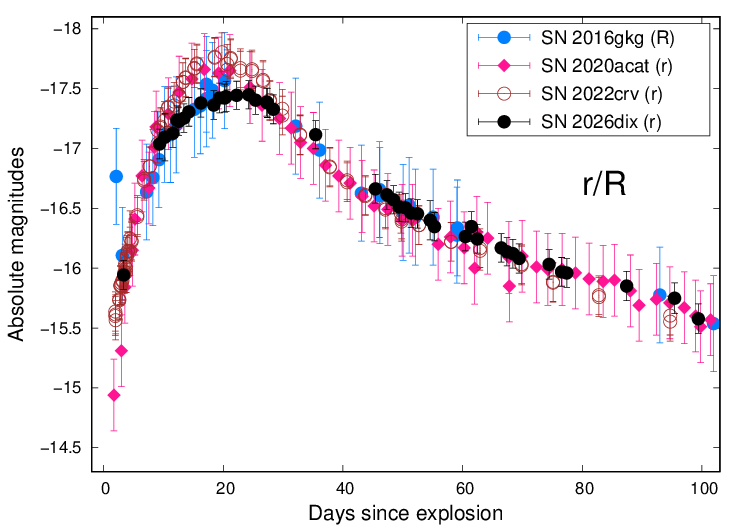}
\includegraphics[width=0.45\textwidth]{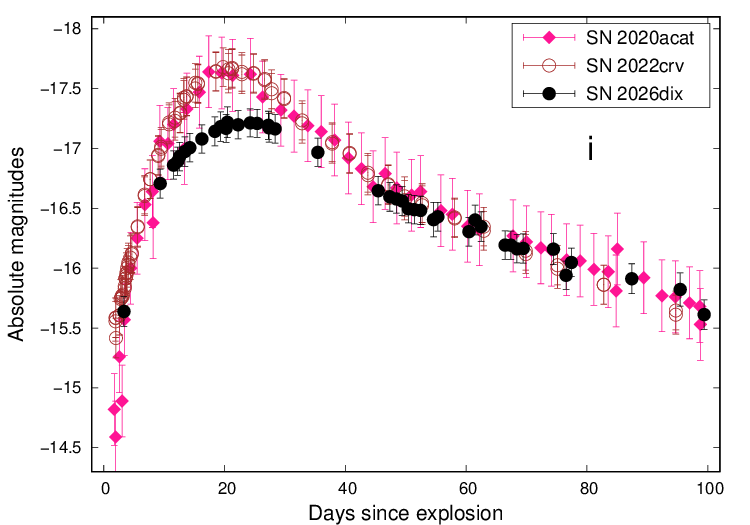}
\caption{Comparison of $BVri$ LCs of SN~2026dix to those of other IIb/Ib transitional SNe 2009mg, 2020acat, and 2022crv, as well as to SN IIb 2016gkg.}
\label{fig:22crv_lc_comp}
\end{figure*}

\clearpage
\section{Comparison of spectra and LCs with Dessart+16 models}

\begin{figure*}[!h]
\centering
\includegraphics[width=0.4\textwidth]{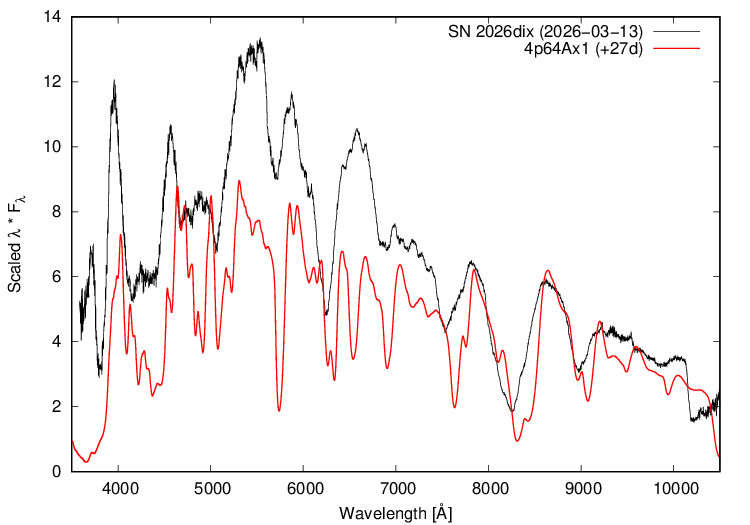}
\includegraphics[width=0.4\textwidth]{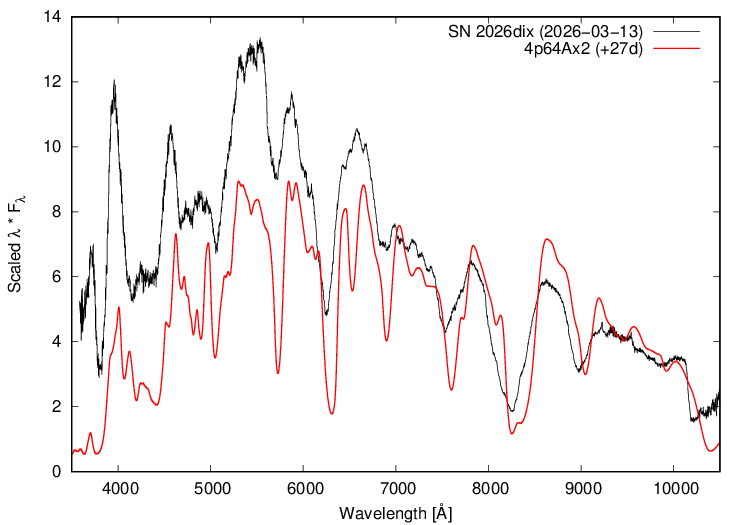}
\includegraphics[width=0.4\textwidth]{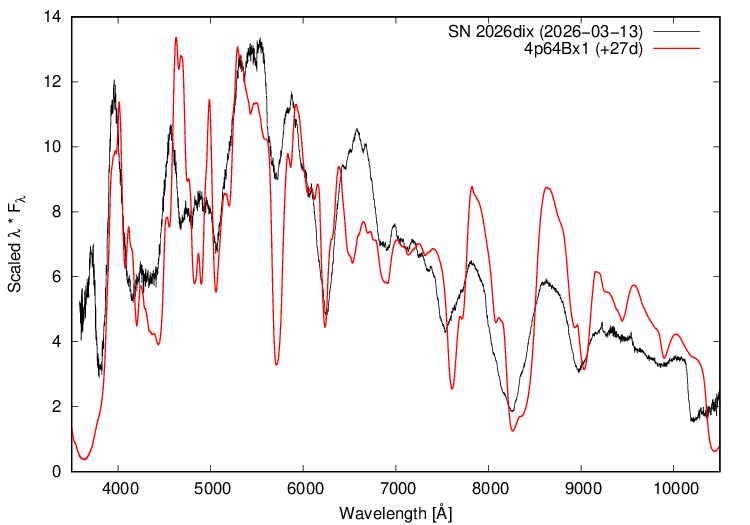}
\includegraphics[width=0.4\textwidth]{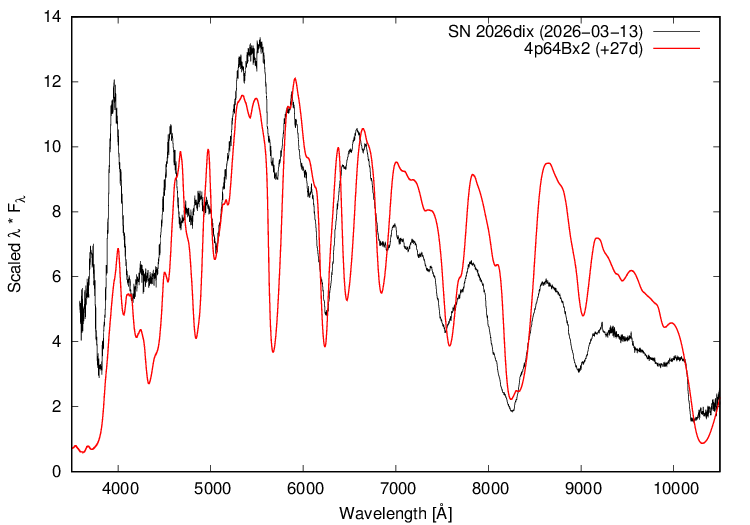}
\includegraphics[width=0.4\textwidth]{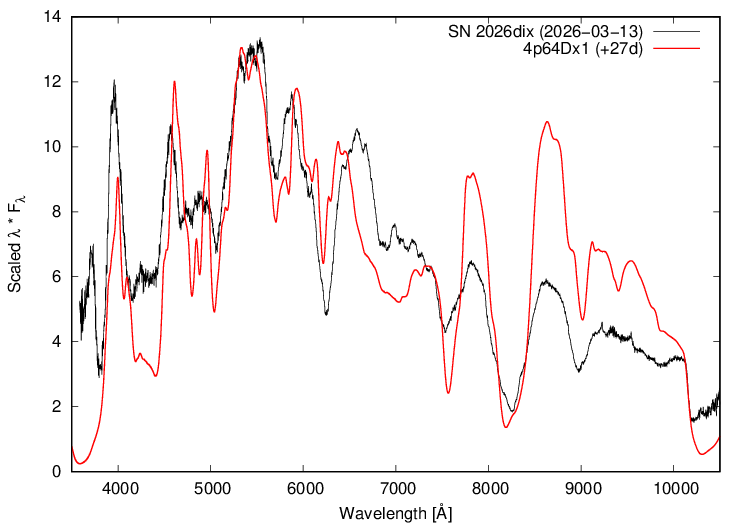}
\includegraphics[width=0.4\textwidth]{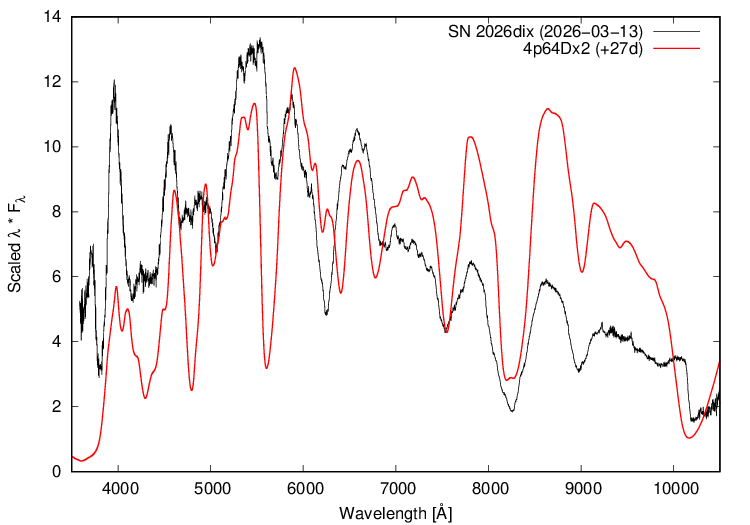}
\includegraphics[width=0.4\textwidth]{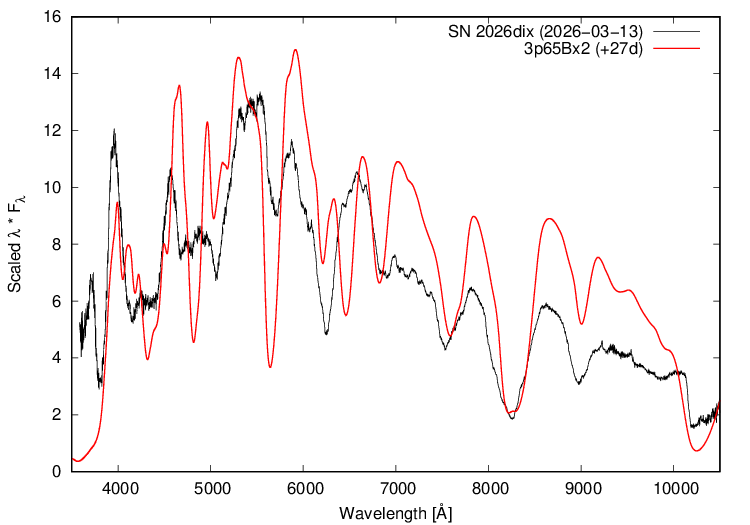}
\includegraphics[width=0.4\textwidth]{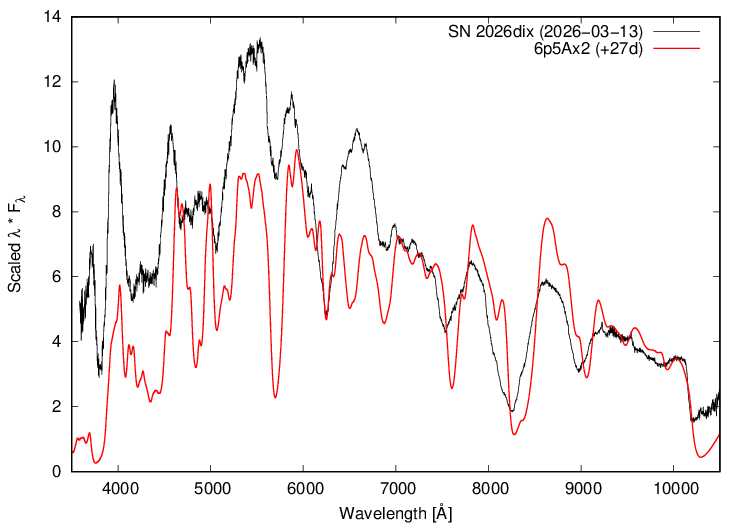}
\caption{Comparison of the near-peak-brightness (2026-03-13) spectrum of SN~2026dix with SN~IIb and SN~Ib model spectra ($3p65$, $4p64$, $6p5$) published by \cite{Dessart_2016}.}
\label{fig:D16_comp_sp_0313}
\end{figure*}

\begin{figure*}
\centering
\includegraphics[width=0.4\textwidth]{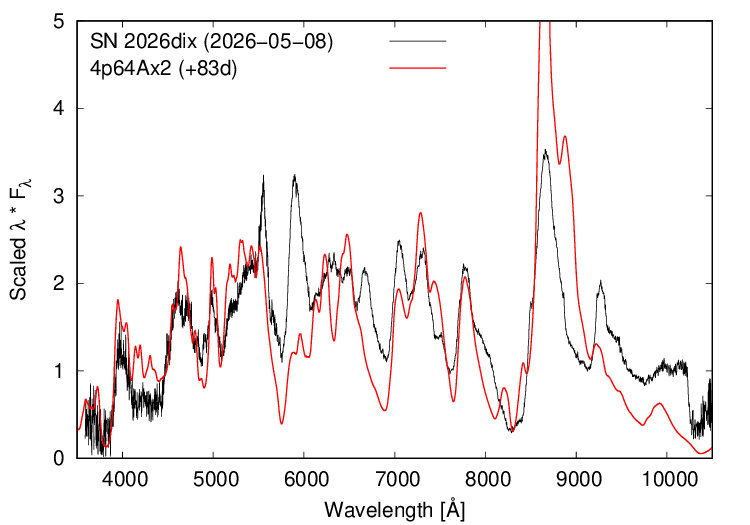}
\includegraphics[width=0.4\textwidth]{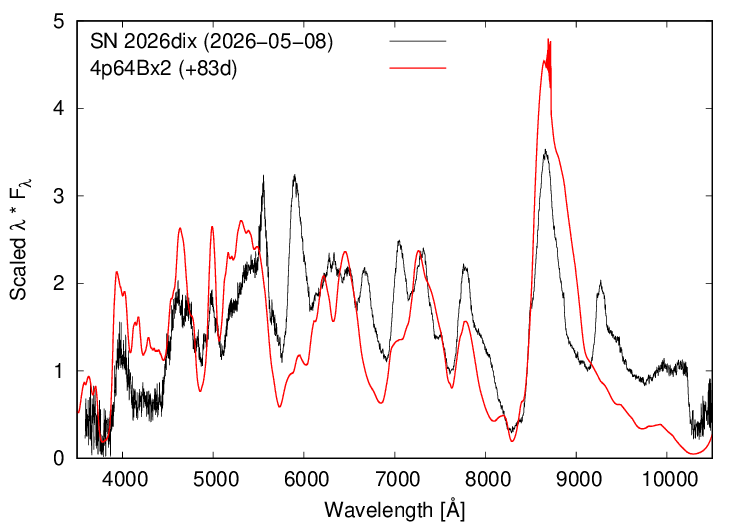}
\includegraphics[width=0.4\textwidth]{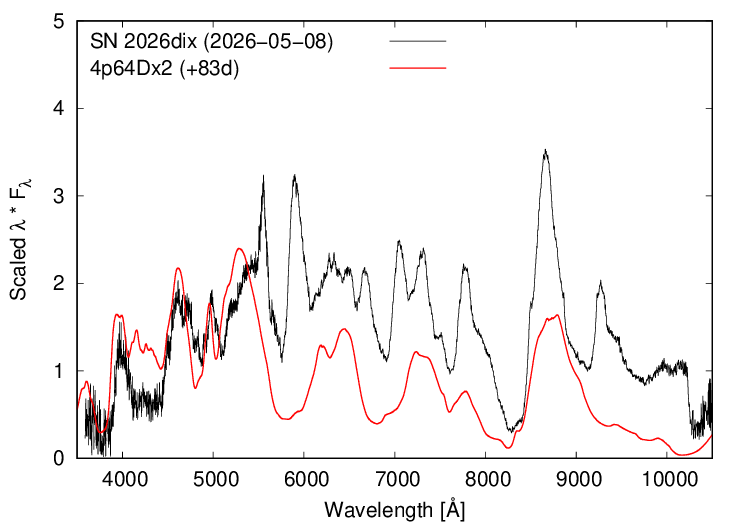}
\includegraphics[width=0.4\textwidth]{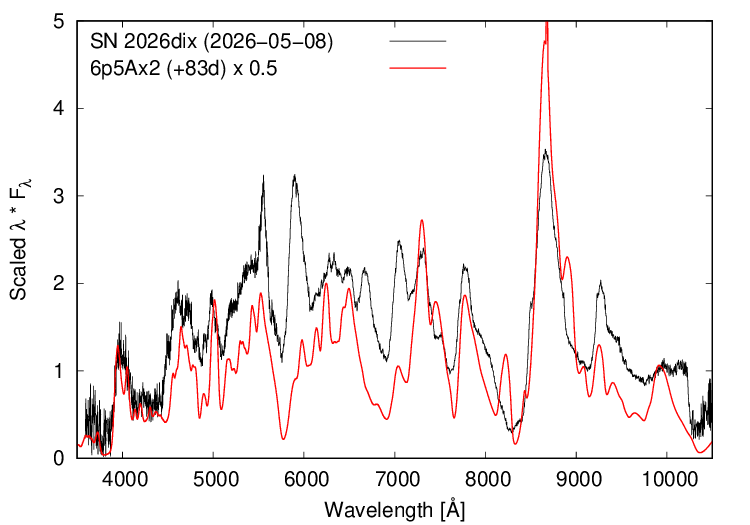}
\caption{Comparison of the 2026-05-08 spectrum of SN~2026dix with SN IIb and SN Ib model spectra ($4p64Ax2$, $4p64Bx2$, $4p64Dx2$, $6p5Ax2$) published by \cite{Dessart_2016}; the $6p5Ax2$ model spectrum is scaled by a factor of 0.5.}
\label{fig:D16_comp_sp_0508}
\end{figure*}

\begin{figure*}
\centering
\includegraphics[width=0.4\textwidth]{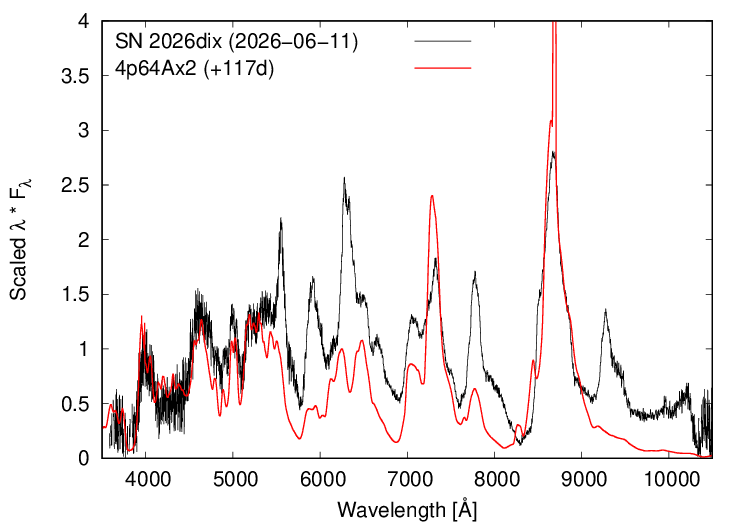}
\includegraphics[width=0.4\textwidth]{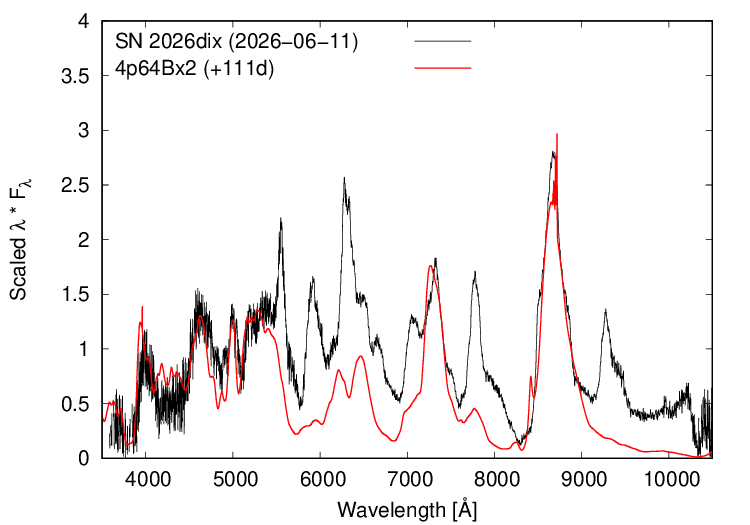}
\includegraphics[width=0.4\textwidth]{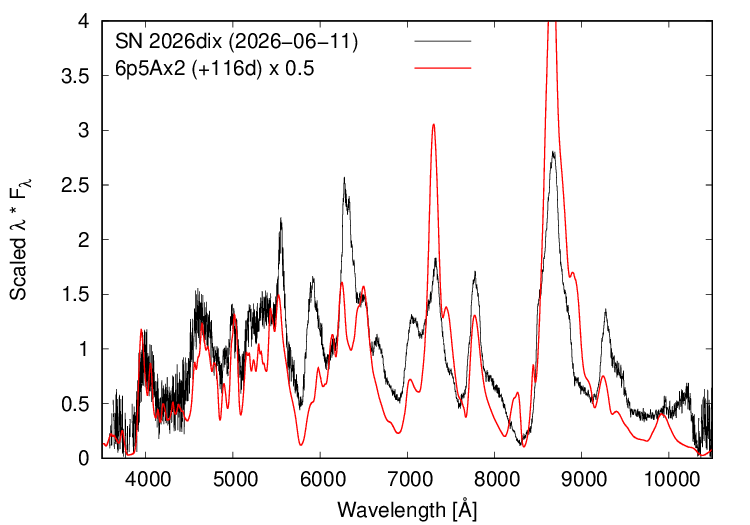}
\caption{Comparison of the 2026-06-11 spectrum of SN~2026dix with SN IIb and SN~Ib model spectra ($4p64Ax2$, $4p64Bx2$, $6p5Ax2$) published by \cite{Dessart_2016}; the $6p5Ax2$ model spectrum is scaled by a factor of 0.5.}
\label{fig:D16_comp_sp_0611}
\end{figure*}

\begin{figure*}
\centering
\includegraphics[width=0.45\textwidth]{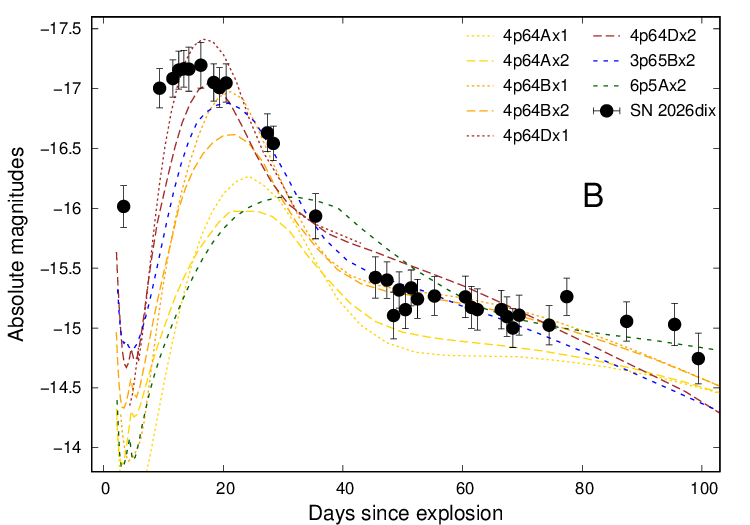}
\includegraphics[width=0.45\textwidth]{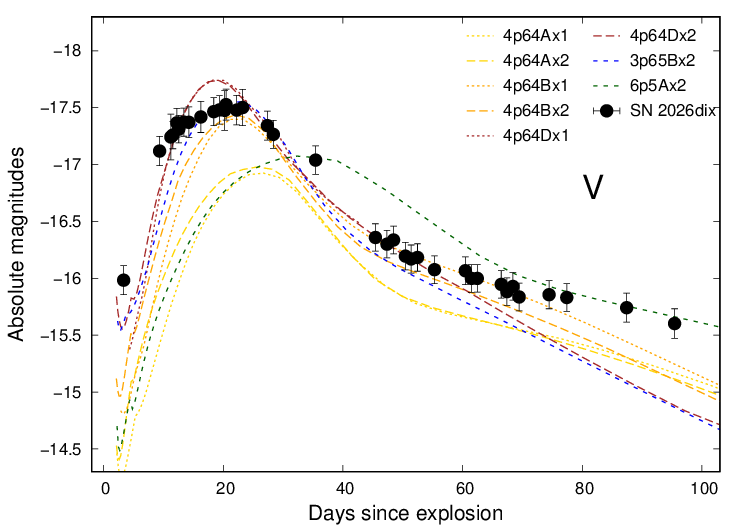}
\includegraphics[width=0.45\textwidth]{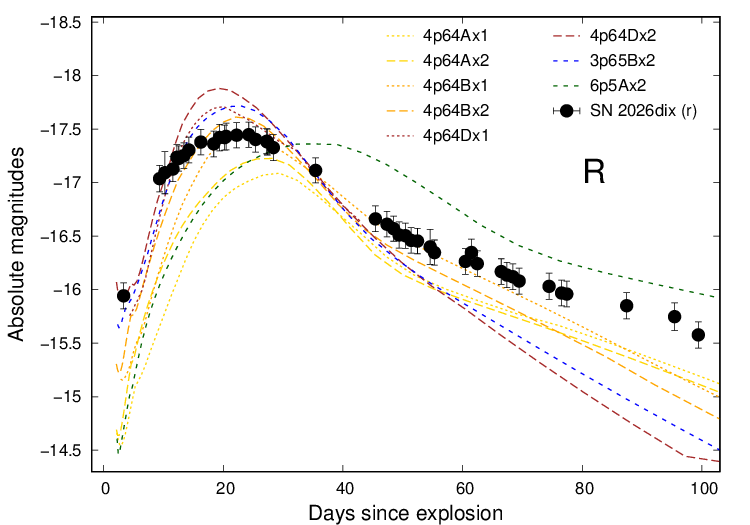}
\includegraphics[width=0.45\textwidth]{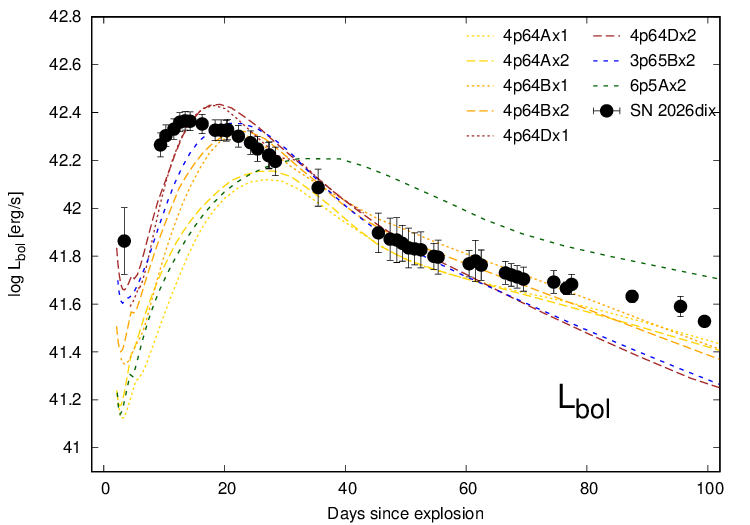}
\caption{Comparison of the absolute {\it B}, {\it V}, {\it R}, and bolometric LCs of SN~2026dix with SN IIb and SN~Ib model LCs ($3p65$, $4p64$, $6p5$) published by \cite{Dessart_2016}.}
\label{fig:D16_comp_lc}
\end{figure*}

\end{appendix}

\end{document}